\documentclass[longauth]{aa}  

\usepackage{graphicx}
\usepackage{txfonts}
\usepackage[colorlinks=true,allcolors=blue]{hyperref}
\usepackage{subcaption}
\usepackage{xcolor}
\usepackage{gensymb}
\usepackage{array}
\usepackage[export]{adjustbox}
\usepackage[utf8]{inputenc} 
\usepackage[T1]{fontenc}
\usepackage{bm}

\newcommand{\oiii}{[O\,\textsc{iii}]}
\newcommand{\nii}{[N\,\textsc{ii}]}
\newcommand{\sii}{[S\,\textsc{ii}]}
\newcommand{\oi}{[O\,\textsc{i}]}

\newcommand{\ha}{H$\alpha$}
\newcommand{\hb}{H$\beta$}
\newcommand{\hg}{H$\gamma$}

\begin{document}

\title{Complex AGN feedback in the Teacup galaxy\thanks{Based on observations made with ESO Telescopes at the La Silla Paranal Observatory under programme ID 0102.B-0107.}}
\subtitle{A powerful ionised galactic outflow, jet-ISM interaction, and evidence for AGN-triggered star formation in a giant bubble}

\author{G. Venturi\inst{1,2,3}\fnmsep\thanks{\href{mailto:giacomo.venturi1@sns.it}{giacomo.venturi1@sns.it}.},
          E. Treister\inst{1},
          C. Finlez\inst{1},
          G. D'Ago\inst{5,6},
          F. Bauer\inst{1,4,7},
          C. M. Harrison\inst{8},
          C. Ramos Almeida\inst{9,10},
          M. Revalski\inst{11},
          F. Ricci\inst{12},
          L. F. Sartori\inst{13},
          A. Girdhar\inst{8,14},  
          W. C. Keel\inst{15},
          D. Tubín\inst{16}
          }
\institute{Instituto de Astrof{\'{\i}}sica, Facultad de F{\'{i}}sica, Pontificia Universidad Cat{\'{o}}lica de Chile, Campus San Joaquín, Av. Vicuña Mackenna 4860, Macul, Santiago, Chile, 7820436
         \and
         Scuola Normale Superiore, Piazza dei Cavalieri 7, I-56126 Pisa, Italy
         \and
             INAF - Osservatorio Astrofisico di Arcetri, Largo E. Fermi 5, I-50125 Firenze, Italy
        \and
        Centro de Astroingenier{\'{\i}}a, Facultad de F{\'{i}}sica, Pontificia Universidad Cat{\'{o}}lica de Chile, Campus San Joaquín, Av. Vicuña Mackenna 4860, Macul, Santiago, Chile, 7820436
        \and
        Institute of Astronomy, University of Cambridge, Madingley Road, Cambridge, CB3 0HA, UK
        \and
        INAF – Osservatorio Astronomico di Capodimonte, Via Moiariello 16, 80131 Naples, Italy
        \and
        Millennium Institute of Astrophysics, Nuncio Monse{\~{n}}or S{\'{o}}tero Sanz 100, Of 104, Providencia, Santiago, Chile
        \and
        School of Mathematics, Statistics and Physics, Newcastle University, NE1 7RU, UK
        \and
        Instituto de Astrofísica de Canarias, Calle Vía Láctea, s/n, 38205, La Laguna, Tenerife, Spain
        \and 
        Departamento de Astrofísica, Universidad de La Laguna, 38206, La Laguna, Tenerife, Spain
        \and
        Space Telescope Science Institute, 3700 San Martin Drive, Baltimore, MD 21218, USA
        \and
        Dipartimento di Matematica e Fisica, Università Roma Tre, Via della Vasca Navale 84, 00146, Roma, Italy
        \and
        ETH Zurich, Institute for Particle Physics and Astrophysics, Wolfgang-Pauli-Strasse 27, 8093, Zurich, Switzerland
        \and
        Ludwig-Maximilians-Universit{\"a}t, Professor-Huber-Platz 2, D-80539 M{\"u}nchen, Germany
        \and
        Department of Physics and Astronomy, University of Alabama, Box 870324, Tuscaloosa, AL 35404, USA
        \and
        Leibniz-Institut für Astrophysik Potsdam (AIP), An der Sternwarte 16, 14482, Potsdam, Germany
        }
        
\date{Received 5 July 2023; accepted 29 July 2023}

\abstract
  {The $z$\,$\sim$\,0.1 type-2 QSO J1430+1339, known as {the `\object{Teacup}'}, is a complex galaxy showing a loop of ionised gas $\sim$10 kpc {in diameter}, co-spatial radio bubbles, a compact ($\sim$1 kpc) jet, and outflow activity. Its {closeness} offers the opportunity to study in detail the intricate interplay between the central supermassive black hole (SMBH) and the material in and around the galaxy, both the interstellar medium (ISM) and circumgalactic medium (CGM).}
  {
  We characterise the spatially resolved properties and effects of the galactic ionised gas outflow and compare them with those of the radio jet and with theoretical predictions to infer its acceleration mechanism.}
  {We used VLT/MUSE optical integral field spectroscopic observations to obtain flux, kinematic, and excitation maps of the extended (up to $\sim$100 kpc) ionised gas and to characterise the properties of stellar populations. We built radial profiles of the outflow properties as a function of distance from the active nucleus, from kiloparsec up to tens of kiloparsec scales, at $\sim$1 kpc resolution.}
  {We detect a velocity dispersion enhancement ($\gtrsim$300 km/s) elongated over several kiloparsecs perpendicular to the radio jet, the active galactic nucleus (AGN) ionisation lobes, and the fast outflow, similar to what is found in other galaxies hosting compact, low-power jets, indicating that the jet strongly perturbs the host ISM during its passage.
  
  We observe a decreasing trend with distance from the nucleus for the outflow properties (mass outflow rate, kinetic rate, momentum rate). The mass outflow rate drops from around 100 $M_\odot$/yr in the inner 1--2 kpc to $\lesssim$0.1 $M_\odot$/yr at 30 kpc.
  The mass outflow rate of the ionised outflow is significantly higher {($\sim$1--8 times})
  than the molecular one, in contrast with what is {often quoted} in AGN.
  {Based on energetic and morphological arguments, the driver of the multi-phase outflow is likely a combination of AGN radiation and the jet, or AGN radiation pressure on dust alone.}
  The  outflow mass-loading factor is $\sim$5--10 and the molecular gas depletion time due to the {multi-phase} outflow is $\lesssim$10$^8$ yr, indicating that the outflow can significantly affect the star formation and the gas reservoir in the galaxy. However, the fraction of the ionised outflow that is able to escape the dark matter halo potential is {likely} negligible.

  We detect blue-coloured continuum emission co-spatial with the ionised gas loop. {Here, stellar populations are younger} ($\lesssim$100--150 Myr) than in the rest of the galaxy ($\sim$0.5--1 Gyr). This constitutes possible evidence for star formation triggered at the edge of the bubble due to the compressing action of the jet and outflow (`positive feedback'), {as predicted by theory.}

  All in all, the Teacup constitutes a rich system in which AGN feedback from outflows and jets, in both its negative and positive flavours, co-exist.
  }
  {}
  
\keywords{Galaxies: quasars: individual: Teacup -- Galaxies: jets -- Galaxies: active -- Techniques: imaging spectroscopy}

\titlerunning{}           
\authorrunning{G. Venturi et al.}

\maketitle

\section{Introduction}
J1430+1339, known as the {\object{Teacup}} (\citealt{Keel:2012a}), is an active galaxy at $z$ $\simeq$ 0.08506 hosting a {highly obscured ($N_\mathrm{H}$ $\sim$ 5$\times$10$^{23}$ cm$^{-2}$; \citealt{Lansbury:2018a})} type-2 quasar (QSO; \citealt{Villar-Martin:2014a,Harrison:2014a}), {with a bolometric luminosity of $L_\mathrm{AGN}$ $\sim$ 10$^{45-46}$ erg/s (\citealt{Harrison:2015a,Lansbury:2018a})}.
{With a mass of $\log (M_\star/M_\odot)$ $\simeq$ 11.15 (\citealt{Ramos-Almeida:2022a}) and a star formation rate (SFR) of $\sim$8--12 $M_\odot$/yr \citep{Jarvis:2020a, Ramos-Almeida:2022a}, corresponding to a specific SFR of $\log((\mathrm{SFR}/M_\star)/\textrm{Gyr}^{-1})$ $\sim$ --1.2 to --1.1, the galaxy lies $\sim$0.6--0.8 dex above the `main sequence' of star-forming galaxies for its redshift and stellar mass (e.g. \citealt{Brinchmann:2004aa,Speagle:2014aa,Popesso:2023aa}).}

Its nickname, the Teacup, is due to the presence of a closed loop of ionised gas to the E of the nucleus, resembling the handle of a teacup, with the main body of the galaxy in continuum emission representing the {cup} itself (Fig. \ref{fig:rgb}). The ionised gas `handle' is particularly prominent in \oiii, extending up to distances of $\sim$10 kpc (\citealt{Keel:2012a, Harrison:2015a}), and it is dominated by ionisation from the AGN photons, as indicated by longslit observations (\citealt{Keel:2015a}). The ionised nebula extends up to {$\sim$140 kpc ($\sim$70 kpc per side; 
\citealt{Villar-Martin:2018a,Villar-Martin:2021a,Moiseev:2023a}).} This lies among the largest ionised nebulae around active galaxies known so far, especially at low $z$ (see \citealt{Villar-Martin:2018a} and references therein).
The galaxy is bulge-dominated and shows disturbed and shell-like features in optical continuum emission, indicative of past merger activity (\citealt{Keel:2015a}).
The source is classified as radio quiet in the $\nu L_\nu(1.4\,\mathrm{GHz})$--$L_\mathrm{\oiii}$ plane, having \oiii\ and radio luminosities of $L_\mathrm{\oiii}$ $\sim$ 5 $\times$ $10^{42}$ erg\,s$^{-1}$ and $L_\mathrm{1.4\,GHz}$ $\sim$ 5 $\times$ $10^{23}$ W\,Hz$^{-1}$, respectively (\citealt{Harrison:2015a,Jarvis:2019a}). Nevertheless, it exhibits lively radio activity, with bipolar lobes in the E-W direction on scales of $\sim$10 kpc per side (\citealt{Harrison:2015a,Jarvis:2019a}). 
Integral field spectroscopic (IFS) observations allowed for a kiloparsec-scale high-velocity component ($v$ $\sim$ --800 km\,s$^{-1}$) of the ionised gas to be identified in both the optical (\citealt{Harrison:2014a, Harrison:2015a}) and the near-IR (\citealt{Ramos-Almeida:2017a}) bands, which has been interpreted as an outflow, after a broad component indicative of an outflow was found in \oiii\ in the SDSS spectrum of the galaxy (FWHM\footnote{Full width at half maximum.} $\sim$ 1000 km\,s$^{-1}$; \citealt{Mullaney:2013a, Villar-Martin:2014a}).

Specifically, \cite{Harrison:2015a} performed a comparison between the optical and radio emissions of the Teacup galaxy, exploiting IFS observations from the European Southern Observatory's (ESO) VIsible MultiObject Spectrograph (VIMOS) at the Very Large Telescope (VLT), \textit{Hubble Space Telescope} (\textit{HST}) imaging, and multi-band Karl G. Jansky Very Large Array (VLA) radio imaging.
The radio lobes appear as bubbles filled with diffuse emission.
The eastern radio bubble brightness peaks on its eastern edge, where it traces the shape of the ionised gas arc.
The ionised gas is much fainter at the location of the western radio bubble, {potentially} explained by a radio-optical ionising flux misalignment or by a lower gas density.
The ionised gas kinematic map shows a $\sim$20$\degree$ misalignment between the kinematic axis and the radio bubbles.
The highest resolution radio images (HPBW\footnote{Half-power beamwidth.} $\sim$ 0.6 kpc) reveal that the central core is constituted by two unresolved sources, a brighter central one and a fainter one at $\sim$0.8 kpc to the NE, named HR-A and HR-B by \cite{Harrison:2015a}, respectively.
These two unresolved sources have been interpreted as likely being a small-scale radio jet, with HR-B being a hot spot (as commonly observed in compact jets; e.g. \citealt{Leipski:2006a, Morganti:2015a}).
\cite{Harrison:2015a} interpret {the $\sim$10 kpc per side radio bubbles} to be the result of the action of the compact radio jet, based on the analogy with other similar cases, such as Mrk 6 (7.5 kpc radio bubbles and a $\sim$1 kpc radio jet; \citealt{Kharb:2006a}), as predicted by jet-driven bubble models (e.g. \citealt{Sutherland:2007a, Wagner:2012a}), or to nuclear AGN winds, as in the energy-conserving expanding bubble models (e.g. \citealt{Faucher-Giguere:2012a, Zubovas:2012aa, Wagner:2013a}).
High-resolution {\it Chandra} X-ray imaging reveals a loop co-spatial with the eastern ionised gas handle and the edges of the radio bubble, whose X-ray spectral properties are consistent with shocked thermal gas (\citealt{Lansbury:2018a}), {supporting a} jet or AGN wind origin for the bubble inflating mechanism.

Near-infrared (NIR) K-band observations with the Spectrograph for INtegral Field Observations in the Near Infrared (SINFONI) at VLT (\citealt{Ramos-Almeida:2017a}) reveal a nuclear blueshifted broad component (FWHM $\sim$ 1600--1800 km\,s$^{-1}$) in both the hydrogen recombination lines (Pa$\alpha$, Br$\delta$, and Br$\gamma$) and the [Si\,\textsc{vi}] $\lambda$1.963$\mu$m coronal line. The larger line width of the ionised gas NIR lines compared to their optical counterparts in the same nuclear region suggests that {the NIR lines, which are less affected by extinction compared to the optical ones, provide} information on the outflowing gas closer to the AGN.
\cite{Ramos-Almeida:2017a} found the presence of ionised gas {from Pa$\alpha$} with FWHM $>$ 250 km\,s$^{-1}$ { across the NE bubble and SW fan,} out to 5.6 kpc from the nucleus, indicative of an extended outflow.
{Here, the above work also reported the tentative detection of an {additional} very broad (FWHM $\sim$ 3000 km\,s$^{-1}$) component.}

Atacama Large Millimeter/submillimeter Array (ALMA) CO(2-1) observations of the central 3$''$\,$\times$\,3$''$ found a 1.3 kpc-scale rotating disc of molecular gas oriented approximately in the N-S direction, with mass $M_\mathrm{H_2}$ $\sim$ 6$\times$10$^9$ $M_\odot$ (consistent with the value inferred in \citealt{Jarvis:2020a} from single-dish APEX observations),
and a molecular outflow in the E-W direction with velocities up to 250 km/s, a mass of $M_\mathrm{out}$ = 3.12$\times$10$^7$ $M_\odot$, and a (de-projected) mass outflow rate of $\dot{M}_\mathrm{out}$ = 15.8 $M_\odot$/yr \citep{Ramos-Almeida:2022a}.
{Combining the same CO(2-1) data with additional new ALMA CO(3-2) observations}, \cite{Audibert:2023a} found evidence for jet-induced molecular gas excitation and turbulence perpendicular to the compact radio jet. {They also explored different outflow scenarios, from more to less conservative, and reported corresponding molecular mass outflow rates between 15 and 41 $M_\odot$/yr.}

In this work, we present new optical integral field observations from the Multi Unit Spectroscopic Explorer (MUSE; \citealt{Bacon:2010aa}) at VLT, which allowed us to simultaneously map the distribution, kinematics, and excitation properties of the ionised gas out to large scales, $\sim$100 kpc (corresponding to the 1$'$\,$\times$\,1$'$ field of view of MUSE at the galaxy distance, $D_L$ $\sim$ 404 Mpc\footnote{From NASA/IPAC Extragalactic Database (NED).}; the physical scale is 1.66 kpc/arcsec). We compared the MUSE observations with the VLA radio images from \cite{Harrison:2015a}, to study the interplay between the radio-emitting material and the ionised gas in the optical.

{In Sect. \ref{sec:data_red+anal} we describe the reduction and analysis of the VLT/MUSE observations, in Sect. \ref{sec:muse_maps} we present and comment on the MUSE maps of ionised gas obtained from the previous step, in Sect. \ref{sec:ionised_outfl} we characterise and discuss the spatially resolved properties of the galactic outflow, in Sect. \ref{sec:enhanced_sigma} we discuss the broad velocity dispersions detected perpendicular to the jet and AGN ionisation axes, in Sect. \ref{sec:stars_handle} we present possible evidence for `positive' AGN feedback, and finally in Sect. \ref{sec:concl} we summarise the conclusions of this work.}
Throughout this study, a $H_0$ = 70 km s$^{-1}$ Mpc$^{-1}$, $\Omega_\mathrm{M}$ = 0.3, and $\Omega_\Lambda$ = 0.7 cosmology is adopted.
We adopt the standard astronomical orientation in the maps (N is up, and E is to the left).

\section{Data reduction and analysis}\label{sec:data_red+anal}
The Teacup was observed with VLT/MUSE in its wide field mode (WFM), which covers $\simeq$1$'$\,$\times$\,1$'$ with a spatial sampling of 0.2$''$/spaxel, and extended spectral mode, spanning the range 4600--9350 \AA.
It was observed as part of two observing programmes, a seeing-limited one (ID 0102.B-0107, PI L. Sartori) and a seeing-enhancer ground layer adaptive optics (GLAO)-assisted one (ID 0103.B-0071, PI C. Harrison).
In this work we only employ the observations from the former programme, the seeing-limited one ($\sim$0.5$''$--0.6$''$ at zenith), 
since visually assessing that their spatial quality was significantly better than that of the GLAO-assisted ones, even after the GLAO correction, due to the worse seeing conditions of these latter ones ($\sim$1.0$''$--1.1$''$ at zenith).
The employed observations consist of six observations of 950 s on-target each, for a total exposure time of 5700 s. {Consecutive} observations were rotated by 90\degree\ relative to one another.

{Regarding the VLA radio observations that we compare with those from VLT/MUSE, we employ the C-band 5.12 GHz image, with a resolution of HPBW = 1.13$''$\,$\times$\,1.04$''$, and the high-resolution (HPBW = 0.37$''$\,$\times$\,0.24$''$) 6.22 GHz image.
For the reduction of these radio data, we refer to \cite{Harrison:2015a} where they were originally presented.}

\subsection{Data reduction}\label{ssec:data_red}
\begin{figure*}
    \centering
        \includegraphics[width=0.45\textwidth,valign=m,trim={6cm 1.5cm 7cm 2cm},clip]{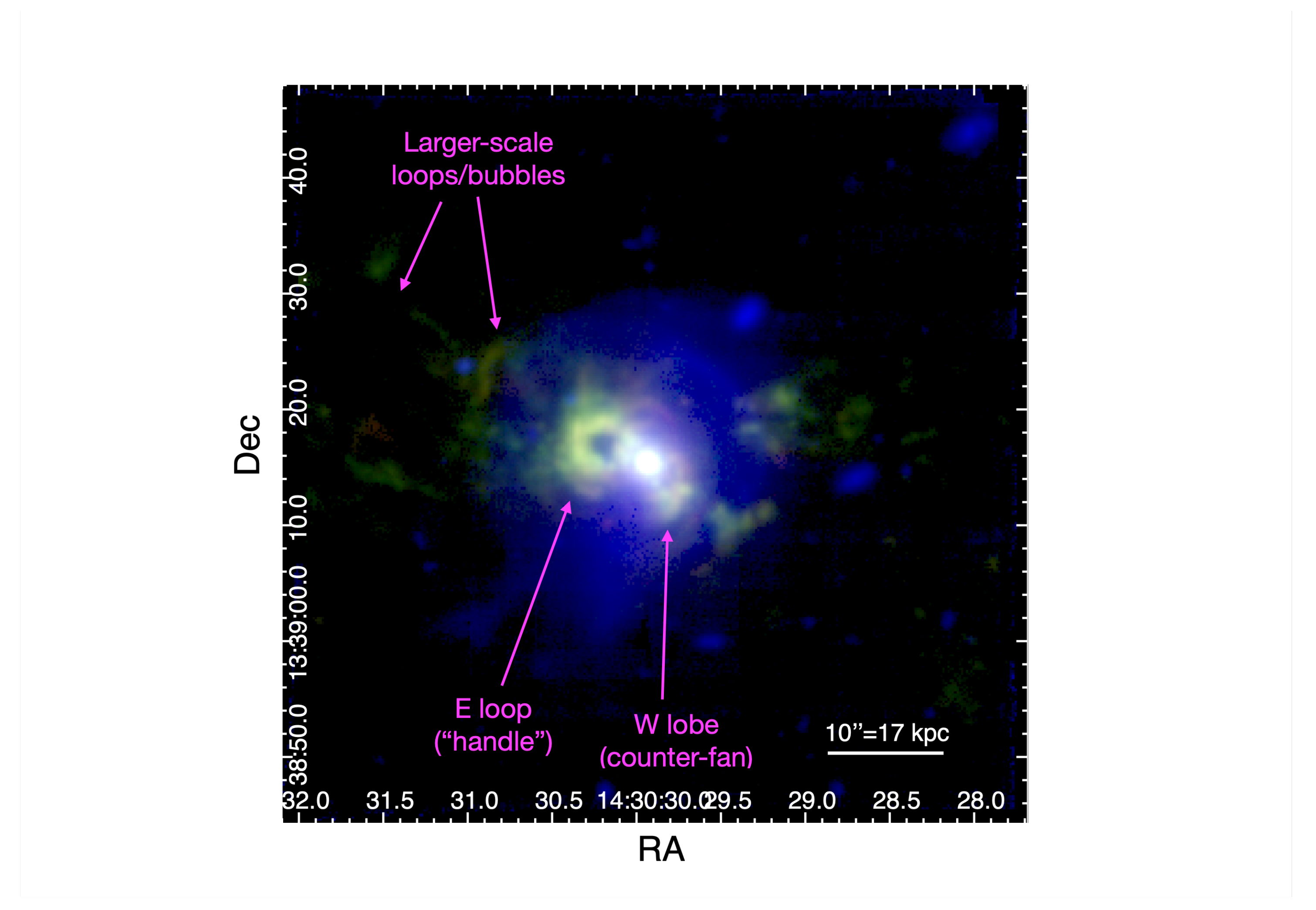}
    \includegraphics[width=0.5\textwidth,valign=m]{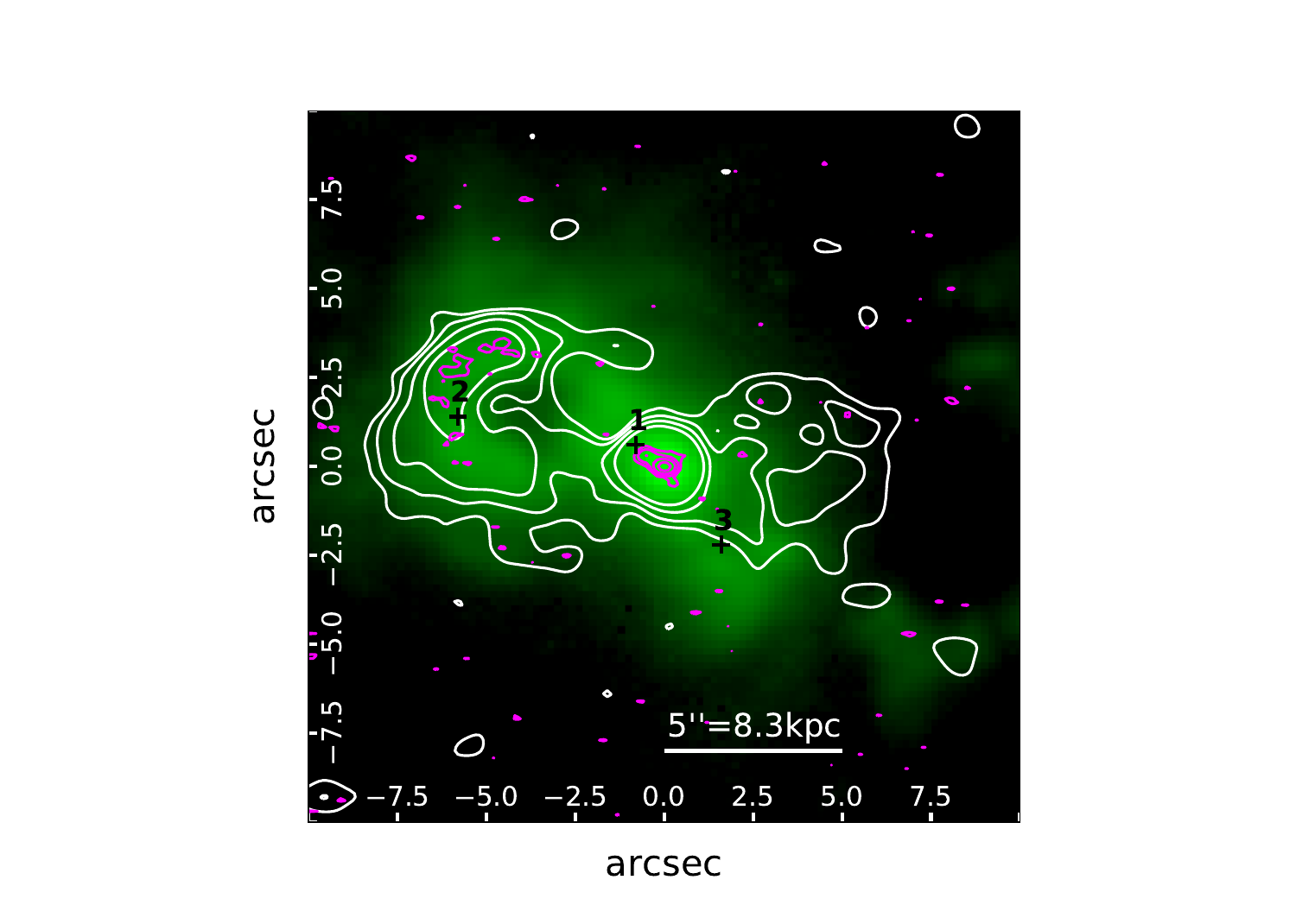}
\caption{Optical (emission lines and continuum) versus radio emission of the Teacup galaxy. Left panel: False-colour image from VLT/MUSE. \oiii\ ionised gas emission is reported in green, \ha\ in red, and {line-free continuum emission averaged between 5500--6700 \AA\ (observed wavelengths)} in blue. The colour intensity scale is the same for \oiii\ and \ha. {The main spatial features mentioned in the text are labelled.}
{Right panel: $20''\times20''$ zoomed-in \oiii\ emission (same as in left panel), with the contours of the VLA 5.12 GHz (white) and highest-resolution 6.22 GHz (magenta) radio images from \cite{Harrison:2015a} overlaid. For the former, contour levels are 35.0, 66.0, 124, and 234 $\mu$Jy/beam; for the latter, they are 40, 102, 261, and 668 $\mu$Jy/beam.
The black numbered `+' symbols mark the regions from which the spectra reported in Fig. \ref{fig:spectra} were extracted.}}
\label{fig:rgb}	
\end{figure*}

We performed the data reduction using the ESO MUSE pipeline (\citealt{Weilbacher:2020a}; version 2.8.1), by employing the software ESO Reflex (Recipe flexible execution workbench; \citealt{Freudling:2013a}), that gives a graphical and automated way to execute the Common Pipeline Library (CPL; \citealt{Banse:2004a}; \citealt{ESOCPL2014}) reduction recipes with EsoRex (ESO Recipe Execution Tool; \citealt{ESOCPL2015}), within the Kepler workflow engine (\citealt{Altintas:2004a}). The data {were corrected} for bias, flat field, illumination, wavelength calibration, flux calibration, geometric reconstruction of the data cube, sky subtraction, and exposure combination.
In order to maximise the on-target time, the observations did not include dedicated sky observations, {such that the sky background} was extracted from empty regions in the science field itself. {Notably}, the default pipeline sky extraction, which automatically creates a spatial mask for sky extraction based on the flux of the collapsed continuum image, did not give optimal results for the analysis of this complex system. The morphology of the continuum and ionised gas line emission is very different in the Teacup (see Fig. \ref{fig:rgb}), and the automatic masking based on the continuum left regions containing significant gas line emission unmasked, that resulted in an extracted sky spectrum containing gas emission lines, in particular \oiii\ $\lambda\lambda$4959,5007, \hb, \ha, \nii\ $\lambda\lambda$6548,6584, and \sii\ $\lambda\lambda$6716,6731. This resulted in subtraction of these emission lines from the cube together with the sky. We thus followed a different approach, by creating a custom sky mask based on \oiii\ $\lambda$5007 {(since this is the strongest and most extended emission line in the Teacup)}. The mask included the 10$\%$ weakest spaxels in \oiii\ flux, after having smoothed the \oiii\ image with a Gaussian kernel having a 3-spaxel standard deviation. The final sky mask was obtained with a logical `AND' between this \oiii\ mask and the mask produced by the pipeline that included 15$\%$ weakest spaxels in the white-light continuum image, so as to select only spaxels weak in both \oiii\ and continuum.
The resulting data cube after having employed this custom sky mask for the sky subtraction showed no over-subtraction of emission lines of the source.

{Unfortunately,} this final data cube contained sky emission line residuals close to the wavelengths of the \sii\ doublet which {potentially} affected its analysis {(the other emission lines of interest were instead not affected by sky residuals), so} we produced another cube to be used exclusively for the analysis of \sii\ (details in Sect. \ref{ssec:data_anal}).
This cube was obtained by running the Zurich Atmosphere Purge (\textsc{ZAP}; \citealt{Soto:2016a}) code on the reduced data cube just described. \textsc{ZAP} is a Python code aimed at removing residual sky contamination from the MUSE data cubes by performing principal components analysis (PCA).
We preferred using the reduced data cube (without \textsc{ZAP} execution) for the rest of the analysis, since PCA introduces further noise in the data and {can} also create artefacts. We checked that \sii\ was not affected by artefacts and that its line profiles were consistent among the two cubes.

\begin{figure}
    \centering
\includegraphics[trim={0 0 0.8cm 0},clip,width=\columnwidth]{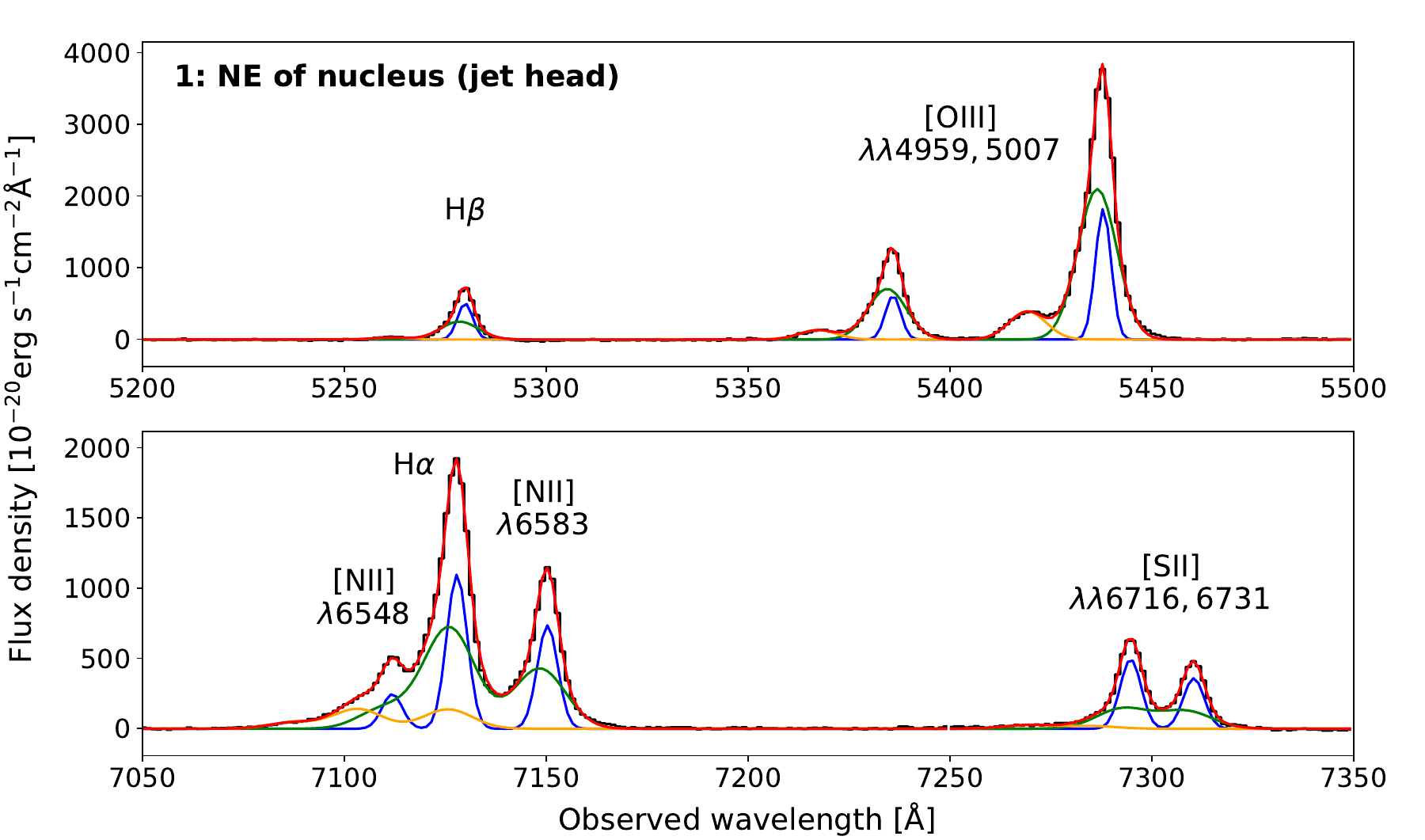}
\includegraphics[trim={0 0 0.8cm 0},clip,width=\columnwidth]{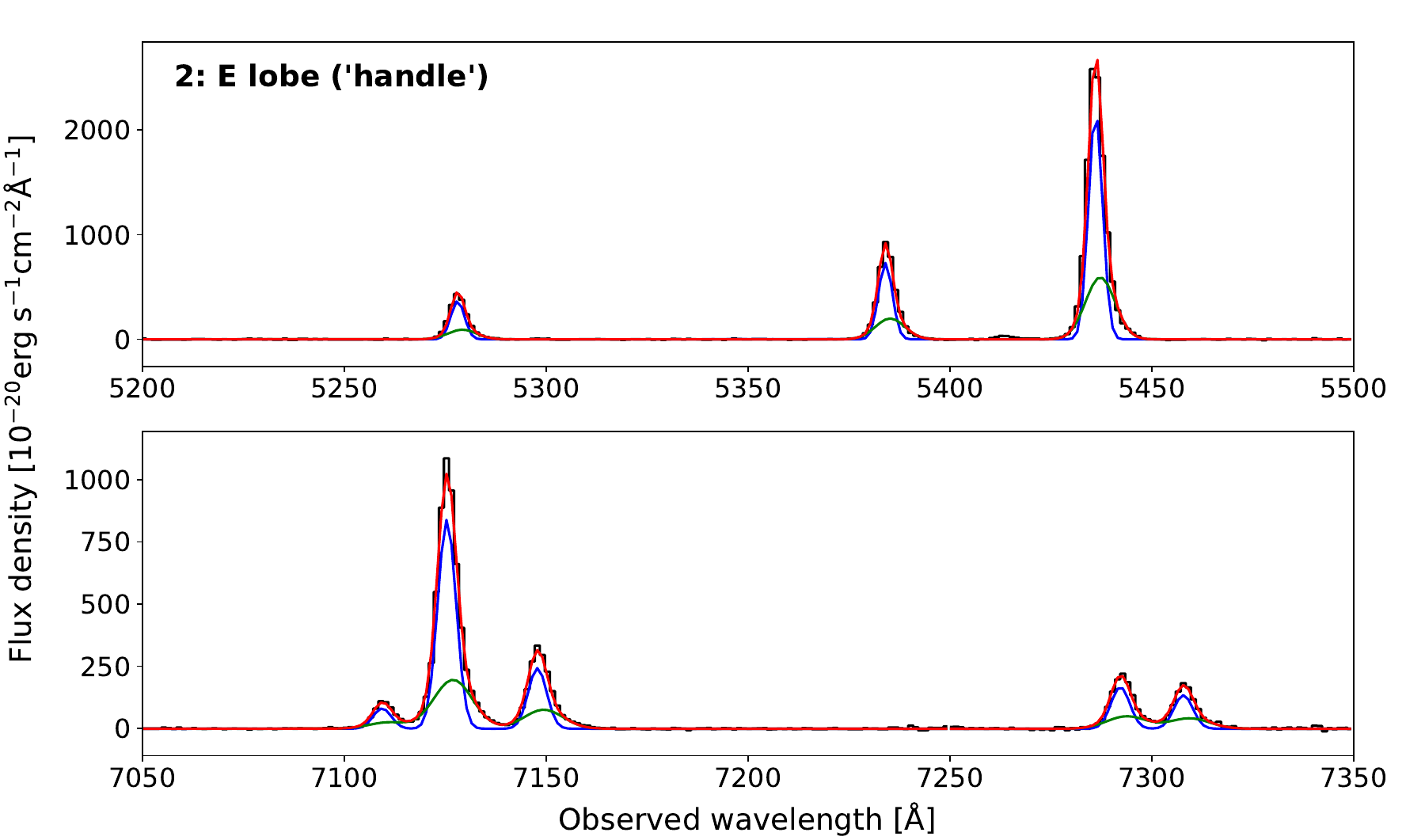}
\includegraphics[trim={0 0 0.8cm 0},clip,width=\columnwidth]{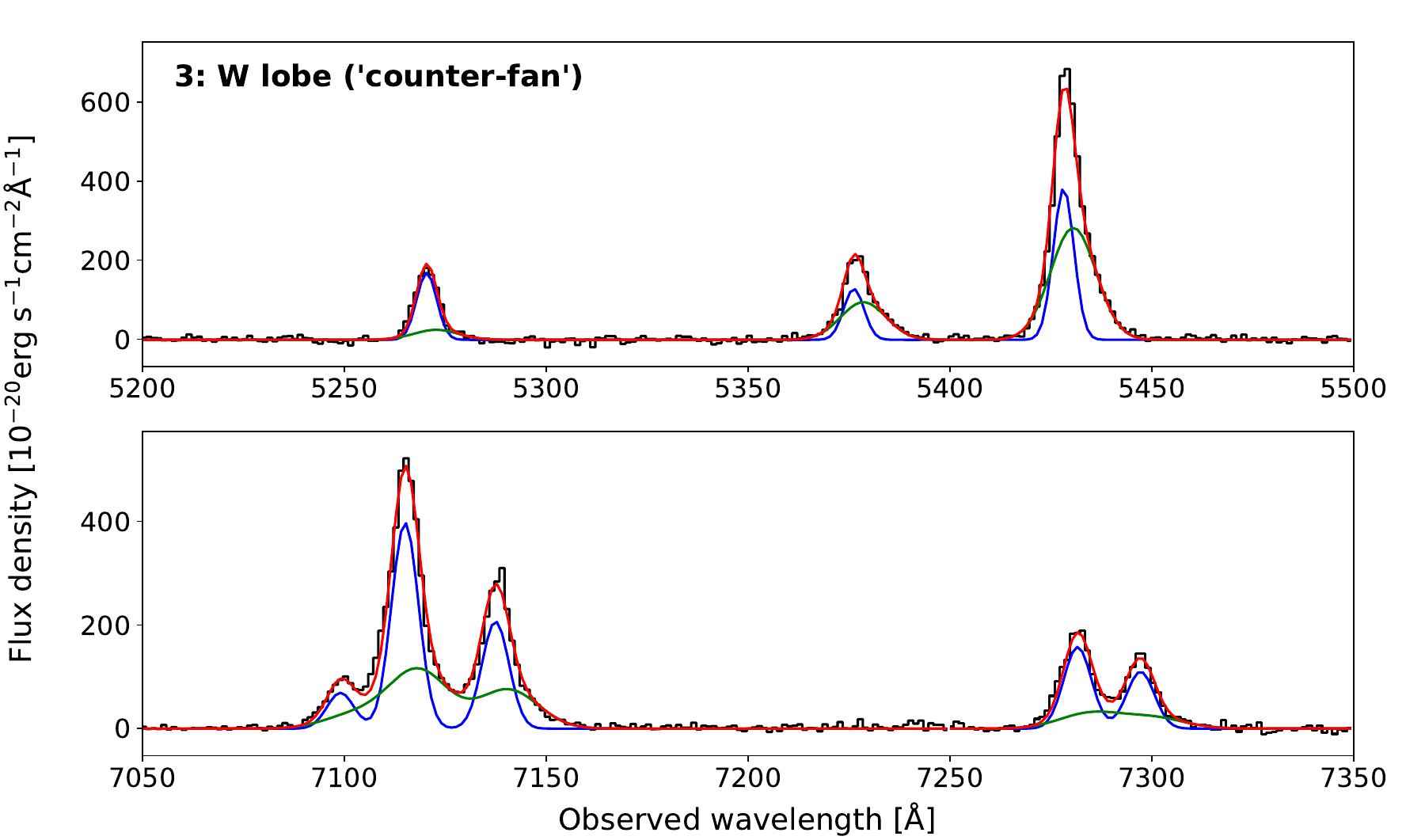}
\caption{Example of modelled spectra from different regions, covering \hb\ and \oiii\ (upper sub-panels) and \nii, \ha, and \sii\ (lower sub-panels). From top to bottom, the spectra come from close to the nucleus, to the NE of it {(number 1)}, from {the eastern lobe or Teacup handle (number 2)}, and from the western lobe {(number 3);  these regions are labelled accordingly in Fig. \ref{fig:rgb}, right panel.} For each emission line, blue, green, and orange curves mark the first (narrower), the second, and the third (broader) Gaussian components employed in the fitting, respectively. 
}
\label{fig:spectra}
\end{figure}

\subsection{Data analysis}\label{ssec:data_anal}
Data analysis was carried out by employing custom Python scripts, with an approach similar to that detailed in \cite{Venturi:2018aa,Venturi:2021a} and \cite{Mingozzi:2019aa}, {summarised below.}

Given that the MUSE spectra of the Teacup show stellar continuum emission underlying the ionised gas emission lines, we first modelled and subtracted it in order to separate these two components and allow for an easier subsequent modelling of the emission lines.
Before doing that, we applied a Voronoi adaptive binning (\citealt{Cappellari:2003aa}) to the data cube to reach a minimum signal-to-noise ratio (S/N) per bin on the stellar continuum. We set an average value of 35 per wavelength channel for the minimum S/N to be achieved per bin, considering the {rest-frame} continuum {in the range 4260--}5530 \AA\, which includes prominent stellar absorption features, and excluding gas emission lines.
We modelled the stellar continuum in each bin across the whole spectral range of MUSE {in its extended mode} (4600--9350 \AA, corresponding to rest-frame stellar templates in the range $\sim$4240--8570 \AA\ given the redshift of the source, $z$ $\simeq$ 0.08506), by making use of the E-MILES single-stellar population (SSP) model spectra (\citealt{Koleva:2012a,Vazdekis:2016a,LaBarbera:2017a}), which cover the spectral range 1680--50000 \AA.
The employed models adopt a unimodal standard Salpeter IMF with slope 1.3 (\citealt{Salpeter:1955aa}) as described in \cite{Vazdekis:1996aa}  and  scaled-solar Padova+00 isochrones (\citealt{Girardi:2000a}), which have M/H metallicities spanning from --1.71 to +0.22 (in log, with respect to the solar value) and ages between 0.063 and 17.8 Gyr.
{The continuum modelling was done using} the penalized pixel-fitting code (\textsc{PPXF}; \citealt{Cappellari:2004aa,Cappellari:2017a}), which convolves the linearly combined stellar templates with a Gaussian function to reproduce the velocity and velocity dispersion of the absorption features. The main gas emission lines, {as well as the residuals of not optimally subtracted sky lines}, were masked during the stellar fitting, {and underlying absorption features were extrapolated over in order to model the full spectral range}.
We employed both an additive and a multiplicative polynomial of degrees 2 and 5, respectively, to account for deformations of the spectra with respect to the templates (e.g. because of reddening).

We then subtracted the modelled stellar continuum on a spaxel basis, by re-scaling the model obtained for a given Voronoi bin to the spectral median of the observed continuum in each spaxel belonging to that bin. {From this continuum-subtracted cube}, we modelled the ionised gas optical emission lines of interest, namely \hg, \oiii\ $\lambda$4363, \hb, \oiii\ $\lambda\lambda$4959,5007, \oi\ $\lambda\lambda$6300,6363, \ha, \nii\ $\lambda\lambda$6548,6584, and \sii\ $\lambda\lambda$6716,6731. Before proceeding with the line modelling, we produced an alternative cube, by applying a Voronoi binning to the continuum-subtracted cube around \hb\ (the signal and noise of the spectral channels in an interval of $\sim$20 \AA\ encompassing the line have been considered), requesting an average S/N per spectral channel of at least 5, in order to retrieve the \hb\ emission from a wider area of the galaxy compared to the unbinned case. We chose \hb\ because it is used for most of the diagnostic maps employing emission line ratios. 
The original continuum-subtracted cube and the continuum-subtracted, Voronoi-binned cube were separately fitted as described below, and the results employed to produce different maps (see Sect. \ref{sec:muse_maps}).

The emission-line fitting was carried out with \textsc{mpfit} (\citealt{Markwardt:2009aa}). We performed two fittings of the emission lines, using one or two Gaussian components per line.
{To reduce the parameter ranges over which the minimisation is made and avoid being trapped in local minima with unphysical bases, we constrained some of the fit parameters.}
The line ratios \ha/\hb\ and \hg/\hb\ were {allowed to vary above a minimum of 2.87 for the former and below a maximum of 0.466 for the latter, which are their theoretical extinction-free values for Case B recombination and} a gas temperature of 10$^4$ K  (\citealt{Osterbrock:2006a}). The \sii\  $\lambda$6716/$\lambda$6731 ratio was allowed to vary in the range of values 0.46--1.43, outside which it stops being sensitive to electron density (\citealt{Osterbrock:2006a}). All the emission lines were tied to have the same velocity shift and velocity dispersion.
These constraints were applied separately to both the one- and two-component fits.
To guide the fit, the starting guess of {centroid} velocity of the first component, meant to reproduce the bulk of the line, was matched with the wavelength of the {\oiii} line peak, {and the parameter was allowed to vary between [--50,+50] km/s around that.} {Based on the spectral shifts of the broad line wings observed in the data,} the starting guess of velocity of the second component, meant to reproduce outflowing motions in the form of asymmetric wings or additional line components, was set to {+300 km/s (--300 km/s)} with respect to the line peak, with the positive (negative) sign in case of a redward (blueward) asymmetry of the line profile, {and the parameter was allowed to vary by up to 500 km/s with respect to the peak velocity towards the asymmetric side of the line.}
To assess which of the two fits (with one or two Gaussians) to adopt in each spaxel, depending on the complexity of the line profiles, we employed the improvement of the reduced $\chi^2$ as a criterion. The two-component fit was selected only in those spaxels whose two-component reduced $\chi^2$ was smaller than 80$\%$ of that resulting from the one-component fit. Moreover, we only selected the two-component fit when the peak S/N of the second component was greater than 3 in the \oiii\ line, to avoid selecting a second component reproducing the noise.
In the nuclear region, the line profiles are very complex and could not be reproduced with two components only {(see e.g. spectrum number 1 in Fig. \ref{fig:spectra}, where a highly blueshifted outflowing component requires the inclusion of an additional Gaussian in the fitting)}. We thus added a third component in a radius of 12 spaxels ($\sim$2.4$''$) from the nucleus, {to account for these additional outflowing components}. The same selection criterion for the number of components based on the reduced $\chi^2$ of the three- versus two-component fits described above was applied, as well as the S/N = 3 cut on the third component of \oiii.

The \sii\ doublet was separately modelled from the data cube we produced after having removed the sky residuals affecting its spectral region by running \textsc{ZAP} (Sect. \ref{ssec:data_red}). We first subtracted the stellar continuum previously obtained for the other data cube, again by re-scaling the model obtained for a given bin to the flux of each spaxel belonging to it. As done before, we obtained a Voronoi-binned cube from the continuum-subtracted one, by using the same binning employed for the original data cube (the one on which we did not apply \textsc{ZAP}). We then modelled the \sii\ doublet on both the binned and the unbinned cube. To do so, in each spaxel we used the same number of Gaussian components used to model the other emission lines in the original cubes (binned and unbinned) and fixed the velocity and velocity dispersion of each Gaussian component of \sii\ to the values obtained for the fit of other emission lines, leaving only the flux free to vary.

The results of all the fitting procedures were then used to produce different maps for the ionised gas, as described in the following in Sect. \ref{sec:muse_maps}. The stellar kinematics, that is not among the main goals of this work, is reported and briefly discussed in Appendix \ref{sec:stellar_kinem}. Some examples of modelled spectra are shown in Fig. \ref{fig:spectra}.

\section{MUSE ionised gas maps}\label{sec:muse_maps}
In this section we present the maps that we have obtained from our analysis of the reduced MUSE data cube of the Teacup galaxy. We applied a cut of S/N = 3 on the emission-line peak to all the maps reported in the following. In case more lines are involved in a given map (e.g. due to a line ratio), the cut is applied to all the lines involved, and the spatial extension of the map is thus limited by the weakest line.
The side of the field of view (FOV) of the MUSE observations corresponds to $\sim$100 kpc at the luminosity distance of the Teacup, $D_L$\,$\sim$\,404 Mpc (1$''$\,$\sim$\,1.66 kpc).

\begin{figure*}
    \centering
	\hfill
	\includegraphics[scale=0.3,trim={2cm 0.5cm 6cm 0.5cm},clip]{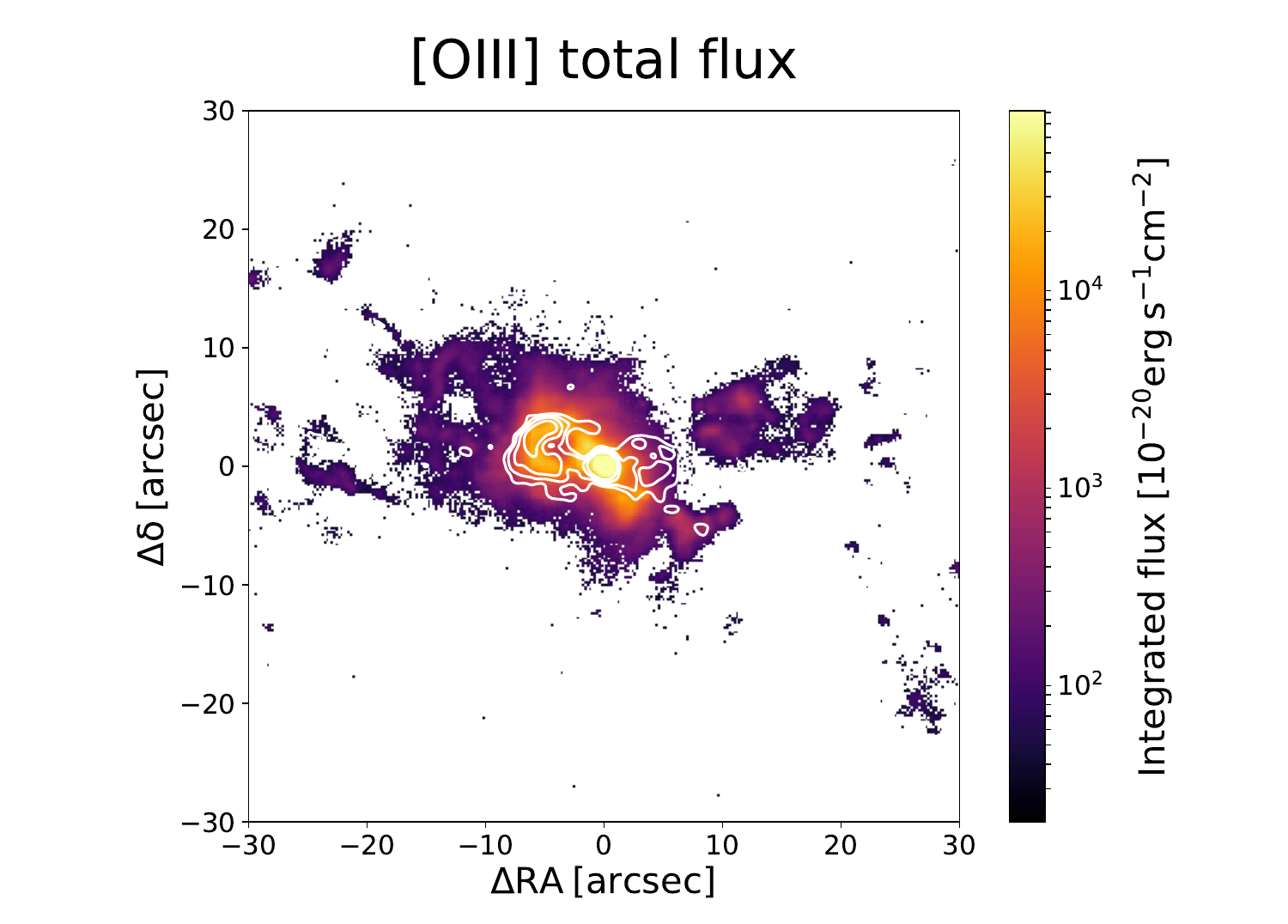}
	\hfill
    \includegraphics[scale=0.3,trim={2cm 0.5cm 6cm 0.5cm},clip]{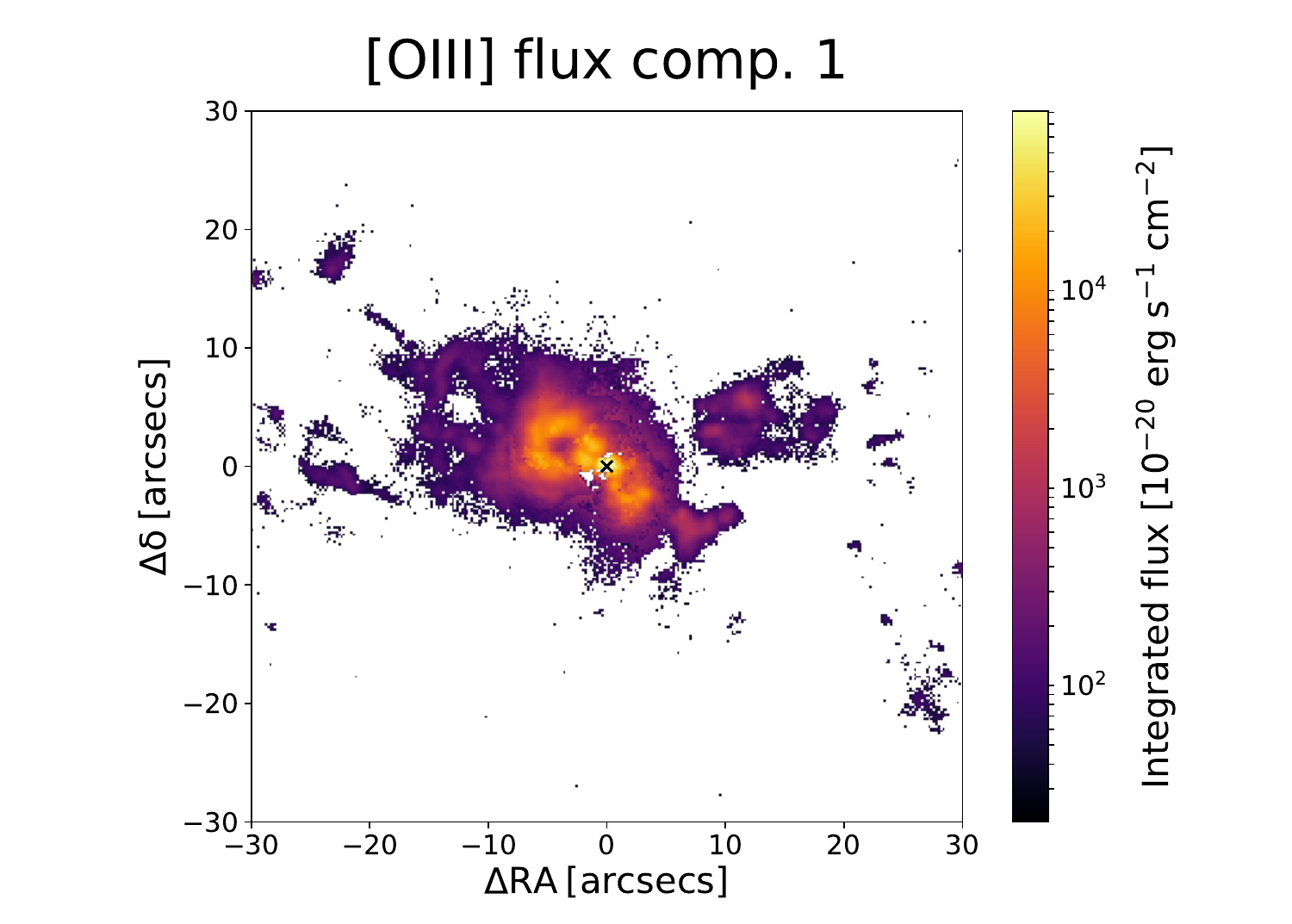}
	\hfill
    \includegraphics[scale=0.3,trim={2cm 0.5cm 2.3cm 0.5cm},clip]{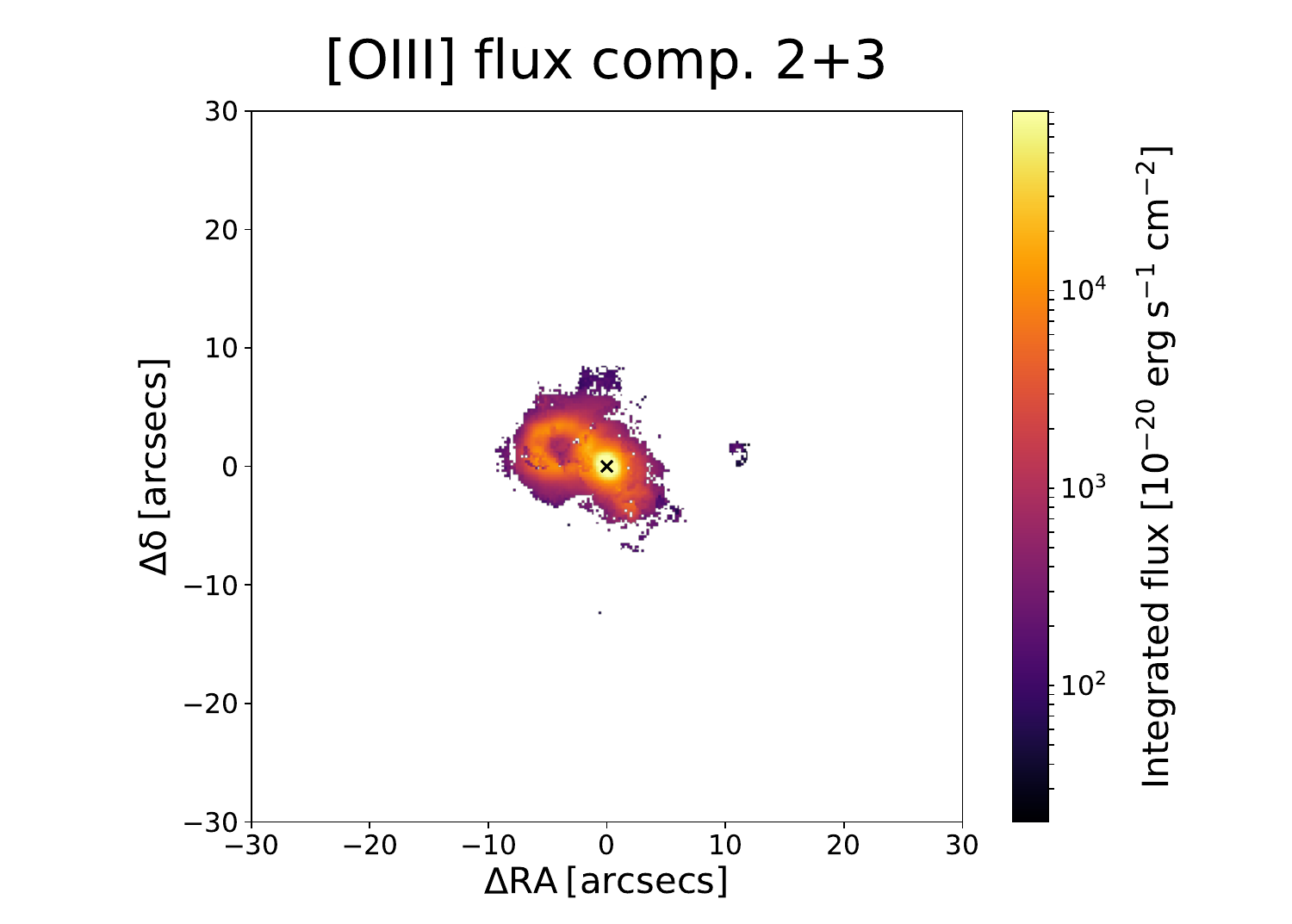}
	\hfill\null\\
	\centering
	\null\hfill
	\includegraphics[scale=0.3,trim={2cm 0.5cm 6cm 0.5cm},clip]{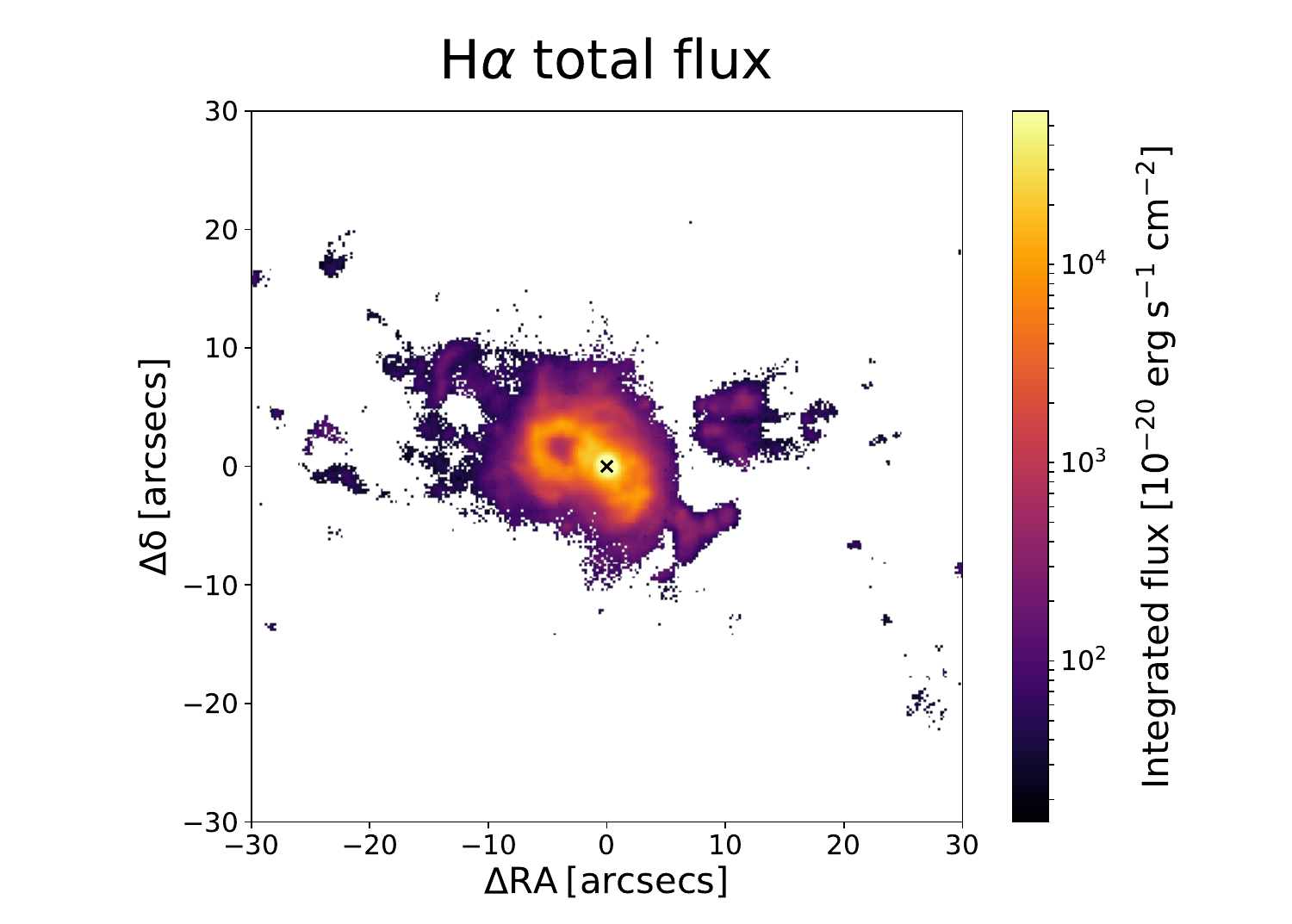}
	\hfill
    \includegraphics[scale=0.3,trim={2cm 0.5cm 6cm 0.5cm},clip]{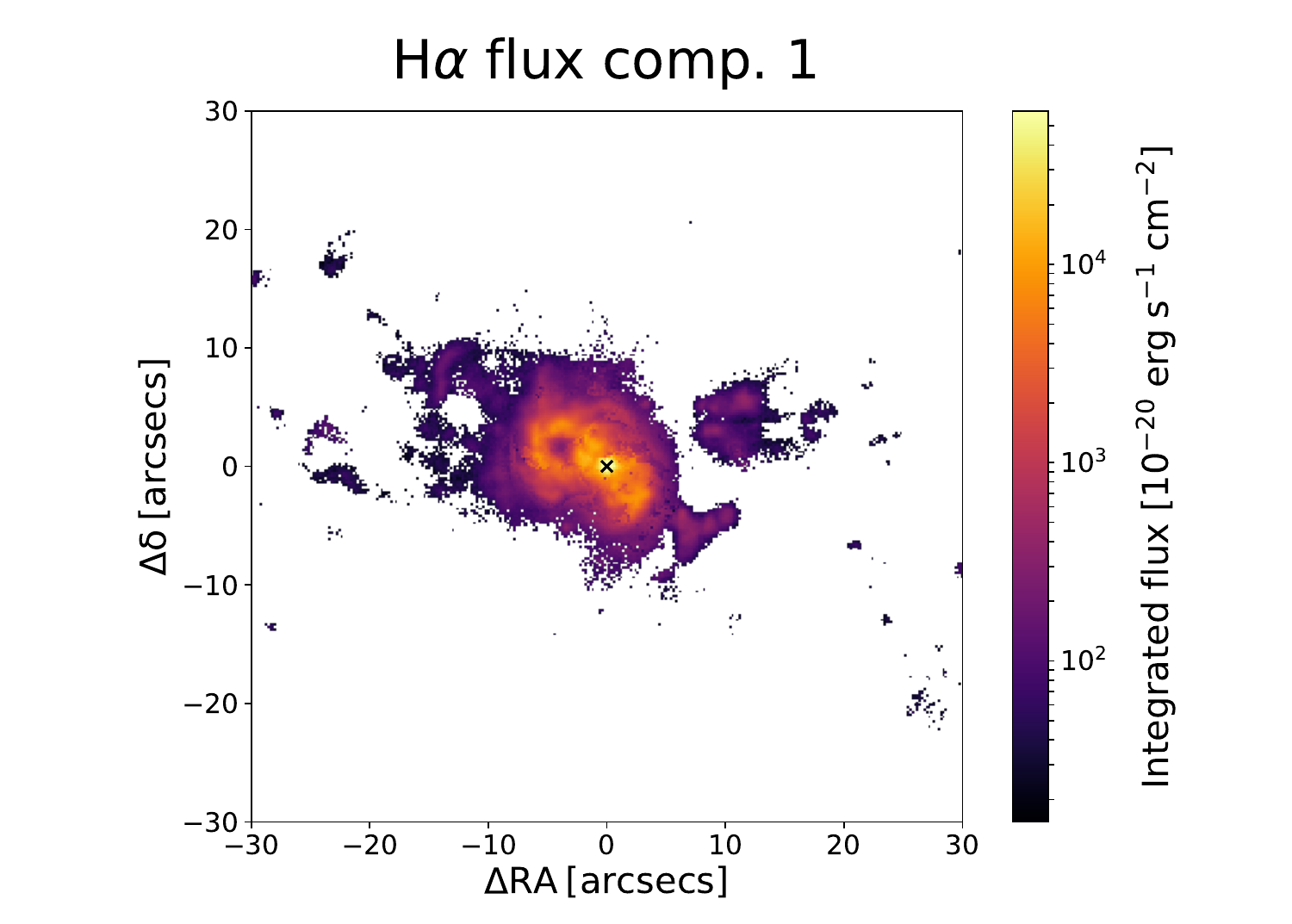}
	\hfill
    \includegraphics[scale=0.3,trim={2cm 0.5cm 2.3cm 0.5cm},clip]{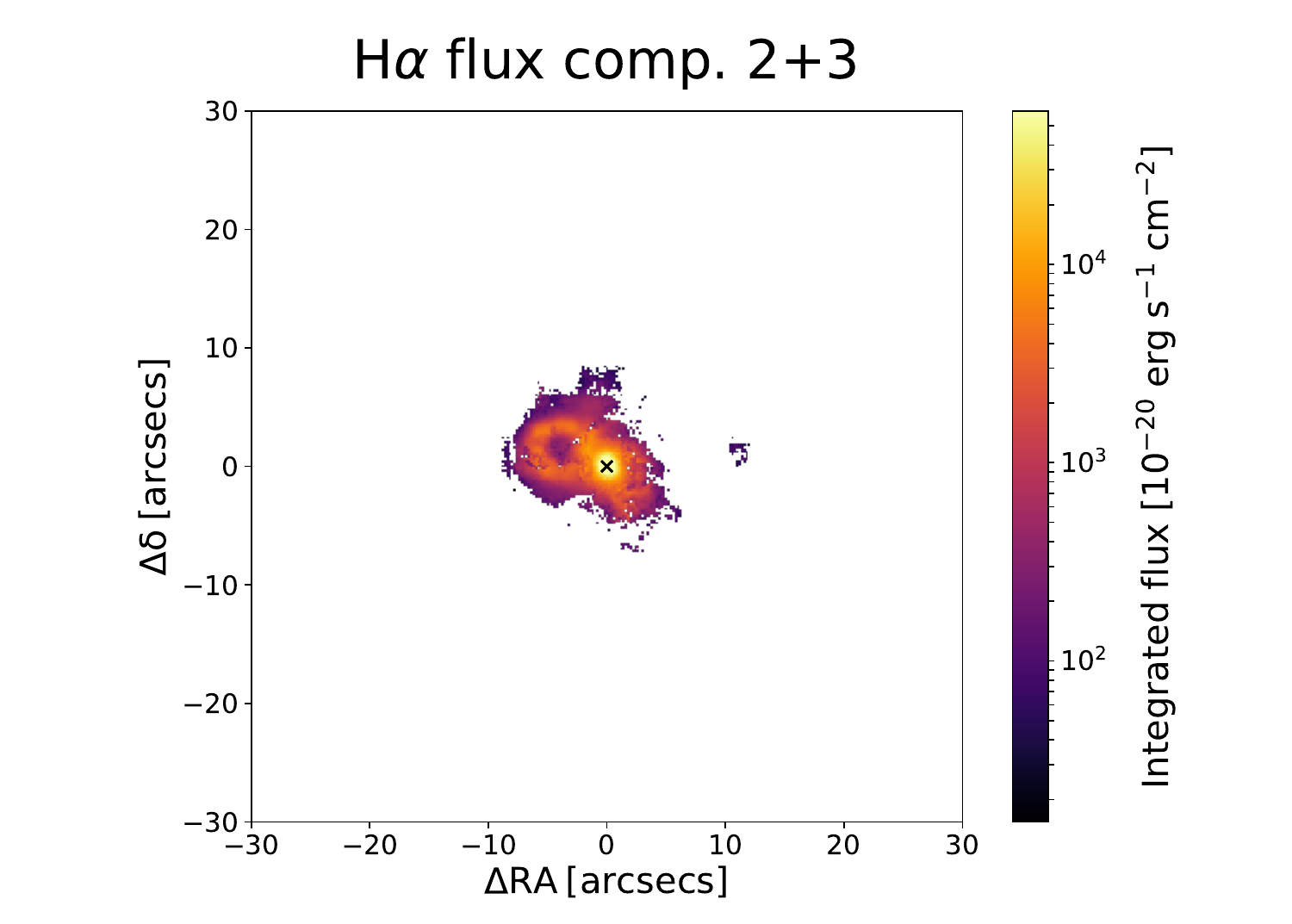}
	\hfill\null
\caption{Maps of the \oiii\ (top panels) and \ha\ (bottom panels) emission line flux distribution in the Teacup galaxy. The fluxes of the total modelled line profile (left panels; same reported in Fig. \ref{fig:rgb}) and of the first, narrower (centre panels) and the outflow (second plus third), broader Gaussian components (right panels) making up the total profile are reported. The colour intensity scale is the same for all the three maps of each line, as reported in the colour bar to the right. The contours in upper-left panel mark the VLA radio emission at 5.12 GHz from \cite{Harrison:2015a}, same as in Fig. \ref{fig:rgb}, right panel.} 
\label{fig:oiii_flux}	
\end{figure*}

\begin{figure*}
    \centering
	\hfill
	\includegraphics[scale=0.3,trim={2cm 0.5cm 6cm 0.5cm},clip]{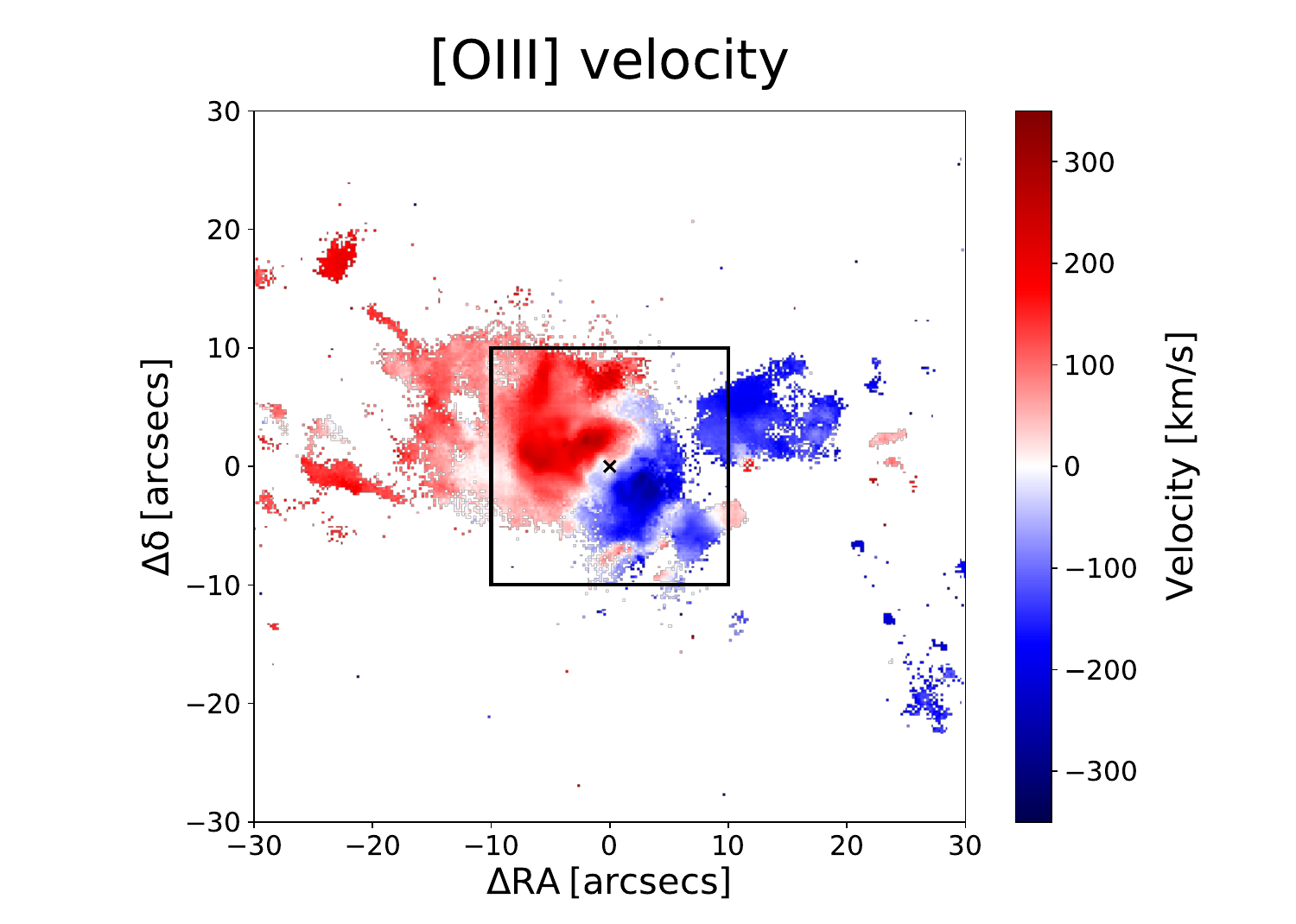}
	\hfill    \includegraphics[scale=0.3,trim={2cm 0.5cm 5.9cm 0.5cm},clip]{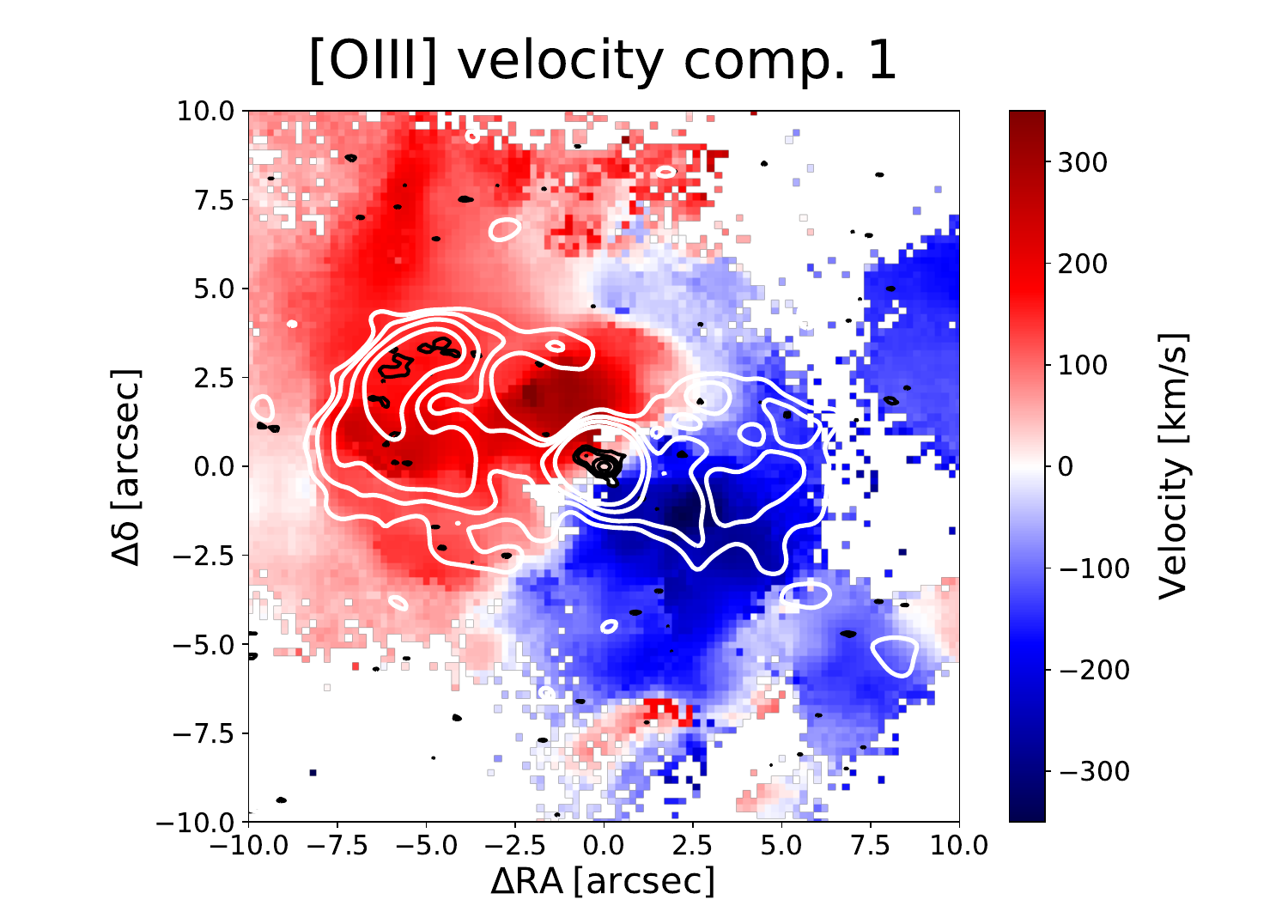}
	\hfill    \includegraphics[scale=0.3,trim={2cm 0.5cm 1.8cm 0.5cm},clip]{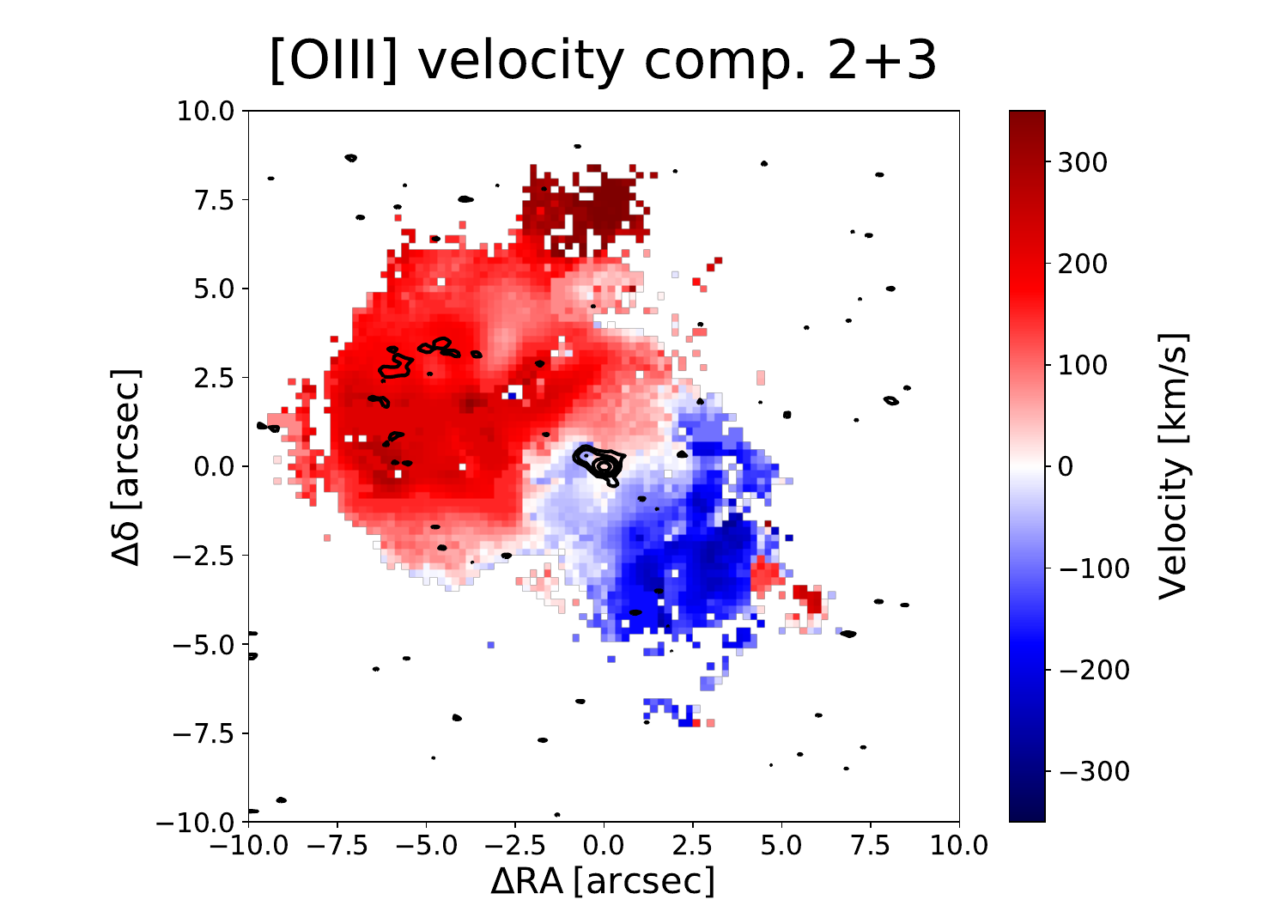}
	\hfill\null \\
	\centering
	\null\hfill
	\includegraphics[scale=0.3,trim={2cm 0.5cm 6cm 0.5cm},clip]{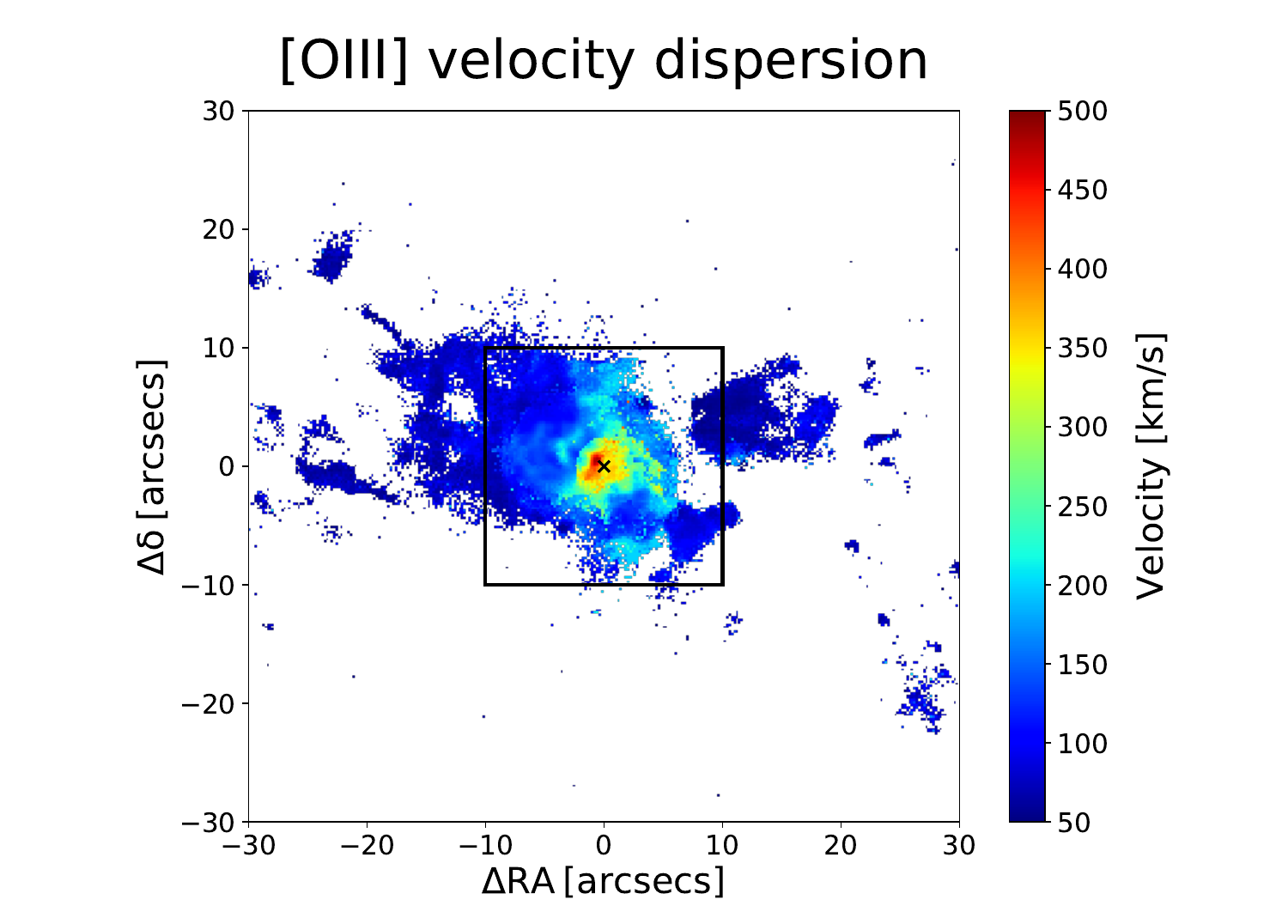}
	\hfill\includegraphics[scale=0.3,trim={1.9cm 0.5cm 5.9cm 0.5cm},clip]{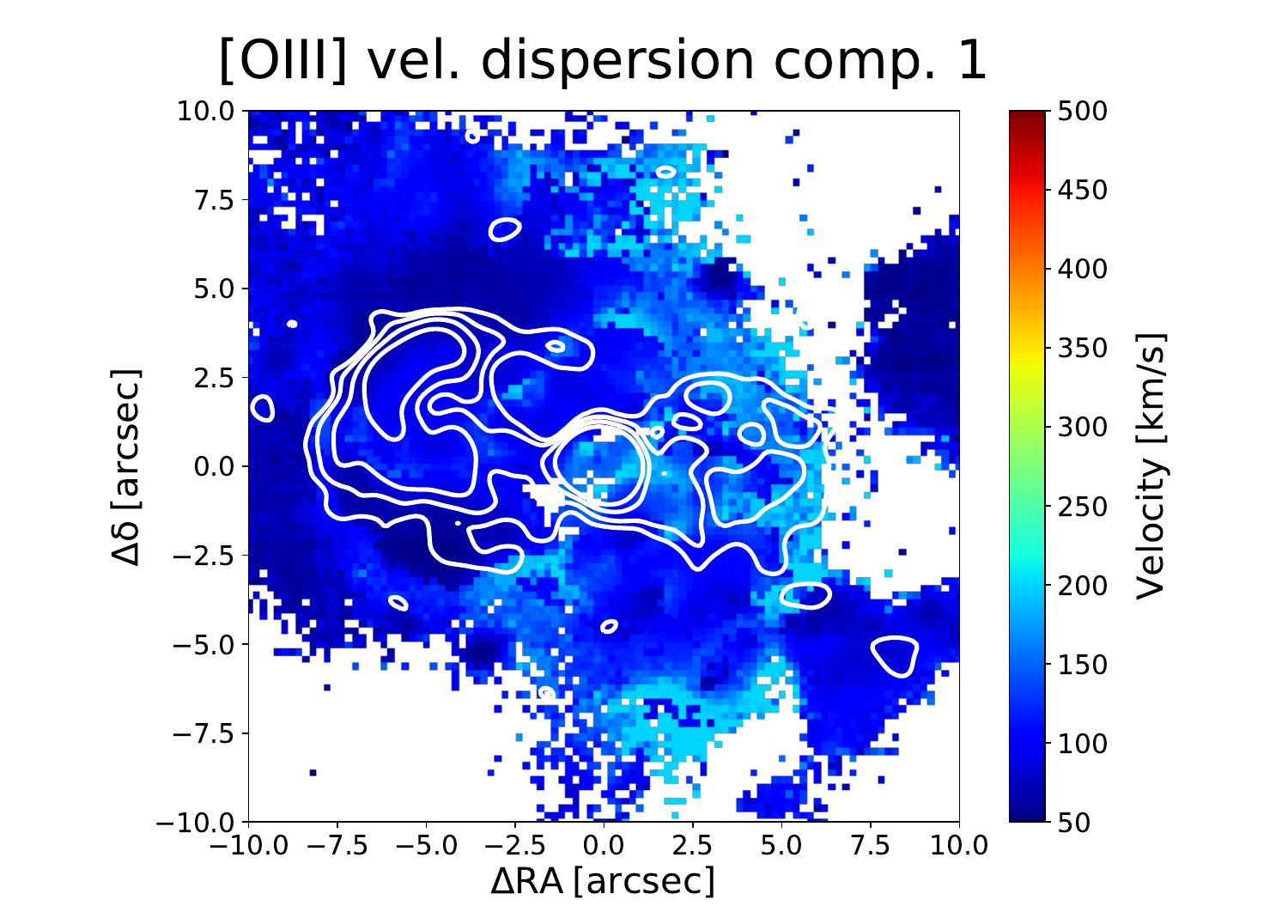}
	\hfill\includegraphics[scale=0.3,trim={2cm 0.5cm 1.8cm 0.5cm},clip]{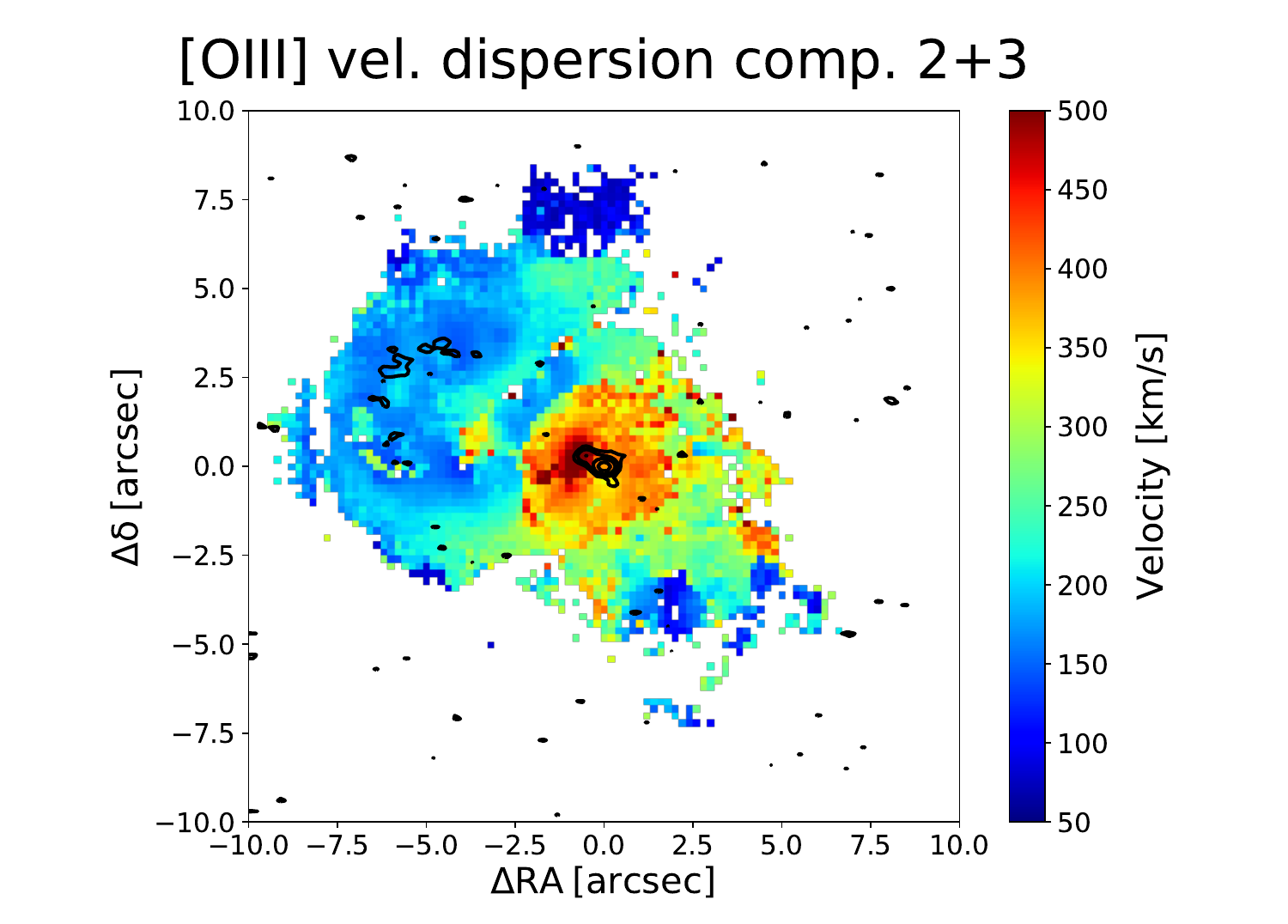}
	\hfill\null\\
 \centering
 	\null\hspace{0.14cm}
	\includegraphics[scale=0.3,trim={2.9cm 0cm 4cm 0.5cm},clip]{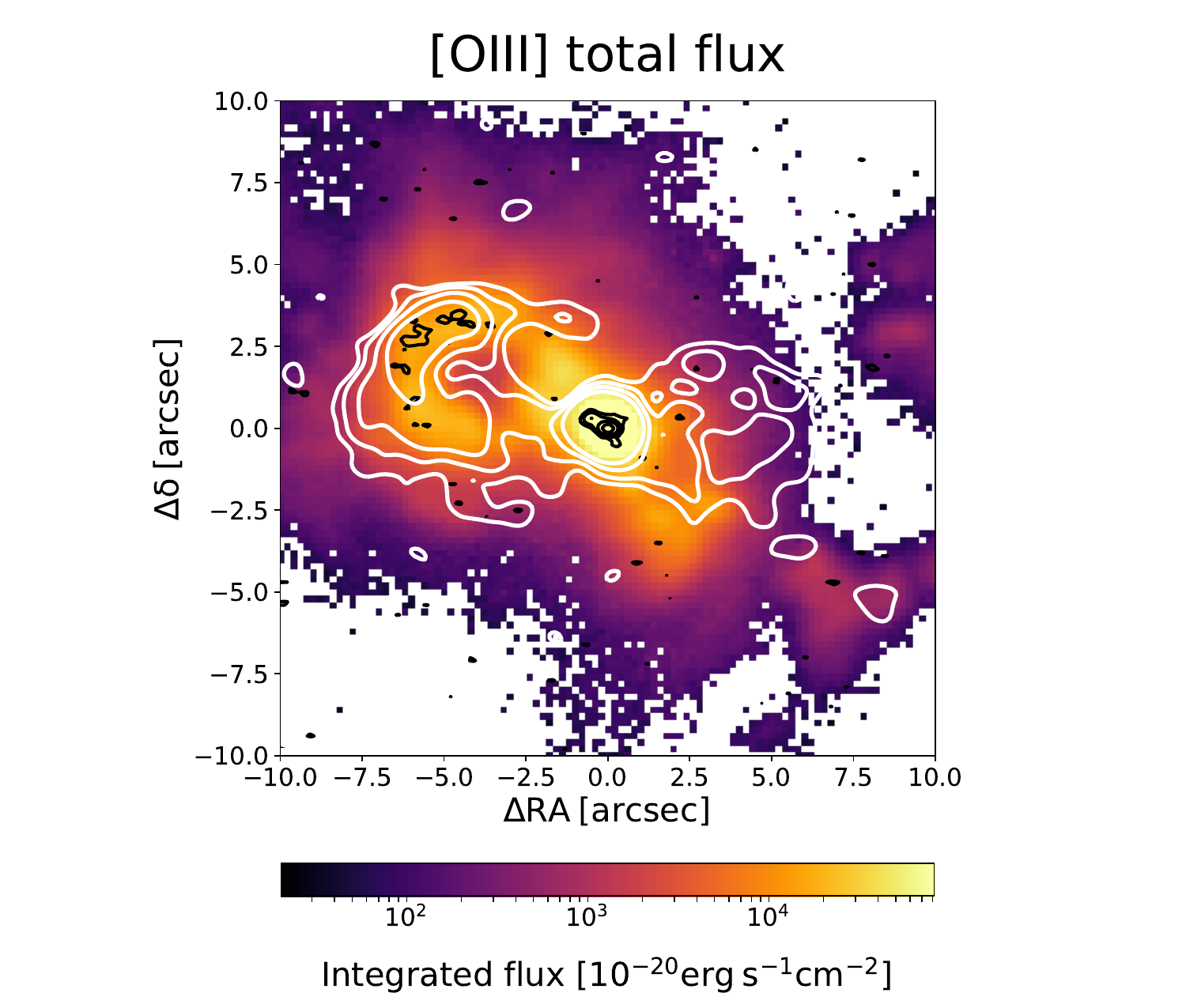}
	\hfill    \includegraphics[scale=0.3,trim={2.9cm 0cm 4cm 0.5cm},clip]{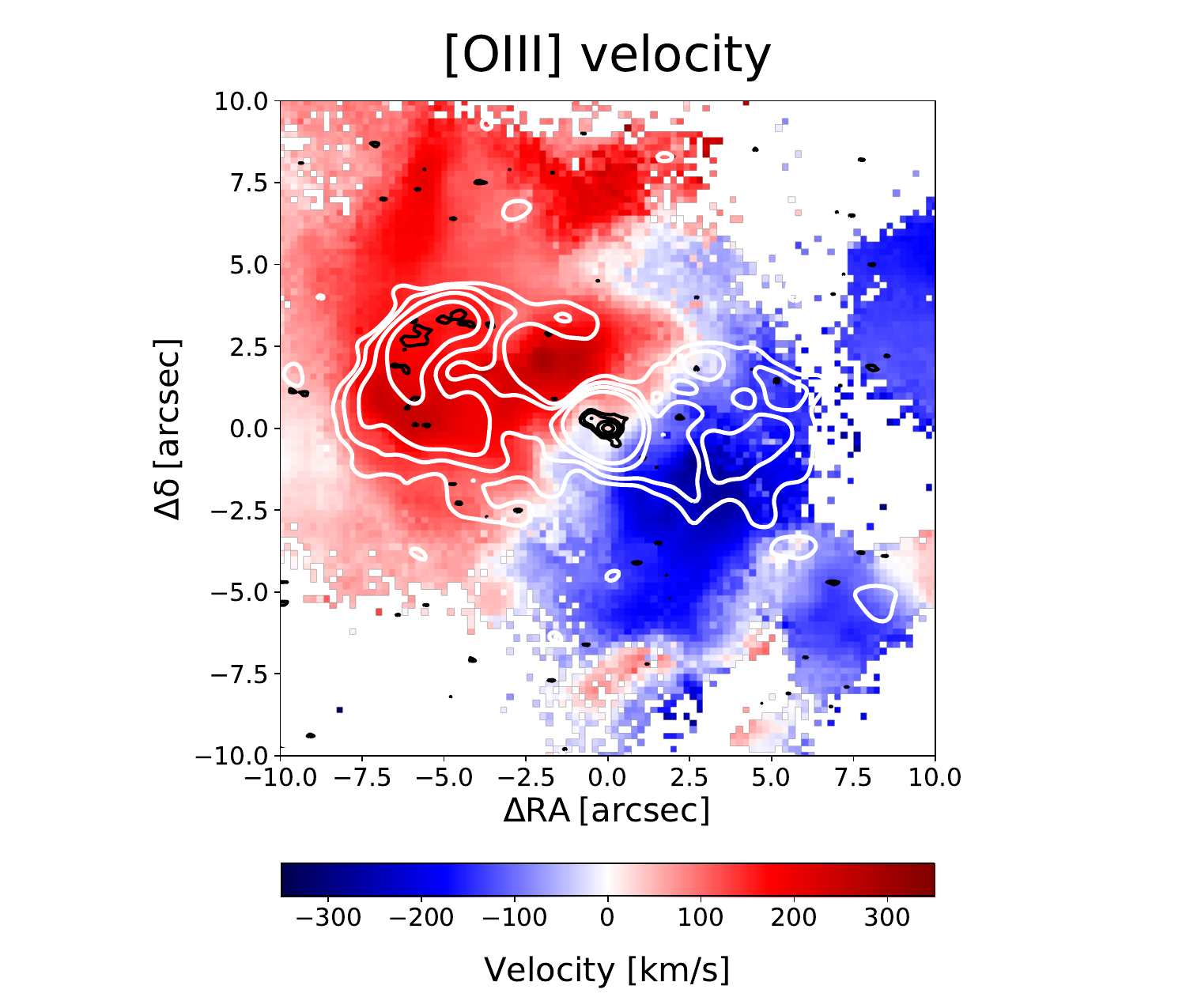}
	\hfill    \includegraphics[scale=0.3,trim={3cm 0cm 0cm 0.5cm},clip]{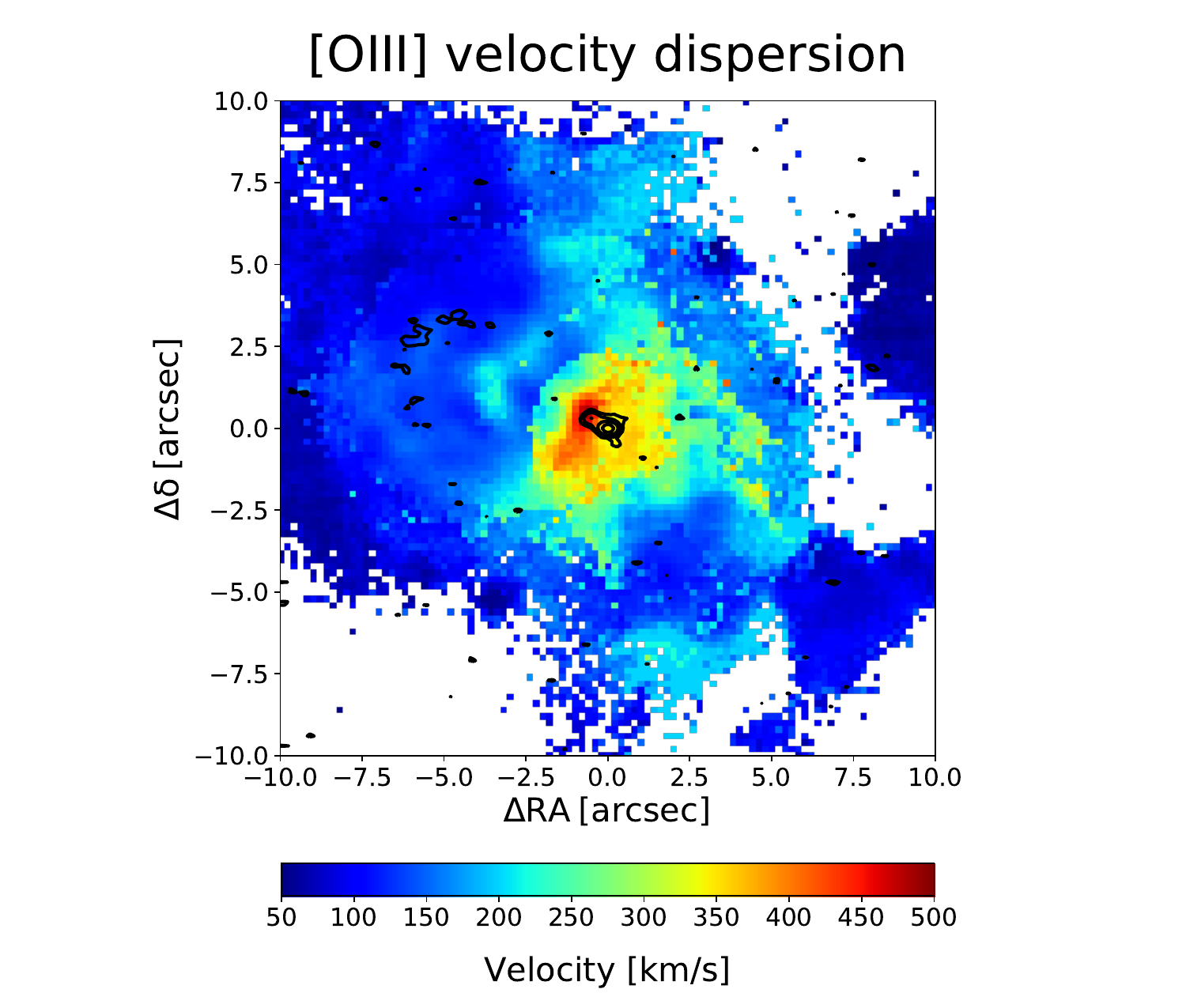}
	\hfill\null\\
\caption{Ionised gas kinematic maps of the Teacup. \oiii\ velocity (upper panels) and velocity dispersion (middle panels), obtained as first- and second-order moments of the velocity, respectively, of the total modelled profile of \oiii\ (left panels), as well as for the first (central panels) and the second plus third Gaussian components (first- and second-order moments of velocity of their summed profile; right panels) employed for the line profile modelling. Central and right panels are zoomed in the central 20$''$\,$\times$\,20$''$ (indicated by the black box in left panels).
Bottom panels: Zoomed versions (also in the central 20$''$\,$\times$\,20$''$) of the \oiii\ flux (left; same as in Fig. \ref{fig:oiii_flux}, top-left panel), velocity (centre; same as in top-left panel) and velocity dispersion (right; same as in mid-left panel) maps for the total modelled emission line profile. 
The contours indicate the same VLA 5.12 GHz and highest-resolution 6.22 GHz radio images from \cite{Harrison:2015a} reported in Fig. \ref{fig:rgb}, right panel, with the same contour levels. 
}
\label{fig:kinmaps}	
\end{figure*}

\subsection{Flux maps}\label{ssec:flux_maps}
In Fig. \ref{fig:rgb} we present a {false}-colour image of the galaxy obtained from the full FOV of MUSE.
The continuum (blue) shows the irregular shape of the galaxy stellar content, resulting from its past merging activity.
The \oiii\ and \ha\ emission, stemming from ionised gas, are reported in green and red, respectively, and the former generally dominates over the latter ({they are shown using the same flux scale}).
Close to the nucleus, at $\sim$10 kpc on its eastern and western sides, the handle of the Teacup and its counter-fan, respectively, are recognisable. Hereafter, we call these two ionised gas structures handle or loop when referring to the eastern one, and counter-fan the western one, in accordance with previous works (\citealt{Keel:2012a, Harrison:2015a, Ramos-Almeida:2017a}), or simply lobes when referring to both of them.
We note that the western optical lobe or counter-fan also shows shell-like emission, though on smaller scales and with a {more poorly defined }shape compared to the eastern handle.
The sensitivity and large FOV of MUSE allow for ionised gas filaments to be revealed and mapped to much larger distances on both sides, up to distances of $\sim$50 kpc per side. 
As part of this extended filamentary emission, we find a second weaker loop reaching about 15$''$ ($\sim$25 kpc) to the E-NE of the nucleus, in the same direction as the Teacup handle, having a more irregular shape than the handle. This second loop might in principle have the same nature as the handle, that is, a bubble of gas inflated by the action of the AGN jet and outflows, but corresponding to a previous expulsion episode.
The large-scale filaments seem to follow an X-shaped geometry centred on the nucleus. This $\sim$100 kpc size structure corresponds to the giant nebula that was spectroscopically identified in \cite{Villar-Martin:2018a}.
{More recent work found arc-shaped ionised gas emission on even larger scales (just outside the FOV of MUSE), to both NE and SW (more prominent and definite in shape to the NE), for a total size of the nebula of $\sim$120--130 kpc (\citealt{Villar-Martin:2021a,Moiseev:2023a}). This appears to be the continuation of the X-shaped filaments detected in the MUSE observations, and would represent an additional large-scale bubble, for a total of three consecutive bubbles (each corresponding to a past AGN outburst), one on $\sim$10 kpc scales (the handle), one on $\sim$25 kpc scales, and one on $\sim$60 kpc scales. These multiple gas shells at increasing distances resulting from repetitive AGN ejections are very similar to those predicted by the TNG50 cosmological simulation as a result of AGN feedback (\citealt{Nelson:2019aa}) and seen from, for instance, NGC 1275 at the centre of the Perseus cluster (e.g. \citealt{Fabian:2003aa,Fabian:2006aa,Sanders:2007aa}).} 

Fig. \ref{fig:oiii_flux} shows maps of the \oiii\ emission line flux, for the total modelled line profile (left; same reported in Fig. \ref{fig:rgb}) as well as for the first, narrower (centre) and second, broader Gaussian component (right) used for the line modelling (Sect. \ref{ssec:data_anal}). The first component follows the same morphology as the total profile. The second component is concentrated in the central regions, including the eastern handle and western counter fan, although it broadly follows the morphology of the total line flux and first component.

The VLA radio contours at 5.12 GHz (same as in \citealt{Harrison:2015a}) are superimposed on the \oiii\ map in Fig. \ref{fig:oiii_flux}. 
This matching shows that the eastern radio bubble peaks at its eastern edge, along the \oiii\ loop, consistent with the results from \cite{Harrison:2015a}.
The MUSE-VLA comparison also clearly shows that there is a misalignment between the axis of the radio emission, oriented in the E-W direction, and that of the optical emission, tilted in the NE-SW direction with respect to the radio one. This can be seen in both the SW counter-fan and the larger scale structure of the gas on both sides.

\subsection{Kinematics}\label{ssec:kinematic_maps}
The kinematic maps of the ionised gas in the Teacup are presented in Fig. \ref{fig:kinmaps}, {with the \oiii\ velocity (upper panels) and velocity dispersion maps ({middle} panels). These are respectively given by the centroid and standard deviation, in case of a single Gaussian component, or by the first- and second-order moments of the velocity of the total modelled line profile, in case of multiple components.}
The maps in the leftmost {of these} panels are obtained from the total modelled profile of \oiii.
The central and right panels show the above two quantities for the first and the second (plus third, when present in the central region) modelled Gaussian components, respectively. {They are zoomed in the central 20$''$\,$\times$\,20$''$, outside which only one Gaussian component was used in the line fitting, making the maps for the first component identical to those for the total line profile.}
{We also show the zoomed versions of the \oiii\ velocity and velocity dispersion for the total line profile (same reported in top-left and mid-left panels for the full FOV) in lower panels, together with the \oiii\ flux} for comparison.
{We considered a systemic velocity of 24486 km/s for the maps, consistent with \cite{Moiseev:2023a}.
We do not report the velocity and velocity dispersion maps for \ha\ since they are virtually identical to those of \oiii.}

The velocity maps show, on the scale of the optical and radio bubbles, a kinematic structure with opposite velocity, that is present in both the first (narrower; $\sigma$ $\lesssim $ 150--200 km s$^{-1}$) and second (broader; $\sigma$ $\gtrsim$ 200 km s$^{-1}$) components, with the first {possibly} tracing a rotational feature {(as modelling of regular circular rotation in \citealt{Moiseev:2023a} seems to suggest; see also \citealt{Villar-Martin:2018a})} while the second the ionised outflow, known to be present in this direction (\citealt{Mullaney:2013a, Villar-Martin:2014a, Harrison:2015a} in \oiii; \citealt{Ramos-Almeida:2017a} in NIR hydrogen recombination lines and [Si\,\textsc{vi}]). 
{In the direction and scales ($\sim$10 kpc) of the ionised and radio bubbles there is no evidence for stellar rotation (see Fig. \ref{fig:stellar_maps}, left panel) or for the presence of a gas disc (which could be responsible for the gas rotation) from the ionised gas morphology (Fig. \ref{fig:oiii_flux}). Moreover, rotating H$_2$ \citep{Ramos-Almeida:2017a} and CO \citep{Ramos-Almeida:2022a, Audibert:2023a} molecular gas is observed in a different direction (N-S, i.e. almost perpendicular) and on much smaller scales ($\sim$2 arcsec, i.e. $\sim$3 kpc). Based on these aspects, another possibility is that all the bipolar ionised gas, including the narrower component, is being evacuated as part of the bubbles.
However, we prefer to adopt a more conservative approach in this work and consider only the broader component(s) as part of the outflow, when we estimate its properties (Sect. \ref{sec:ionised_outfl}).}

On scales larger than the optical and radio bubbles, there {are two sudden discontinuities} in velocity, which drops to smaller values (in absolute value) {at the outer edge of the bubbles ($\sim$10 kpc)} to rise again at $\sim$20-25 kpc ($\sim$15$''$) from the nucleus. This is more visible on the eastern (redshifted) side.
This large-scale pattern can be appreciated even better in the velocity maps obtained from the {stellar continuum}-subtracted, Voronoi-binned cube, reported in Fig. \ref{fig:kinmaps_binned} in Appendix.
This structure, on the scale of the $\sim$100 kpc giant bubble, has opposite-sign velocities {(receding to the E, approaching to the W)} and is detected only in the first component, with small velocity dispersions ($\lesssim$100 km s$^{-1}$). It might thus {either} represent a separate rotating component in addition to that observed on the scale of the bubbles, possibly a leftover of the past galaxy merger event, {or the distinct kinematics (either rotating and/or still expanding) of the largest-scale bubble expelled in a past AGN outburst (see Sect. \ref{ssec:flux_maps}; \citealt{Villar-Martin:2021a,Moiseev:2023a}).}

The velocity dispersion map of the total profile ({mid-left and bottom-right panels} of Fig. \ref{fig:kinmaps}) shows that this is maximal within a radius of about 10 kpc around the nucleus, with values from $\sim$200 km s$^{-1}$ up to $\sim$500 km s$^{-1}$ close to the nucleus. Such high values stem from the second plus third components (mid-right panel), though also the first component (mid-central panel) has some residual enhancements ($\sim$150--200 km s$^{-1}$) in the central region compared to the average velocity dispersion elsewhere.
Interestingly, the highest values of the velocity dispersion ($\gtrsim$250 km s$^{-1}$) in the total profile and second (plus third) component velocity dispersion maps are not isotropically distributed around the nucleus, but {mainly appear} in the direction perpendicular to the ionised and radio bubbles, {rather than} in the direction of the bipolar outflow as normally expected in AGN {(e.g. \citealt{Venturi:2018aa,Lopez-Coba:2020aa,Juneau:2022aa,Kakkad:2023aa}). Specifically, they reach the highest values ($\sim$500 km/s) at the location of the jet head, the eastern hotspot HR-B in the highest-resolution radio image at 6.22 GHz.} We discuss this aspect in more detail in Sect. \ref{sec:enhanced_sigma}.

{Tentative detection of a very broad (FWHM $\sim$ 3000 km/s) Pa$\alpha$ component to the NE and SW of the nucleus was reported by \cite{Ramos-Almeida:2017a}, which we do not detect in the (higher S/N) MUSE data. We briefly discuss this in Appendix \ref{ssec:broad_Pa}.}

\subsection{Radio-optical misalignment}

Interestingly, the comparison of the {lower-resolution radio emission at 5.12 GHz} with the \oiii\ velocity maps {of the total profile and first-component (bottom-central and top-central panels in Fig. \ref{fig:kinmaps}, respectively)} shows that the optical kinematic axis is clearly misaligned with respect to the axis of the radio bubbles (by $\sim$20\degree, as found by \citealt{Harrison:2015a}). The optical-radio misalignment is even more evident in the velocity field than in the flux map ({Fig. \ref{fig:kinmaps}, bottom-left panel and} Fig. \ref{fig:oiii_flux}, top-left panel).

On the other hand, the highest-resolution (beam HPBW = 0.37$''$\,$\times$\,0.24$''$) VLA radio image at 6.22 GHz follows {more closely} the optical flux and kinematic axes (Fig. \ref{fig:kinmaps}, bottom-central, top-central, and top-right panels) and is thus misaligned with respect to the large-scale radio structure. 
{On scales of $\sim$2.5$''$ from the nucleus, however, the velocity map of the first (narrower) component shows a kinematic structure with opposite velocities (redshifted to the NE and blueshifted to the SW, exceeding 300 km/s) that is misaligned from both the radio bubbles at 5.12 GHz and the highest resolution radio hotspots at 6.22 GHz and whose centre appears to be offset from the nucleus. This might represent an additional fast rotating structure, possibly resulting from the past merging activity. 
The kinematics of the narrower component is very complex and understanding it would require accurate kinematic modelling, which goes beyond the scope of this work.}

As mentioned before, the two unresolved knots constituting the high-resolution radio structure are likely a compact radio jet. 
{The jet is very well aligned with the velocity of the second (plus third) broader components (top-right panel), which trace a bipolar outflow.}
Being aligned with the {ionised} outflow, the jet might contribute to the outflow driving process. Since also the {\oiii} flux axis is aligned with that of the jet, this means that the accretion disc (to which the jet is expected to be perpendicular; e.g. \citealt{Blandford:2019a}) and the obscuring toroidal structure expected to surround the nucleus down to $\sim$pc scales (defining the opening angle of the AGN ionising field on galactic scales; e.g. \citealt{Bianchi:2012a, Ramos-Almeida:2017b}) are co-planar.
This is indeed not always the case, since a misalignment between the radio jet and the \oiii\ emission, associated with the AGN ionisation field, is sometimes observed (e.g. \citealt{Goosmann:2011a, Fischer:2013a, Fischer:2014a, May:2020a}).
Possible causes for the observed misalignment between the large-scale radio emission and the small-scale radio jet, the optical lobes, and optical kinematics are jet precession (e.g. \citealt{Kharb:2014a, Kharb:2018a}) or jet-bending due to jet-cloud interactions (as e.g. in the case of NGC 1068; \citealt{Gallimore:1996a, Gallimore:1996b}). 

No clear evidence for the presence of a gas cloud or complex of clouds emerges from the MUSE observations on the scale of the radio knots ($\sim$0.5$''$--1$''$), nor of a sudden bend of the jet trajectory, that could favour the jet bending scenario due to interaction with gas clouds. 
Unfortunately, the resolution of MUSE observations (FWHM $\sim$ 1$''$) does not allow for this possibility to be tested further.
The higher resolution {\it HST} images show no evidence for dust lanes in the jet trajectory (dust lanes are only observed to the W of the nucleus; \citealt{Keel:2015a}) or for an impact on clouds \citep{Harrison:2015a} which could bend the jet.
We note however that the scales of the jet-cloud interaction in NGC 1068 and of the high-resolution jet in the Teacup are very different, being $\sim$20 pc for the former case and $\sim$800 pc for the latter.
In any case, the jet bending scenario seems unlikely, given that the large-scale radio bubbles are misaligned with respect to the small-scale jet on both the E and W sides, that would require a bending by the same angle due to a jet-cloud interaction on both sides of the nucleus.

On the other hand, jet precession could in principle be a viable explanation for the misalignment. However, it seems very unlikely that the jet during its precession is currently pointing exactly in the direction of the optical flux and kinematic axes, whose direction is defined on tens of kpc.
Therefore, all in all, the origin of the misalignment between the large-scale radio emission and the small-scale radio jet, the optical lobes, and optical kinematics is not clear.

\begin{figure*}
    \centering
	\hfill
	\includegraphics[scale=0.3,trim={2cm 0.5cm 2.3cm 0.5cm},clip]{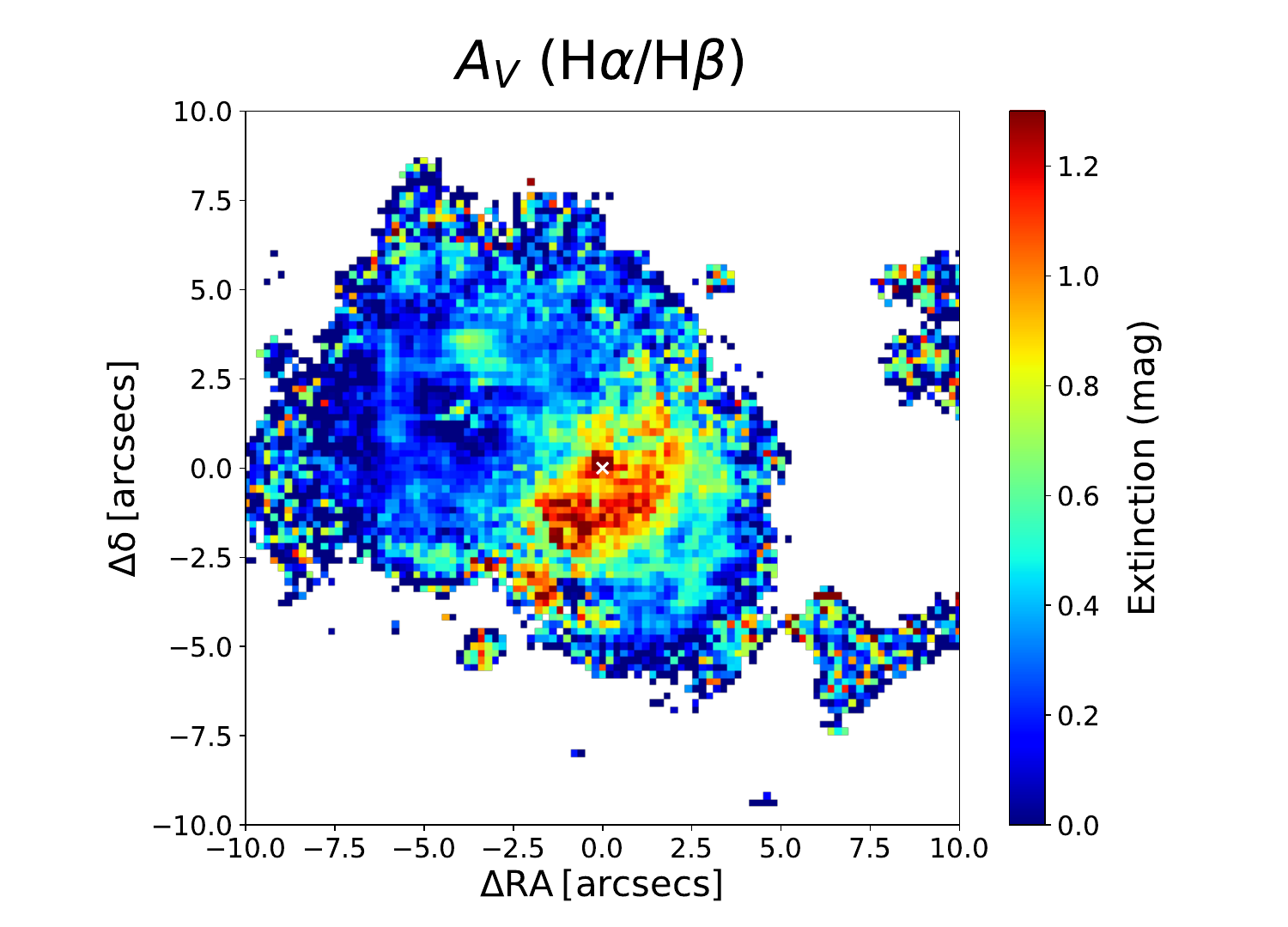}
	\hfill
    \includegraphics[scale=0.3,trim={2cm 0.5cm 2.3cm 0.5cm},clip]{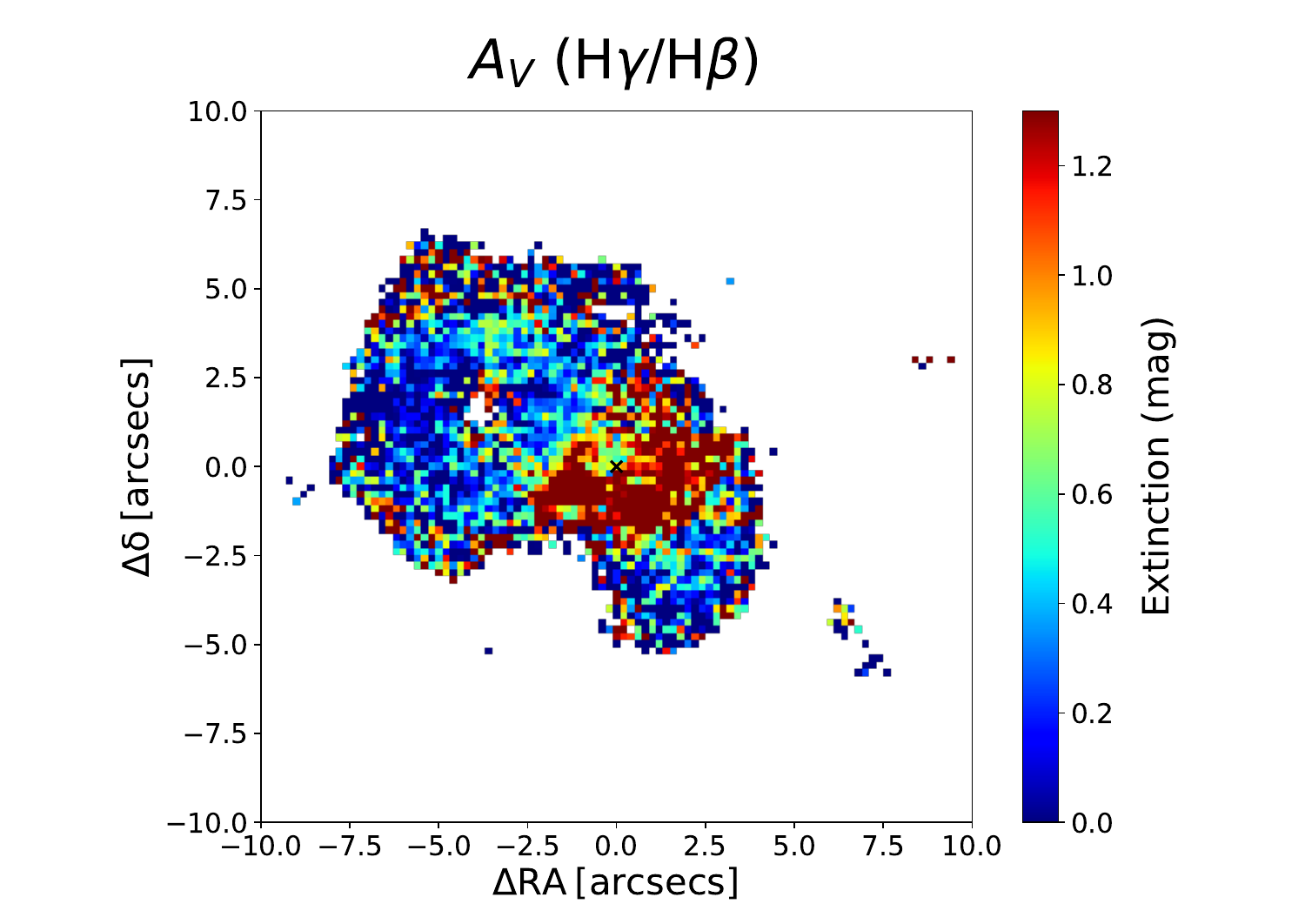}
    \hfill\null\\
    \centering
	\null\hfill
    \includegraphics[scale=0.3,trim={2cm 0.5cm 2.3cm 0.5cm},clip]{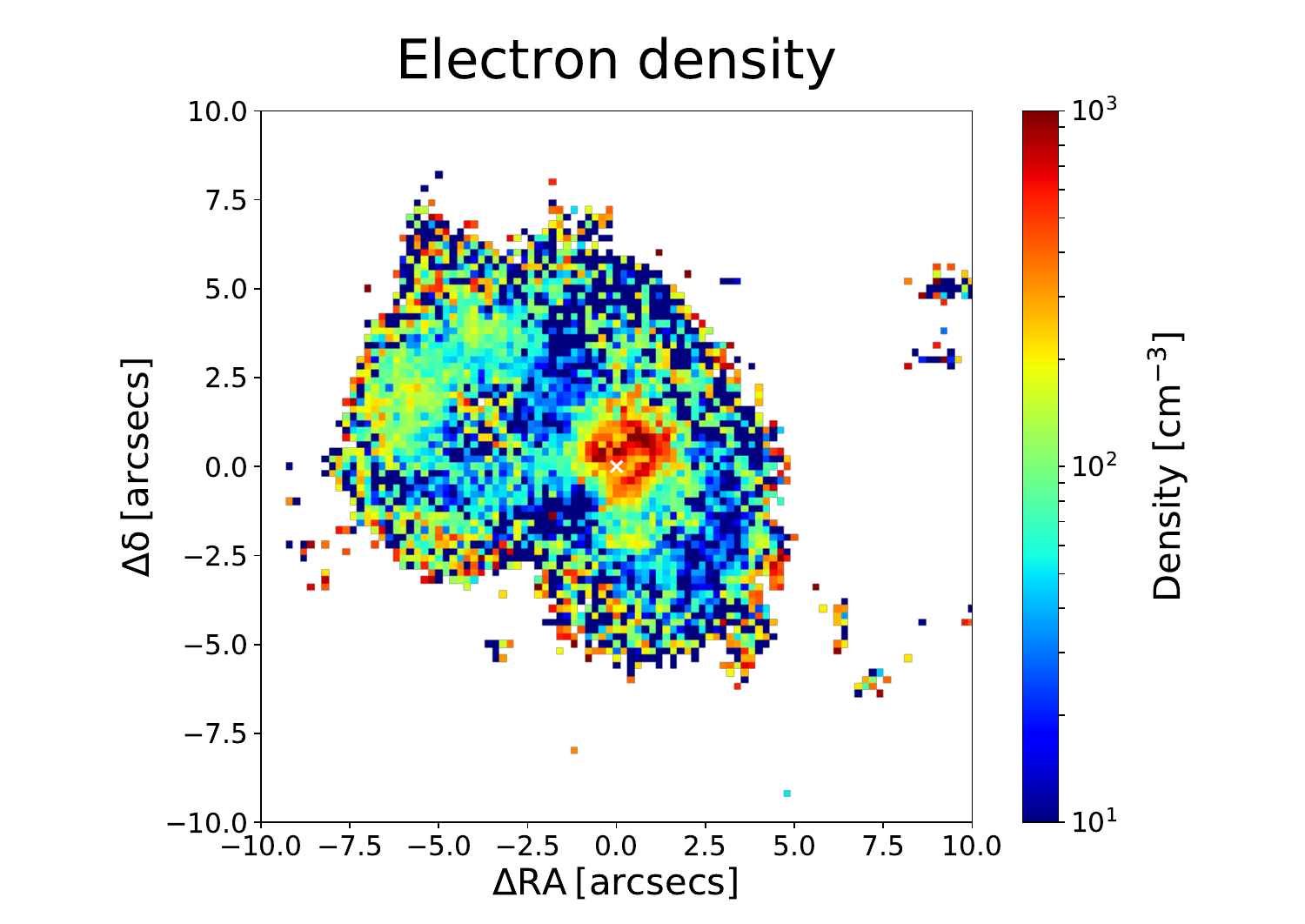}
	\hfill
    \includegraphics[scale=0.3,trim={2cm 0.5cm 2.3cm 0.5cm},clip]{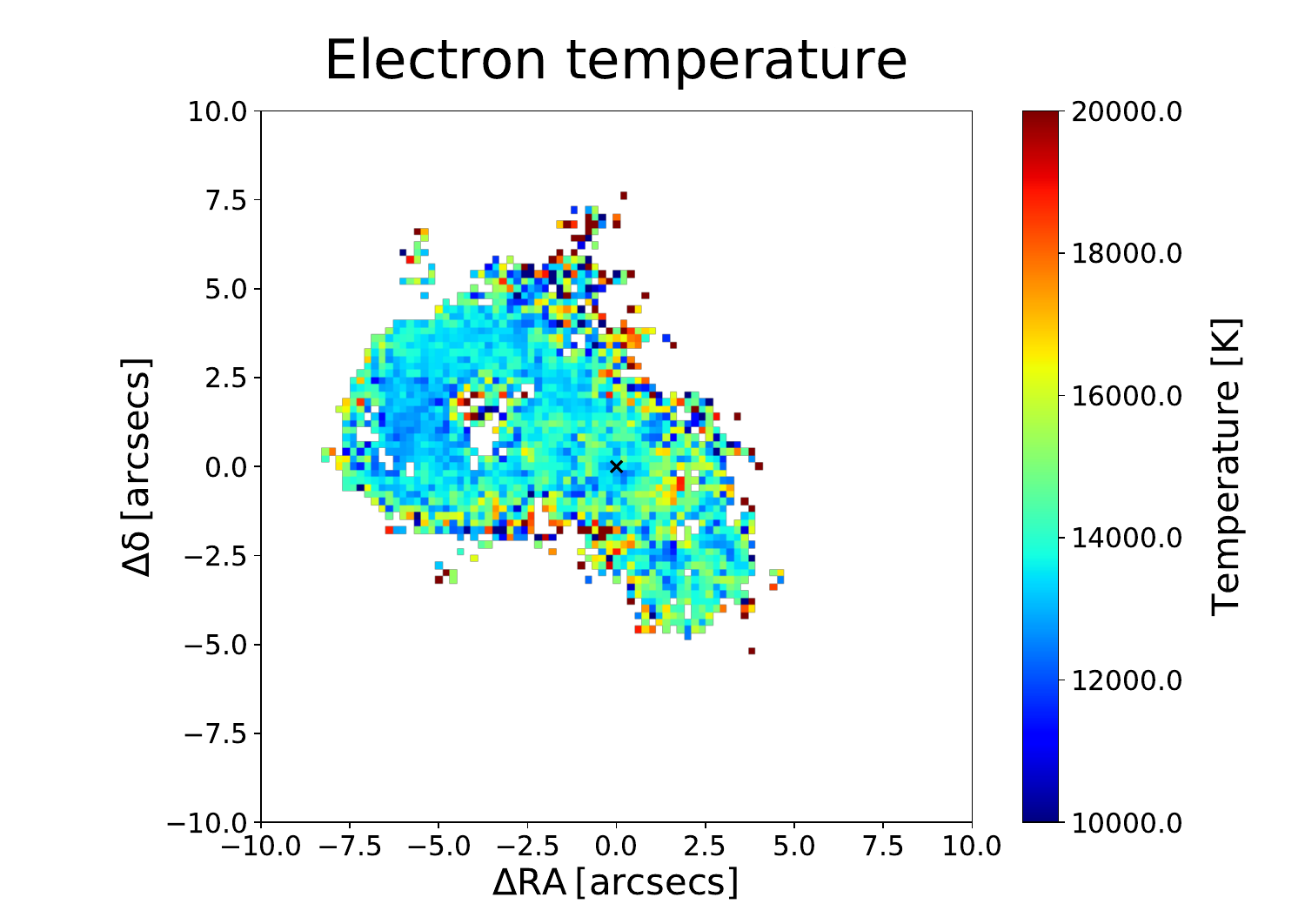}
	\hfill\null
\caption{Maps of physical properties of the ionised gas in the Teacup. Extinction $A_V$ from \ha/\hb\ (top-left) and \hg/\hb\ (top-right), electron density (from \sii\ $\lambda$6716/$\lambda$6731; bottom-left), and electron temperature (from \oiii\ ($\lambda$5007+$\lambda$4959)/$\lambda$4363; bottom-right). The line fluxes involved in the line ratios employed for the maps are those of the total modelled emission line profiles (with one, two, or three Gaussians). The maps are zoomed {in the central 20$''$\,$\times$\,20$''$}.}
\label{fig:ext_dens}	
\end{figure*}

\subsection{Extinction, density, and temperature}\label{ssec:gas_props}
In Fig. \ref{fig:ext_dens} we show the maps for some of the physical properties of the gas, namely the {implied} extinction $A_V$ (from both \ha/\hb\ and \hg/\hb), the electron density (from \sii\ $\lambda$6716/$\lambda$6731; \citealt{Osterbrock:2006a}), and the electron temperature (from \oiii\ ($\lambda$5007+$\lambda$4959)/$\lambda$4363).
We adopted a \cite{Calzetti:2000a} attenuation law for galactic diffuse ISM ($R_V$ = 3.12) and intrinsic ratios of \ha/\hb\ = 2.86 and \hg/\hb\ = 0.466 (for an electron temperature of $T_e$ = 10$^4$ K; \citealt{Osterbrock:2006a}) to evaluate the extinction $A_V$.
For the electron temperature, $T_e$, we employed the relation from \cite{Proxauf:2014a}, updated from that in \cite{Osterbrock:2006a}.

All these maps have been obtained from the total modelled emission line profiles. We do not report the maps for the single components as the lines involved (\hb, \sii\ doublet, and \oiii\ $\lambda$4363) are limited only to the very central regions in their second component due to low S/N.

As can be seen in Fig. \ref{fig:ext_dens}, top panels, the extinction is maximal around the nucleus, in an elongated area in the NW-SE direction, where it reaches values above $A_V$ $\sim$ 1.2. 
This is in agreement with the $r$--$i$ colour map of the central 4$''\times$4$''$ shown in Fig. E.1 of \cite{Ramos-Almeida:2022a}, and with the presence of strong dust lanes westwards of the nucleus (\citealt{Keel:2015a}). 
On larger scales, lower values of the extinction, around 0.6, are also found in some regions in the direction of the \oiii\ E loop and W counter-fan. The values of extinction obtained from \hg/\hb\ tend to be higher around the centre compared to those from \ha/\hb\ by up to a factor of $\sim$2. This can possibly be due to a discrepancy between the adopted attenuation law (\citealt{Calzetti:2000a}) and the actual one experienced by the gas in the galaxy, {or to deviations from pure Case B ionisation.}
However, a comparative analysis of different dust attenuation laws goes beyond the scope of this work.

The electron density (Fig. \ref{fig:ext_dens}, bottom-left panel) peaks in a limited region close to the nucleus, where it reaches values $\gtrsim$500 cm$^{-3}$. It quickly drops away from the nucleus, down to values close to or below the threshold of 10 cm$^{-3}$ (below which the \sii\ doublet ratio becomes virtually insensitive to density variations) at 1$''$--2$''$ from the nucleus.
Interestingly, the density rises again at the eastern edge of the \oiii\ loop, with values $>$100 cm$^{-3}$.

Finally, the electron temperature (Fig. \ref{fig:ext_dens}, bottom-right panel) is fairly constant (neglecting the noisy pixels with high values at the edges), being comprised in the range $T_e$ $\sim$ 1.3--1.4 $\times$ 10$^4$~K. The temperature $T_e$ = 10$^4$ K assumed for the intrinsic values of the line ratios of \ha/\hb\ and \hg/\hb\ is thus a fairly good approximation of the actual temperature.

\begin{figure*}
\includegraphics[width=\textwidth]{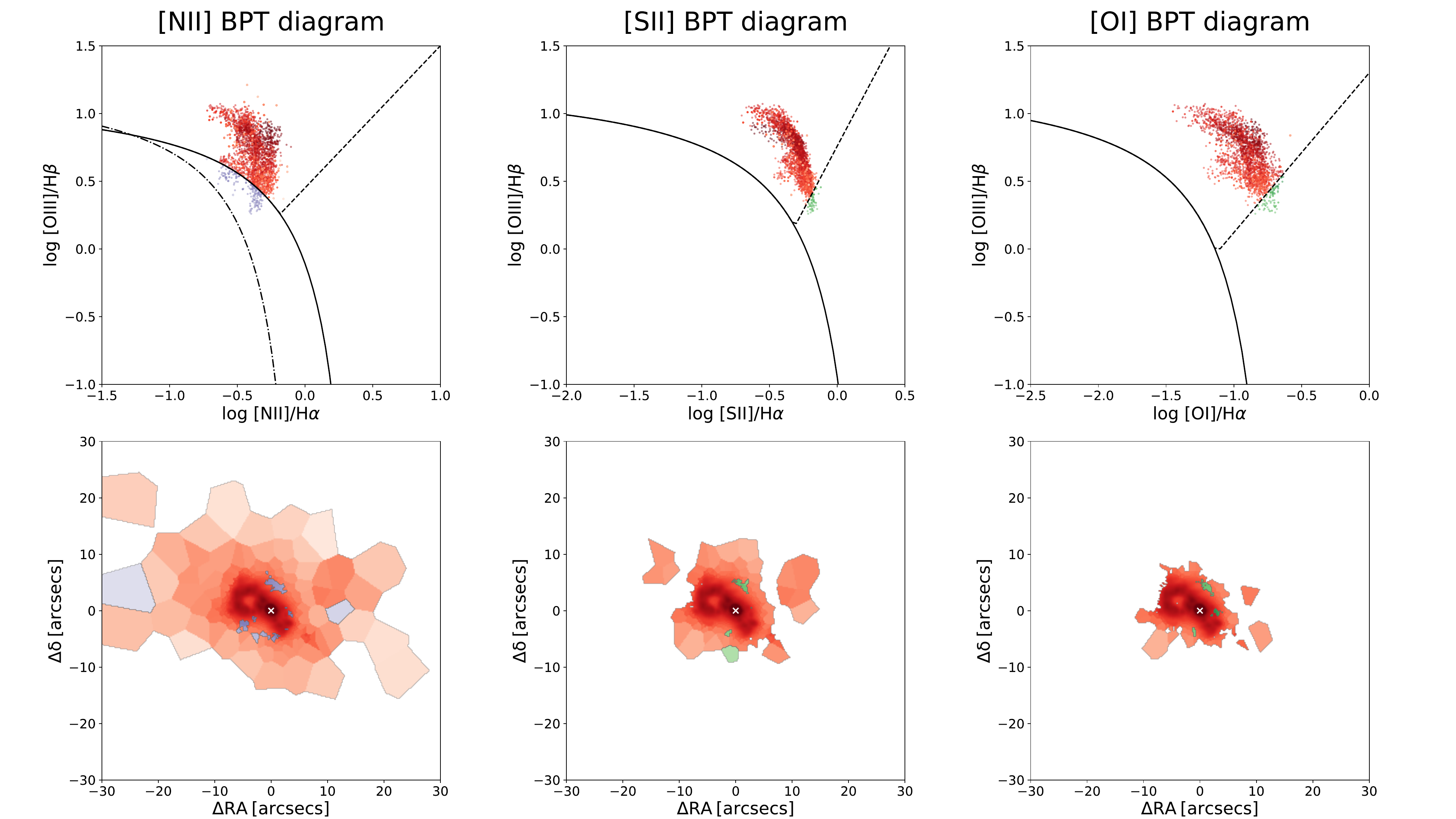}
\caption{Spatially resolved diagnostic diagrams {of gas excitation for} the Teacup, obtained from the star-subtracted Voronoi-binned data cube. \oiii$\ \lambda$5007/\hb\ vs. \nii$\ \lambda$6584/\ha\ (top-left), vs. \sii\ ($\lambda$6716+$\lambda$6731)/\ha\ (top-centre), and vs. \oi\ $\lambda$6300/\ha\ (top-right) diagrams, together with their respective spatial distributions (bottom panels). The fluxes of the total modelled emission line profiles have been employed for the diagrams.
The solid curve defines the theoretical upper bound for pure star formation from \cite{Kewley:2001a}, while the dashed one in the \nii-{diagram} is the \cite{Kauffmann:2003aa} empirical classification. The dot-dashed line represents the demarcation line between Seyfert galaxies and shocks or LI(N)ERs from \cite{Schawinski:2007a} in the \nii-{diagram} and from \cite{Kewley:2006a} in the \sii- and \oi-{diagrams}. 
{Regions dominated by AGN ionisation are then marked in red, composite regions in the \nii-{diagram} in purple, shock-ionised or LI(N)ER regions in green, and star formation-dominated regions would be marked in blue (when present).}
}
\label{fig:bpt}
\end{figure*}
	
\begin{figure*}
\includegraphics[width=\textwidth,trim={0 3.5cm 0 7cm},clip]{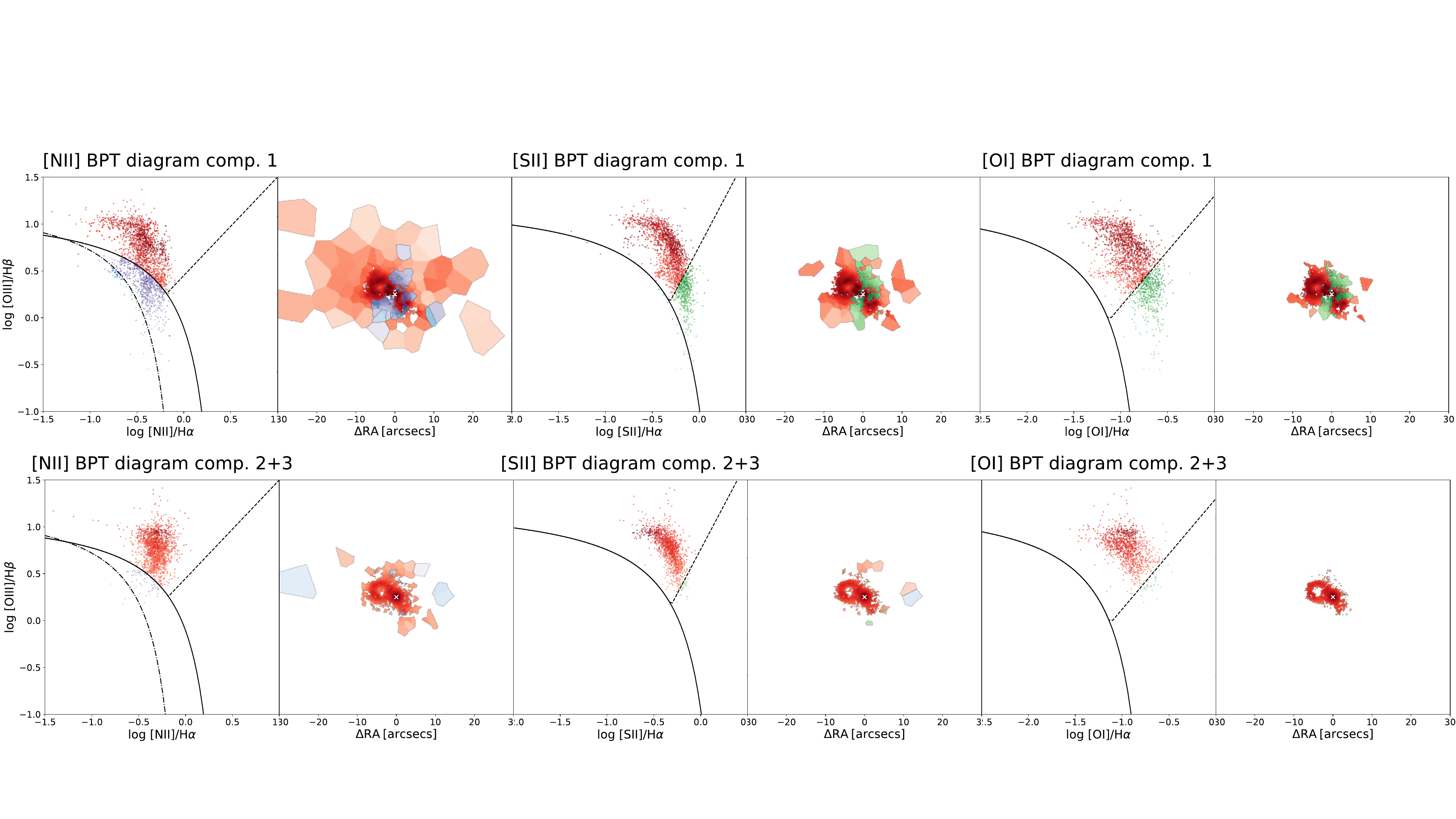}
\caption{Same as in Fig. \ref{fig:bpt}, but separately for narrower (first) and outflow (second plus third) modelled components.}
\label{fig:bpt_comp}
\end{figure*}

\subsection{Gas excitation}
In Fig. \ref{fig:bpt} we report the spatially resolved {emission-line ratio} diagnostic diagrams (\citealt{Baldwin:1981aa, Veilleux:1987aa}) for the Teacup, used to trace the dominant gas ionisation mechanism throughout and around the galaxy. The diagrams exploit ratios between emission lines close in wavelength, making these ratios virtually unaffected by dust extinction.

The diagrams show that the gas ionisation from AGN photons dominates over almost the whole system (red colour), from the nucleus to the optical lobes and up to the outskirts of the 100-kpc giant gas nebula. However, shock-like ionisation is present in a region close to the nucleus perpendicular to the \oiii\ lobes (green colour), in the same direction where enhanced velocity dispersion is observed ($\gtrsim$300 km s$^{-1}$; see Fig. \ref{fig:kinmaps},  bottom-right and mid-right panels).
This is also observed in the single-component diagrams reported in Fig. \ref{fig:bpt_comp}, though only in the first (narrower) component and not in the second (broader) one.
In the \nii-{diagram} this region does not fall within the area of the diagram associated with shocks, but in the composite region (purple colour), however, the \nii/\ha\ map in Fig. \ref{fig:line_ratios} shows that this region is characterised by higher values of the line ratio than in the direction of the optical lobes, which may be ascribed to a shock contribution to the gas ionisation. \sii/\ha\ and \oi/\ha\ (also reported in Fig. \ref{fig:line_ratios}) are more sensitive than \nii/\ha\ to the presence of shocks (see e.g. \citealt{Allen:2008aa, Sharp:2010aa}), which {can} explain why there is this discrepancy between the \nii-{diagram} and the other diagrams.
Moreover, the curves in these {diagnostic} diagrams are in fact not strict separations between the different ionisation regimes.
{Shocks with velocities between $\sim$200--400 km/s, comparable with the enhanced velocity dispersion observed, can indeed reproduce the observed ratios in the composite region of the \nii-diagram as well (e.g. \citealt{Allen:2008aa, Ho:2014aa}).}
We further discuss these aspects in Sect. \ref{sec:enhanced_sigma}.

\begin{figure*}
\centering
\begin{minipage}[c]{0.49\textwidth}
\centering
\includegraphics[scale=0.3,trim={1cm 0cm 2cm 0.5cm},clip]{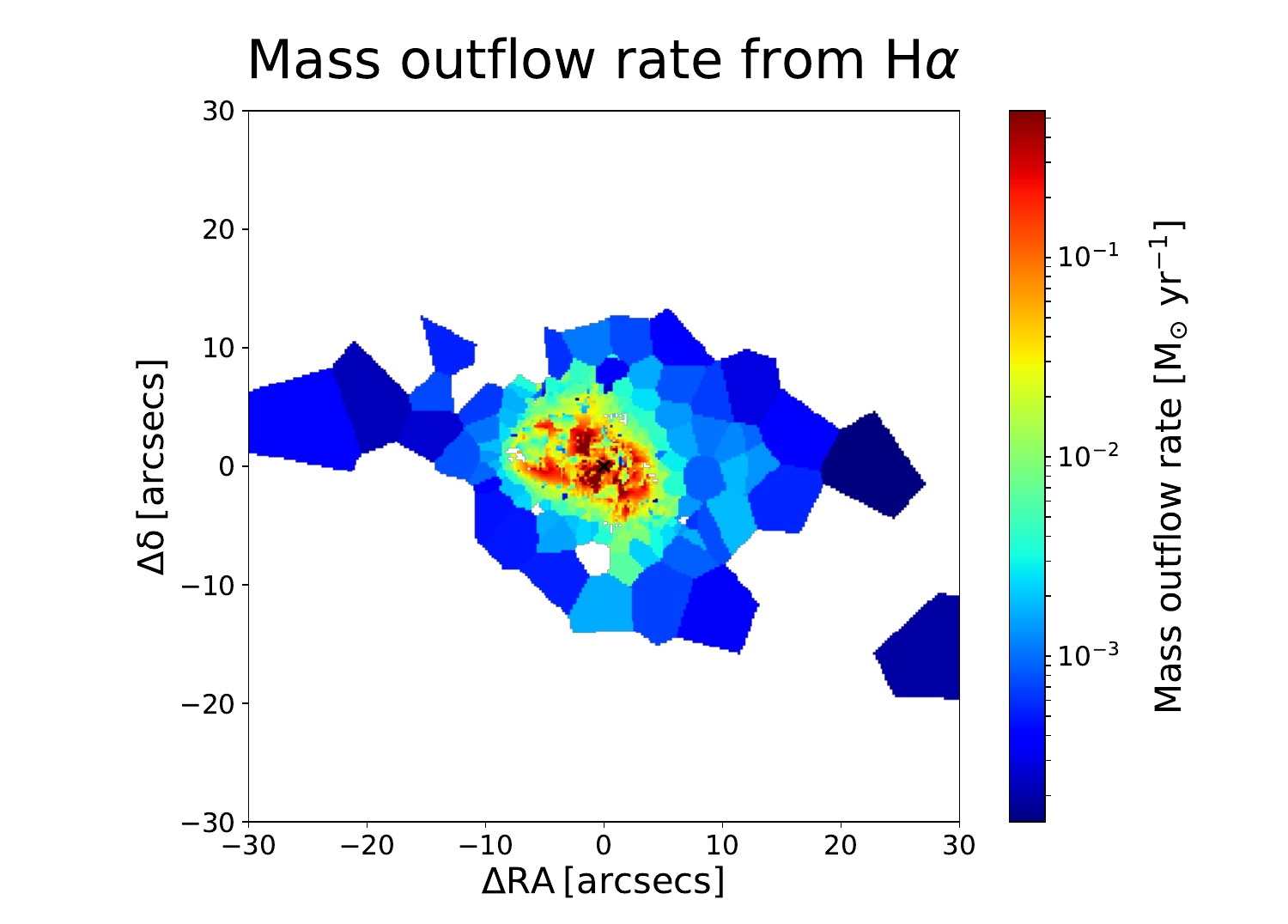}\\
\includegraphics[scale=0.6,trim={0cm 0.3cm -0.9cm -0.5cm},clip]{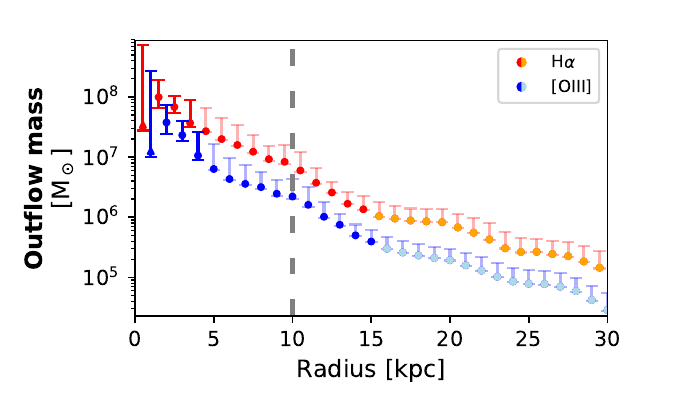}
\end{minipage}
\hfill
\begin{minipage}[c]{0.5\textwidth}
\centering
\includegraphics[scale=0.61,trim={-1.3cm 0.6cm 0 1cm},clip]{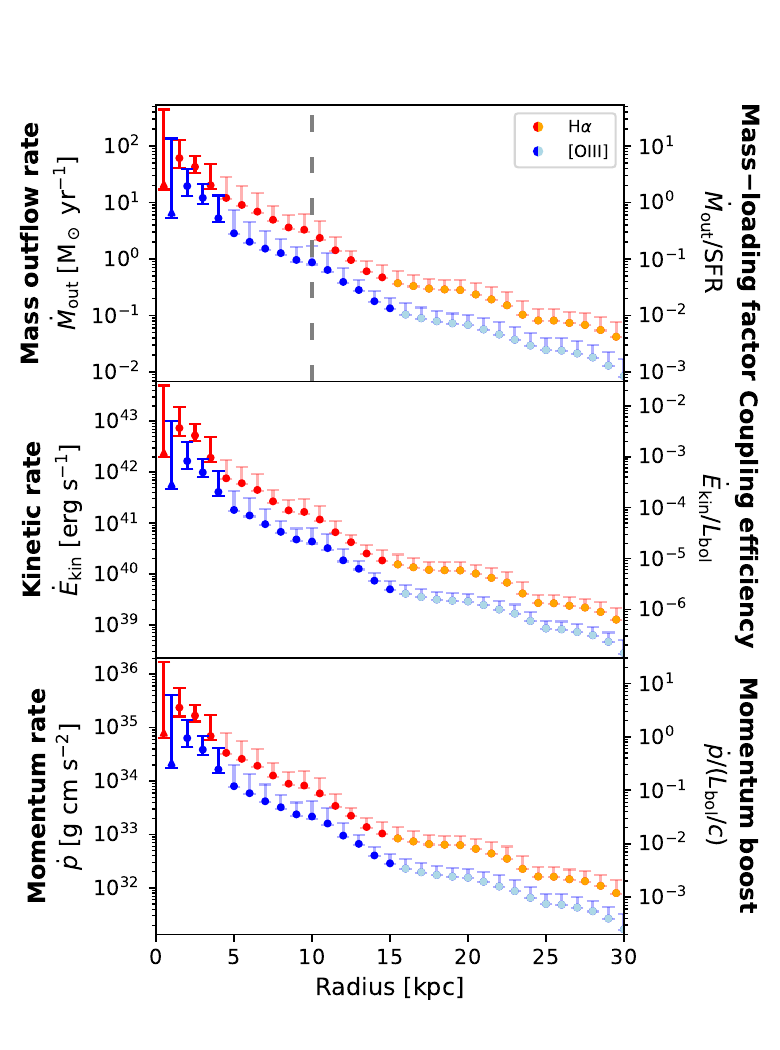}
\end{minipage}\\
\includegraphics[scale=0.61,trim={0 0.2cm 0 0},clip]{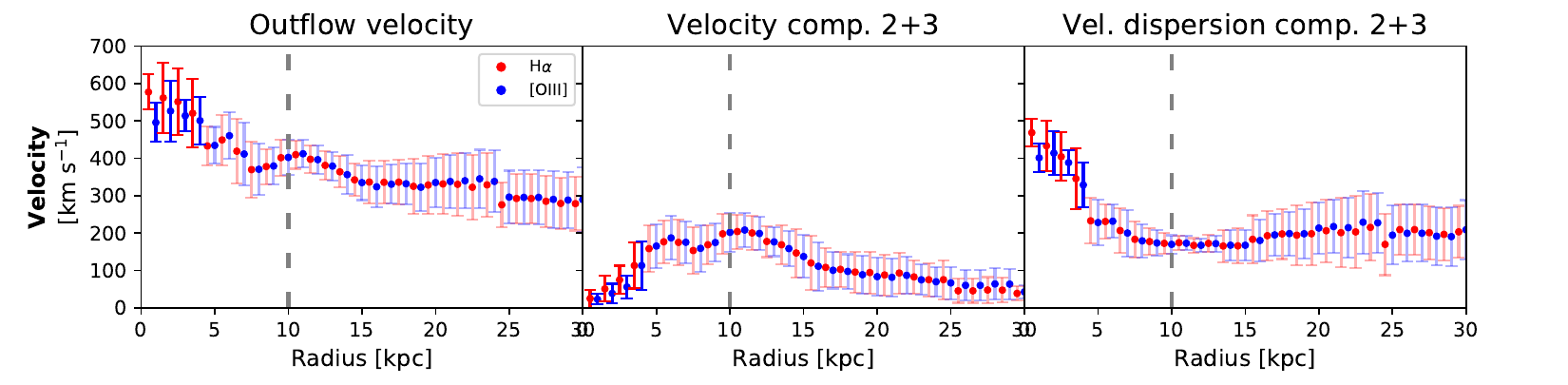}
\caption{Spatially resolved properties of the galactic ionised outflow in the Teacup. Top-left: Mass outflow rate map, obtained {from the outflow (second plus third) modelled component(s) of \ha}. 
Top-right: Radial profiles of mass outflow rate (top), kinetic rate (middle), and momentum rate (bottom) as a function of distance from the nucleus, from \ha\ (red and orange) and \oiii\ (blue and light blue) outflow components. {These quantities were obtained by summing up the single-spaxel values (calculated through Eqs. \ref{eq:moutrate}, \ref{eq:kinrate}, and \ref{eq:momrate}) contained in each radial bin, and re-scaling by $\Delta R$/$\Delta R_\mathrm{radbin}$, being $\Delta R_\mathrm{radbin}$ the radial bin width (=1 kpc) and $\Delta R$ the spaxel size.} Orange and light blue points mark those external radial bins for which we adopted a single median gas density, since no density measurement was available (see Sect. \ref{sec:ionised_outfl} for details). A different symbol is used for the innermost radial bin, which comprises a few spaxels whose outflow density could not be constrained (see Sect. \ref{sec:ionised_outfl}). Lighter error bars are used when the assigned statistical uncertainties were estimated from representative regions instead of on a spaxel basis (see Sect. \ref{sec:ionised_outfl}). The vertical dashed lines mark the distance of the handle. 
{On the rightmost $y$ axis, we also report mass-loading factor (top), coupling efficiency (middle), and momentum boost (bottom) of the outflow, for which we consider SFR $\sim$ 10 $M_\odot$ and $L_\mathrm{bol}$ $\sim$ 5$\times$10$^{45}$ erg/s (see text).}
Mid-left: Radial profiles of ionised outflow mass, from the flux of the outflow component(s) of \ha\ and \oiii. 
Bottom: Radial profiles of outflow velocity (left; $v_\mathrm{out}$ = $v_{2+3}$ + FWHM$_{2+3}$/2 $\simeq$ $1.18 \sigma_{2+3}$), combination of velocity, $v_{2+3}$ (centre), and velocity dispersion, $\sigma_{2+3}$ (right), of second plus third modelled components, {for both \oiii\ and \ha}. 
The reported values are the flux-weighted (by the outflow component(s)) averages in each radial bin.
We stress that the reported error bars are not statistical errors, but are a combination of the statistical errors and the standard deviation of the values, to show the extent of variation of the velocities in each radial bin.}
\label{fig:moutrate}	
\end{figure*}


\section{Characterisation of the extended ionised outflow}\label{sec:ionised_outfl}
The spatially resolved nature of MUSE observations, together with its wide FOV (1$'\times$1$'$), allows for the properties of the ionised outflow in the Teacup to be characterised on a wide range of scales (from $\sim$1 to 100 kpc) as a function of distance from the active nucleus.
We consider the second and third, broader {($\sigma$ $\gtrsim$ 150--200 km/s)} Gaussian components from our line modelling as pertaining to the outflowing gas and we adopt the physical properties found for these two components to calculate the properties of the outflow.
{As discussed in Sect. \ref{ssec:kinematic_maps}, while all the ionised gas (including the first, narrower component) might be in outflow as part of the bubbles, we adopt a more conservative approach and consider only the broader components to estimate the outflow properties.}

The summed flux of the two broader components is used to compute properties like outflow density and extinction, while velocity and velocity dispersion are obtained as the first- and second-order moments of the velocity of the summed profile of the two broader components, respectively.
We employ the maps produced from the {stellar continuum}-subtracted Voronoi-binned cube, in order to have a sufficiently high S/N also in the fainter regions and thus extend this analysis to larger distances from the nucleus.

First, we estimate the mass of gas contained in the outflow by converting the extinction-corrected flux of the outflow components of \ha\ and \oiii, using the two following relations (\citealt{Carniani:2015aa,Fiore:2017aa}):
\begin{equation}
    M_{\mathrm{out,H}\alpha}/\mathrm{M}_\odot = 0.6 \times 10^9 
    \left( \dfrac{L_{\mathrm{H}\alpha}}{10^{44}~\mathrm{erg~s}^{-1}} \right) 
    \left( \dfrac{n_\mathrm{e}}{500~\mathrm{cm}^{-3}} \right)^{-1} 
\label{eq:mass_ha}
\end{equation}
and
\begin{equation}
    M_\mathrm{out,\oiii}/\mathrm{M}_\odot =
    0.8 \times 10^8 
    \left( \dfrac{L_\mathrm{\oiii}}{10^{44}~\mathrm{erg~s}^{-1}} \right) 
    \left( \dfrac{n_\mathrm{e}}{500~\mathrm{cm}^{-3}} \right)^{-1} 
\label{eq:mass_oiii}
\end{equation}
which assume fully ionised gas with electron temperature $T_\mathrm{e}$ = 10$^4$~K {(fairly in agreement with the one estimated from the \oiii\ diagnostic ratio in Sect. \ref{ssec:gas_props})}, $\langle n_e \rangle^2 / \langle n_e^2 \rangle$ = 1 and, for the latter relation, a solar oxygen abundance.

We stress that to estimate extinction and density in each bin we also adopted the outflow (second plus third) modelled components, not the total line fluxes. We used \ha/\hb\ for tracing extinction instead of \hg/\hb\ or a combination of the two because the former gives more reliable ratios between the faint outflow components due to the higher S/N of \hb\ compared to \hg.
The maps of extinction and density of the outflow components (reported in the top-left and bottom-left panels of Fig. \ref{fig:Avdens_c2+grid}, respectively) have some problems. {Some deviant values are present} due to the fact that the wings of \hb\ and \sii\ can be faint and the fitting procedure may have not been optimal in all cases. Therefore, in order to obtain more robust spatially resolved estimates of extinction and density for the second component, as needed to calculate the mass of ionised gas, we masked the most deviant values ($>$3$\sigma$ from the mean at a given distance from the nucleus). We also masked isolated external bins (usually at the edge of the S/N threshold), for which the validity of the inferred value could not be assessed from neighbouring bin values. 
Finally, we applied a median filter with a kernel size of 3 spaxels to the \ha/\hb\ and \sii\ ratio maps, before calculating extinction and density, in order to mitigate high spatial frequency variations below the spatial resolution of the observations. 

To get robust estimates of the outflow quantities in the central region, where the strongest outflows are present, as well as to obtain reliable uncertainties on them, we performed Monte Carlo simulations with 10000 iterations of the three-component emission line fitting (as described in Sect. \ref{ssec:data_anal}) in each spaxel within a radius of 10 spaxel from the nucleus; in each iteration, the flux in each spectral channel was varied by the error on the flux times a random normal distribution having standard deviation equal to 1. In some spaxels, some parameters accumulated at {their extremes allowed by the fitting procedure for most of the iterations, especially the \sii\ ratio, accumulating at its upper limit (lower limit for the density).}  In these cases, we loosened {the constraints in} the Monte Carlo simulations by allowing the spectra to vary by an amount equal to the error on the flux times a random normal distribution having standard deviation equal to 3.
We could not run a Monte Carlo simulation in each spaxel of the data cube because the computational time would have been excessive.

The density could not be constrained in a handful of spaxels (seven, {for a total of $\sim$0.8$''$ in diameter}) around the nucleus, even with the Monte Carlo simulations, due to the \sii\ doublet line profiles being very broad and blended, making the fit highly degenerate. We therefore assigned a typical density value of the central regions to these degenerate spaxels, to allow for the calculation of outflow mass and derived quantities. This value has been estimated by first calculating the median density in each 1-spaxel wide radial annulus comprised between 3 to 8 spaxels from the nucleus, and then obtaining the 50th-percentile of these annular median densities. We stress however that in these seven nuclear spaxels the density is basically unconstrained and this has a deep impact on the total nuclear mass since they host the highest outflow line fluxes. We return to this aspect later in this Sect.

The resulting maps of extinction and density of the outflow are reported in top-right and bottom-right panels of Fig. \ref{fig:Avdens_c2+grid}, respectively. Here, we set to 10 cm$^{-3}$ those densities falling below this value, {where} the \sii\ ratio gets insensitive to density.
{Outflow extinction and electron density range between $A_V$ $\sim$ 0--1.6 mag and $n_e$ $\lesssim$10 to $\sim$5000~cm$^{-3}$, respectively.}
{In Appendix \ref{sec:appendix_dens} we explore alternative independent methods to estimate the electron density.}

In order to get a value of extinction and density in the outer regions, where no values are available, and be able to calculate the gas mass in those regions, we assigned them the median values calculated over the available areas (excluding the central region, where these quantities peak and is therefore not representative of the more external areas). The medians were calculated at the level of the \ha/\hb\ and \sii\ line ratios, and then converted to their relative physical quantities. This resulted in a median extinction and density of $A_V$ $\simeq$ 0 and $n_e$ $\simeq$ 12 cm$^{-3}$.

We thus employed these polished maps of extinction and density of the outflow components to obtain the mass of ionised gas in each spaxel using Eqs. \ref{eq:mass_ha} and \ref{eq:mass_oiii}.
We stress that electron densities of $\sim$10 cm$^{-3}$ are to be considered as upper limits, since this is the threshold below which the \sii\ ratio is not sensitive to density anymore. Therefore, mass outflow rates resulting from densities of $\sim$10 cm$^{-3}$ are to be considered as lower limits.
Given that electron densities of the ISM can be as low as $\sim$1 cm$^{-3}$ (e.g. \citealt{Osterbrock:2006a}),
the difference can be at most by a factor of $\sim$10.

For the mass outflow rate calculation, we follow the same approach adopted in \cite{Venturi:2018aa} (see also e.g. \citealt{Maiolino:2012aa, Gonzalez-Alfonso:2017a, Harrison:2018aa, Revalski:2021a} for a thorough discussion on mass outflow rate calculation). The mass outflow rate through a spherical surface of radius $r$ subtended by a solid angle $\Omega$ is $\dot{M}(r) = \Omega r^2 \rho(r) v(r)$. Assuming that, within each shell with radial thickness $\Delta R$, density and outflow velocity are constant, the radial average of the outflow rate within $\Delta R$ is
\begin{equation}
    \dot{M}_\mathrm{out} = \dfrac{M_\mathrm{out} v_\mathrm{out}}{\Delta R}, \label{eq:moutrate}
\end{equation}
where $M_\mathrm{out}$ is the outflow mass (obtained from either \ha\ or \oiii), $v_\mathrm{out}$ the outflow velocity, 
and $\Delta R$ the physical width of each spaxel (0.2$''$, i.e. $\sim$0.3 kpc).
For the outflow velocity, we adopted a combination of velocity and width of the outflow Gaussian component(s) from the emission-line modelling, in accordance with many other works (e.g. \citealt{Rupke:2005aa, Fiore:2017aa, Fluetsch:2019aa}). The idea behind this is that the observed line profile of the outflowing gas is given by the combination of velocities directed at different angles with respect to the line of sight, and therefore, due to projection effects, only the highest observed velocities represent the actual velocity of the outflowing material, when this is directed in the observer's line of sight.
Specifically, we adopt the definition used for instance in \cite{Rupke:2005aa} and \cite{Fluetsch:2019aa}, that is,
\begin{equation}
  v_\mathrm{out} = v_2 + \mathrm{FWHM}_2/2 \simeq v_2 + 1.18\, \sigma_2,  
\end{equation}
 where $v_2$, FWHM$_2$, and $\sigma_2$ are the centroid velocity, the FWHM, and the velocity dispersion of the second modelled component, respectively. In the central region, where also a third Gaussian component was used, the first- and second-order moments of velocity calculated over the summed line profile of second plus third components ($v_{2+3}$ and $\sigma_{2+3}$, respectively), have to be substituted in the equations above.

The map of the mass outflow rate thus obtained from the outflow (i.e. second, plus third when present) modelled Gaussian component(s) of \ha\ is reported in Fig. \ref{fig:moutrate}, top-left panel. 
{The analogous map obtained from \oiii\ outflow component(s) is very similar (except for having smaller values overall, as we discuss below), therefore we do not show it.}
We then produced radial profiles of the mass outflow rate as a function of distance from the nucleus, that are displayed in the top-right panels of Fig. \ref{fig:moutrate}. The profiles are sampled in bins of 1 kpc.
These radial profiles have been produced by summing, in each radial bin, the single-spaxel mass outflow rates from the maps in left panels, and re-scaling by $\Delta R$/$\Delta R_\mathrm{radbin}$, being $\Delta R_\mathrm{radbin}$ = 1 kpc.

We first note that the mass outflow rate obtained from \ha\ is larger than that obtained from \oiii\ by about a factor of 3--4. This is not surprising, as also \cite{Carniani:2015aa} and \cite{Fiore:2017aa} report a higher outflow mass from \hb\ than \oiii, by a factor of $\sim$2 and $\sim$3, respectively. \cite{Carniani:2015aa} ascribe it as due to the different volumes from which the two lines are emitted, differing exactly by a factor $\sim$2 in their case study. 
The discrepancy is a bit larger in the case of the Teacup. The reason for this discrepancy is unclear, one possibility could be that the radiation field in the Teacup is less energetic, giving a weaker \oiii\ emission for a given Balmer emission line flux ({the latter instead directly relating} to the number of ionising photons and not to the hardness of the radiation field; e.g. \citealt{Osterbrock:2006a}).
{In the rest of the work, we focus on the outflow quantities obtained from \ha, which is a more robust tracer of the outflow mass compared to \oiii, not subject to metallicity variations and to the uncertainties mentioned above. We leave the radial profiles of outflow mass (and derived quantities) obtained from \oiii\ for comparison.}

We stress that the mass outflow rate in the innermost radial bin is mostly unconstrained, due to density being unconstrained in seven nuclear spaxels, as mentioned earlier in this Sect. As said, the outflow line fluxes in these spaxels are among the highest, therefore, even if they are only very few, their impact on the mass outflow rate of the innermost radial bin is dominant. We took this into account by assigning errors spanning the whole density range ($\sim$1--50000 cm$^{-3}$) to the nuclear problematic spaxels and highlighted this aspect by using a different symbol for the innermost radial bin in the figure.

The mass outflow rate radial profiles exhibit a decreasing trend with distance, {dropping by about three orders of magnitude over 30 kpc, from $\sim$10$^2$ $M_\odot$/yr to $\lesssim$10$^{-1}$ $M_\odot$/yr. 
Radially decreasing mass outflow rates are} observed in the case of other ionised outflows in AGN (e.g. \citealt{Venturi:2018aa, Revalski:2018aa, Revalski:2021a}) and in principle could be both ascribed to a slowing down of the outflow with distance and to the fact that the AGN ionising flux drops as $r^{-2}$.

We note that in the case of the cited studies the mass outflow rate peaks at distances in the range $\sim$0.1--1 kpc, not on the nucleus. In the case of the Teacup, it is unclear whether the mass outflow rate peaks on the nucleus or at larger distances, since it is highly unconstrained in the innermost radial bin, as discussed above. Moreover, the spatial resolution of the MUSE WFM observations analysed in this work, $\sim$1$''$, corresponding to $\sim$1.7 kpc, is coarser than the scales of the peaks reported in the above works. 
Models of outflows driven by AGN radiation pressure predict that the mass outflow rate is expected to peak at about 1--2 kpc from the nucleus and then decrease with distance (the latter aspect is observed {in this study and in those} mentioned above) as the outflow propagates (\citealt{Costa:2018aa,Meena:2023a}).
Higher spatial resolution observations would therefore be required to investigate whether the mass outflow rate of the Teacup peaks on the nucleus or not. 
At larger distances, the mass outflow rate exhibits a small secondary peak at $\sim$10 kpc from the nucleus, corresponding to the {size of the handle of the Teacup}, indicated as a vertical dashed line in the radial profiles.

In the remaining top-right panels of Fig. \ref{fig:moutrate} we report the radial profiles of kinetic rate (middle) and momentum rate (bottom) of the ionised outflow. These are obtained as follows:
\begin{equation}
    \dot{E}_\mathrm{kin} = \dot{M}_\mathrm{out} v_\mathrm{out}^2 /2
    \label{eq:kinrate}
\end{equation}
and
\begin{equation}
    \dot{p}_\mathrm{out} = \dot{M}_\mathrm{out} v_\mathrm{out}
    \label{eq:momrate}
\end{equation}
for kinetic and momentum rate, respectively.
They also show a decreasing trend with distance from the nucleus and a secondary peak at $\sim$10 kpc, at the scale of the handle, as in the case of the mass outflow rate.

By inspecting in Fig. \ref{fig:moutrate} the radial profiles of outflow mass (mid-left), outflow velocity (bottom-left), as well as velocity, $v_{2(+3)}$ (bottom-centre), and velocity dispersion, $\sigma_{2(+3)}$ (bottom-right), of the second (plus third, when adopted) modelled spectral components, which define the outflow velocity ($v_\mathrm{out}$ $\simeq$ $v_{2(+3)}$ + $1.18 \sigma_{2(+3)}$), it is possible to understand which of these quantities determines the shape of the radial profiles of mass outflow rate, kinetic rate, and momentum rate. We thus see that the outflow mass and the velocity dispersion of the second plus third component(s), rather than their velocity shift, drive the shape of the radial profiles of the above outflow properties.

If we adopt instead the more extreme definition from {\cite{Rupke:2013a}, \cite{Feruglio:2015aa}, \cite{Fiore:2017aa}, and many others}, $v_\mathrm{out}$ = $v_2$ + $2 \sigma_2$, the resulting mass outflow, kinetic, and momentum rates are a factor of $\sim$ 1.7, 5, and 3 higher, respectively, than those obtained with the definition we adopted.
Another different approach would have been to employ the centroid velocity of the second (plus third) component(s) for the outflow velocity, without any combination with the velocity dispersion, $v_\mathrm{out}$ = $v_{2(+3)}$. The assumption behind this would be that the broadening of the emission line profiles is given by a real distribution of velocities of the outflow and not by projection effects due to the outflow propagating along different angles with respect to the observer's line of sight. However, considering that the observed $v_{2(+3)}$ are quite small ($\lesssim$100 km/s) where the line profiles are broadest, around the centre (up to $\sigma_{2(+3)}$ $\sim$ 400--500 km/s; see Fig. \ref{fig:kinmaps}, top-right and mid-right panels, and Fig. \ref{fig:moutrate}, bottom-centre and bottom-right panels), we find this alternative assumption unlikely and we have thus not adopted this approach.
We point out that, in this case, the values of mass outflow, kinetic, and momentum rate would be a factor of $\sim$3--20, 10--3000, and 5--200 times lower, respectively, compared to those resulting from the adopted approach, with the largest differences occurring at the smaller radii from the nucleus, where the two approaches give very different outflow velocities ($\gtrsim$500 km/s in case of the adopted definition versus $\lesssim$100 km/s in case of the alternative $v_\mathrm{out}$ = $v_{2(+3)}$ assumption; Fig. \ref{fig:moutrate}, bottom panels).

\subsection{Outflow energetics: Jet versus AGN acceleration}\label{ssec:outfl_accel}
The comparison between the outflow and the radio jet energetics allows us to obtain insights on the outflow driving mechanisms.
\cite{Harrison:2015a} infer a jet kinetic luminosity of $\dot{P}_\mathrm{jet}$ $\sim$ 2.5$\times$10$^{42}$ erg/s from the radio luminosity of the two high-resolution knots HR-A and HR-B, assuming that 1$\%$ of the total jet energy is converted into radio luminosity, {while \cite{Audibert:2023a} find $\dot{P}_\mathrm{jet}$ $\sim$ 1--3$\times$10$^{43}$ erg/s from the empirical relations of \cite{Birzan:2008aa} and \cite{Cavagnolo:2010aa}. Therefore, we adopt a range of jet powers, $\dot{P}_\mathrm{jet}$ $\sim$ $10^{42.5-43.5}$ erg/s for our comparisons.}

To check whether the jet is capable of accelerating the ionised outflow, its power should be compared with the outflow kinetic rate on similar scales 
($\sim$1 kpc), which correspond to those of the innermost radial bin (Fig. \ref{fig:moutrate}). However, this is mostly unconstrained since it is affected by the degeneracy in estimating the density in a few nuclear spaxels.
We therefore compare the jet power with the kinetic rate in the second innermost radial bin. This ranges between 5$\times$10$^{42}$--2$\times$10$^{43}$ erg/s considering the uncertainties. 
{This would require a transfer of $\sim$15~$\%$ up to 100~$\%$ of the jet kinetic power to the ISM. Transfer efficiencies of the jet energy to the kinetic energy of the ISM of $\lesssim$30~$\%$ are expected from simulations (e.g. \citealt{Wagner:2011a,Mukherjee:2016a}), which only partially overlap with the above range.
Therefore, purely from the energetic point of view, the jet alone is able to drive the outflow only when simultaneously considering the lower end of the inferred outflow kinetic rates and the upper end of the jet powers.}

We can also investigate if the AGN alone is able to accelerate the outflow.
Different outflow acceleration modes are predicted by theory. In the thermal feedback scenario, a fast wind accelerated by the AGN on accretion-disc scales drives the expansion of a hot, shocked bubble in the galaxy ISM (e.g. \citealt{King:2003aa,King:2015aa,Costa:2014aa}). Depending on the cooling efficiency, the bubble may either radiate away most of its energy and push the ambient ISM through its ram pressure (`momentum-conserving' outflow) or retain its energy while it expands, efficiently sweeping up the ISM it encounters (`energy-conserving' outflow). In the radiation pressure-driven scenario, the AGN is able to directly accelerate gas located on galactic scales through radiation pressure on dust, providing large momentum boosts of the outflow, $\dot{p}/(L_\mathrm{AGN}/c)$ (e.g. \citealt{Thompson:2015aa,Ishibashi:2015aa,Ishibashi:2018aa,Costa:2018aa}).

Therefore, we compare the above ionised outflow kinetic rate of the Teacup with its AGN bolometric luminosity, $L_\mathrm{AGN}$.
{We estimate $L_\mathrm{AGN}$ by using the 2--10 keV X-ray luminosity from \cite{Lansbury:2018a} ($L_\mathrm{X} \sim 0.8-1.4 \times 10^{44}$ erg/s) and the X-ray dependant bolometric correction $k_\mathrm{bol}(L_\mathrm{X})$ from \cite{Duras:2020a} for type-2 AGN, including its intrinsic spread of 0.27 dex. We thus obtain $L_\mathrm{AGN} \sim 7.2\times10^{44}-5.6\times10^{45}$ erg/s. We adopt an approximate mean value of $L_\mathrm{AGN}$ $\sim$ 2 $\times$ 10$^{45}$ erg/s, which matches the estimate from \cite{Harrison:2014a} from mid-to-far-IR SED fitting. 
We also adopt this value for the kinetic coupling efficiency and momentum boost radial profiles in Fig. \ref{fig:moutrate} (rightmost $y$ axes).}
We thus obtain a kinetic coupling efficiency of $\dot{E}_\mathrm{kin} / L_\mathrm{AGN}$ {$\sim$0.003--0.009,  where, as said above, we have considered the range of kinetic rates in the second innermost radial bin defined by the uncertainties. When considering the lower and upper estimates of the AGN bolometric luminosity, $\dot{E}_\mathrm{kin} / L_\mathrm{AGN}$ varies from $\sim$0.007--0.03 (for $L_\mathrm{AGN} \sim 7.2\times10^{44}$ erg/s) to $\dot{E}_\mathrm{kin} / L_\mathrm{AGN}$ $\sim$0.0009--0.003 (for $L_\mathrm{AGN} \sim 5.6\times10^{45}$ erg/s).}

{Despite the large uncertainty on $L_\mathrm{AGN}$ which prevents firm conclusions to be drawn, we can see that these ranges of kinetic coupling efficiencies are only partially consistent with the theoretical predictions for an energy-conserving outflow, $\sim$0.005--0.05 (e.g. \citealt{Hopkins:2010aa,Zubovas:2012aa,Costa:2014aa,King:2015aa}), while they are more generally compatible with a radiation-pressure driven one, for which kinetic coupling efficiencies of $\sim$0.001--0.01 are expected \citep[e.g.][]{Ishibashi:2018aa,Costa:2018aa}.}

The momentum boost, considering again the ionised outflow momentum rate in the second innermost radial bin, which ranges between $\dot{p}$ $\sim$ 2--6$\times$10$^{35}$ g\,cm/s$^2$ considering the uncertainties (Fig. \ref{fig:moutrate}), is $\dot{p}/(L_\mathrm{AGN}/c)$ $\sim$ 2--8 (for $L_\mathrm{AGN}$ $\sim$ 2 $\times$ 10$^{45}$~erg/s), $\sim$7--20 ($\sim7.2\times10^{44}$~erg/s), and $\sim$0.9--3 ($\sim5.6\times10^{45}$~erg/s).
{These are generally} too low compared to expectations for energy-conserving outflows ($\sim$20) and too
high for a momentum-conserving one ($\sim$1).
{It is instead generally} consistent with the range of momentum boosts of $\sim$1--10 predicted by models of outflows driven by direct radiation pressure on dust within scales of a few kpc (the first $\sim$5 Myr of propagation; \citealt{Ishibashi:2018aa, Costa:2018aa}), which are compatible with the innermost kpc scales we are considering here.

All in all, the large uncertainties involved in the above comparisons prevent firm conclusions to be drawn on the outflow driving mechanism.
It seems however that direct AGN radiation pressure on dust 
is more generally compatible with the observations compared to a thermal acceleration mechanism, either energy- or momentum-conserving.
From purely energetic arguments, the jet alone is capable of driving the outflow only when considering the lowermost kinetic rates and the uppermost jet powers.
Another possibility is that the ionised outflow is driven by the combined action of the jet and the AGN, through either direct radiation pressure or the action of an energy- or momentum-conserving bubble.

We note that the above estimates of the ionised outflow properties for the Teacup are 
around the high end of those found for ionised outflows in AGN of similar luminosity {($L_\mathrm{AGN} \lesssim 5 \times  10^{45}$ erg/s)}, based on current observational literature (\citealt{Carniani:2015aa, Fiore:2017aa, Bischetti:2019a}).
This may be due to the fact that these literature estimates are based on integrated spectra or barely resolved observations, for which an outflow mass integrated over several kpc and a single outflow radius are employed in Eq. \ref{eq:moutrate}. Such a radius may correspond to the largest distance reached by the outflow, when available, or by the ionised gas emission from imaging, or to the entire spectral aperture. This may consequently lead to a lower mass outflow rate than in the spatially resolved case, when considering the inner regions from the nucleus (corresponding to a small outflow radius), if the outflow mass peaks there as in the case of the Teacup.

{We then attempted to obtain an integrated mass outflow rate by modelling the spectrum extracted from a circular aperture centred on the nucleus. However, due to the mixing of multiple velocity components stemming from different regions resulting in very broad, blended \sii\ line profiles, the modelling is degenerate and cannot constrain the outflow density and thus its mass, as in the case of the handful of nuclear spaxels discussed in Sect. \ref{sec:ionised_outfl}. 
This issue persisted with different aperture sizes. Hence, we adopted a different approach and obtained a pseudo-integrated mass outflow rate by summing up the single-spaxel mass outflow rates within a circular aperture re-scaling the result by $\Delta R/R_\mathrm{ap}$,  with $\Delta R$ the width of each spaxel and $R_\mathrm{ap}$ the radius of the aperture. By using apertures with radii of 1.5 (SDSS-like), 5, 10, and 20 arcsec (corresponding to about 2.5, 8.3, 16.6, and 33.2 kpc, respectively), we obtained $\dot{M}_\mathrm{out} \sim$ 43, 21, 11, and 6 $M_\odot$/yr, respectively.
The value obtained from the SDSS-like aperture is consistent with those measured on similar scales for the second and third innermost radial bins in the mass outflow rate radial profile (Fig. \ref{fig:moutrate}, top-right panel) and the aperture size is compatible with the AGN outflow radii reported in the literature \citep{Carniani:2015aa, Fiore:2017aa, Fluetsch:2019aa, Fluetsch:2021a}.
In light of this, we adopt the mass outflow rate in the second innermost radial bin to carry out quantitative comparisons since it is more robust than the SDSS-like pseudo-integrated one, which includes the nuclear spaxels where the outflow electron density, and thus the outflow mass and mass outflow rate, could not be constrained.
}

{The higher ionised mass outflow rates of the Teacup compared to those observed in other AGN might be related to the fact that it was originally selected from SDSS images precisely because of its spatially resolved line emission, combining high surface brightness with large angular extent \citep{Keel:2012a}. This selection bias may have enhanced the probability of detecting a more powerful and extended ionised outflow.}


We also note that the momentum boosts we find for the ionised outflow in the Teacup are much lower than those reported for the same object and gas phase by \cite{Harrison:2014a} of $\dot{p}/(L_\mathrm{AGN}/c)$ $\sim$ 40. However, in their work, the mass outflow rate and related quantities are the result of the mean between two different methods, one employing the same approach used in this work (i.e. a relation similar to Eq. \ref{eq:moutrate}), the other considering an energy conserving bubble in a uniform medium (e.g. \citealt{Heckman:1990aa,Nesvadba:2006a,Veilleux:2005aa}), which gives values larger by a factor of $\gtrsim 100$--$1000$ compared to the former method.

\subsection{Multi-phase outflow}
The recent study of the molecular phase of the outflow in \cite{Ramos-Almeida:2022a} allows for a comparison with the ionised outflow component presented in this work. They detect a molecular outflow in the inner $\sim$0.5$''$ {($\sim$800 pc)} from the nucleus, with maximum velocities of $\sim$250 and $-$180 km/s for the approaching and the receding parts of the outflow, respectively. These are much lower (by a factor $\gtrsim$2--3) than the velocities up to more than 600 km/s found for the ionised component. A lower outflow velocity for the molecular phase 
is not surprising (see e.g. \citealt{Carniani:2015aa, Fluetsch:2019aa}) and can be ascribed to the stronger resistance to acceleration by the dense molecular gas.
\cite{Ramos-Almeida:2022a} report 
a de-projected mass outflow rate of $\dot{M}_\mathrm{out}$ = 15.8 $M_\odot$/yr for the cold molecular phase from ALMA CO observations on the same spatial scales. 
{In a more recent work, \cite{Audibert:2023a} considered four different outflow scenarios to calculate the molecular outflow mass of the Teacup, using the same CO(2-1) dataset as \cite{Ramos-Almeida:2022a}, and they reported mass outflow rates ranging between 15 and 41 $M_\odot$/yr.}

The mass outflow rate of the ionised gas that we obtain in the second innermost radial bin from Fig. \ref{fig:moutrate} (being the innermost one basically unconstrained; see Sect. \ref{sec:ionised_outfl}), ranging between $\sim$40--130 $M_\odot$/yr considering the uncertainties, {is  larger than that of the molecular gas, by a factor of $\sim$3--8 or $\sim$1--3 when considering the lowest (15 $M_\odot$/yr) or the uppermost (41 $M_\odot$/yr) values for the molecular phase, respectively.}
Therefore, the addition of the molecular budget to the ionised one does not substantially change the picture that we have obtained from the ionised phase alone, in terms of outflow acceleration mechanisms and impact on the host galaxy.

The discrepancy found in the Teacup between the two gas phases is surprising, given that the molecular phase is usually found to dominate the mass outflow rate budget, exceeding the ionised outflow one by a factor of $\sim$10--10$^3$ {at $L_\mathrm{AGN} \lesssim 10^{46}$ erg/s} (\citealt{Carniani:2015aa, Fiore:2017aa, Bischetti:2019a, Fluetsch:2019aa, Fluetsch:2021a}).
{We stress, however, that the limited number of molecular outflows so far quoted in literature may be biased to the most extreme values, because they are the easiest to detect in CO-bright galaxies.
We point out that, in the case of the Teacup, the ionised and molecular outflow properties} are obtained with different methods, since \cite{Ramos-Almeida:2022a} extract the outflowing flux {by integrating only the high-velocity gas in a slit of 0.2$''$ oriented along the CO disc minor axis (E-W direction),} while we employ the flux of the second, broad component of the fit (calculated all over the regions at the same distance from the nucleus, not only in a 0.2$''$-wide slit). This might explain the large discrepancy in favour of the ionised outflow. 
Moreover, the molecular outflow resides in the inner $\sim$0.5$''$ ($\sim$0.8 kpc), while the ionised mass outflow rate of the second innermost radial bin we have considered for the comparison resides on scales of $\sim$2 kpc. 
Higher-resolution observations for the ionised gas are therefore needed for a matching with the molecular outflow on the same spatial scales.

We can then evaluate the effect of the outflow on the star formation activity and molecular gas reservoir in the Teacup.
Considering the SFR of the Teacup {(between $\sim$8--12 $M_\odot$/yr, from FIR emission; \citealt{Jarvis:2020a,Ramos-Almeida:2022a})}, the mass-loading factor ($\eta$ = $\dot{M}_\mathrm{out}$/SFR) of the ionised gas is {$\sim$3--30} ($\sim$5--{30} when also considering the molecular outflow) at the second innermost radial bin. This means that 
the outflow consumes the gas reservoir much faster than star formation processes. Mass-loading factors above 1, and even exceeding 10, are often found in AGN (e.g. \citealt{Cicone:2014aa, Fiore:2017aa, Fluetsch:2019aa}).

We can calculate the molecular gas depletion time due to outflows, that is, the time needed to remove from the galaxy all the molecular gas reservoir available for star formation given the current mass outflow rate (assuming that the {cold} gas content is not replenished with fresh supply), as $\tau_\mathrm{depl} = M_\mathrm{gas}/\dot{M}_\mathrm{out}$, {where $\dot{M}_\mathrm{out}$ is the mass outflow rate of the multi-phase (ionised plus molecular) outflow}. Here we consider $M_\mathrm{gas}$ $\sim$ $M_\mathrm{H_2}$, for which we employ the total molecular gas mass of $M_\mathrm{H_2}$ $\sim$ 6$\times$10$^9$ $M_\odot$ from \cite{Jarvis:2020a} and \cite{Ramos-Almeida:2022a}. This gives $\tau_\mathrm{depl}$ $\sim$ 4$\times$10$^7$--10$^8$ yr, consistent with the values found for AGN of similar luminosity ($L_\mathrm{AGN}$ $\sim$ {10$^{45-46}$ erg/s}; \citealt{Cicone:2014aa, Fluetsch:2019aa}), {while the depletion time due to star formation ($M_\mathrm{gas}$/SFR) is larger, $\sim 5-8 \times 10^8$ yr (consistent with the values expected for galaxies above the star-forming main sequence at the redshift of the Teacup; e.g. \citealt{Schinnerer:2016aa,Tacchella:2016aa,Tacconi:2018aa}).} 
Such short depletion timescales indicate that the outflow may be effective in depleting the molecular gas reservoir. 
We stress, however, that these are the actual mass-loading factor and depletion timescale only if the outflow continues at the present rate (actually, at the rate given by the second innermost bin considered, at a scale of $\sim$2 kpc). 

The ionised gas emission observed on large scales actually indicates that the AGN in the Teacup was more luminous in the past ($\sim$10$^5$ yr timescales) and has faded since then, by a factor of $\sim$10 considering a present-day bolometric luminosity of $L_\mathrm{AGN}$ $\sim 2\times 10^{45}$ erg/s (see \citealt{Gagne:2014a, Keel:2017a, Villar-Martin:2018a} and discussion in \citealt{Lansbury:2018a}). Therefore, {the nuclear outflow would have also likely been higher in the past than today and consequently the past gas depletion times may have been even lower than that estimated above from present-day nuclear outflow rate}, suggesting that the outflow in the Teacup could {have been even more effective than today in depleting} the molecular gas reservoir in the galaxy.

To estimate to which extent the outflow has actually affected the host galaxy, we compare the mass of gas participating in the outflow with the molecular gas reservoir available from star formation. We obtain the total ionised mass in outflow by summing together the outflow mass values found from our spatially resolved analysis. This gives an ionised outflow mass of $\sim$3.6$\times$10$^8$ $M_\odot$, or a total outflow mass of $\sim$4$\times$10$^8$ $M_\odot$ when also considering the molecular mass ($\sim$3$\times$10$^7$ $M_\odot$; \citealt{Ramos-Almeida:2022a}). The outflow mass is then about 5$\%$ the total molecular gas mass in the galaxy, $\sim$ 6$\times$10$^9$ $M_\odot$. 

We can evaluate what fraction of the outflow is actually able to escape the galaxy gravitational potential. We calculate the escape velocity by considering a Hernquist potential (\citealt{Hernquist:1990a}) for the stellar component and a Navarro-Frenk-White (NFW) potential (\citealt{Navarro:1997a}) for the dark matter (DM) halo,
given by the following density profiles, respectively: 
\begin{equation}
    \rho(r) = \frac{M_\star}{2 \pi} \frac{a}{r} \frac{a}{(r+a)^3}
\end{equation}
and
\begin{equation}
    \rho(r) = \frac{\rho_\mathrm{crit} \delta_\mathrm{c}}{(r/r_\mathrm{s}) (1+r/r_\mathrm{s})^2} .
\end{equation}
In the former, $M_\star$ is the stellar mass of the Teacup, for which we adopt $\log (M_\star/M_\odot)$ $\simeq$ 11.15 (\citealt{Ramos-Almeida:2022a}), and $r_\mathrm{eff}$ $\sim$ 3.7 kpc its effective (half-light) radius from the SDSS $g$-band de Vaucouleurs profile fit (\citealt{SDSS:DR17}), where $r_\mathrm{eff} \simeq 1.8135 a$.
In the latter, $\rho_\mathrm{crit}$ is the critical density of the Universe, $\delta_\mathrm{c}$ the characteristic overdensity for the halo, and $r_\mathrm{s}$ = $r_\mathrm{200}/c$ the characteristic radius, $c$ being the concentration parameter. The virial DM halo mass ($M_{200}$) is inferred from the stellar mass via the stellar-to-halo mass relation (SHMR) of \cite{Moster:2013a}, inverted following \cite{Posti:2019a}, and the concentration $c$ is obtained through the $M_{200}-c$ relation of \cite{Dutton:2014a}. We thus obtain $\log (M_{200}/M_\odot)$ $\sim$ 13.2.
We calculate the escape velocity for each spaxel in the FOV, each having a given distance $r$ from the centre of the galaxy. This results to be in the range $\sim$1200--1500 km/s for the radial distances considered ($<$30 kpc).

We then obtain the portion of flux of the outflow component (the second and third we have used in the line fitting; see Sect. \ref{sec:ionised_outfl}) having a velocity, either blue- or red-shifted, larger than the escape velocity, $|v|$ $>$ $v_\mathrm{esc}$. From the sum over all the spaxels of the fluxes of this high-velocity part of the outflow component ($F_{|v| \,>\, v_\mathrm{esc}}$) and of the fluxes of the entire outflow component ($F_\mathrm{out}$), we derive the outflow escape fraction as $F_{|v| \,>\, v_\mathrm{esc}} / F_\mathrm{out}$ $\sim$ 0.4 $\%$. 
This is the fraction of the outflowing material that is fast enough to be able to escape the galaxy gravitational potential. Considering that the outflow mass is about 5$\%$ the molecular gas mass in the galaxy (see above), this means that the outflowing gas that is actually able to escape the DM halo amounts to about 0.02 $\%$ of the molecular gas mass. 

We stress however that the actual escape fraction may be higher, since the outflow velocities measured from the line profiles are subject to projection effects and the fraction of outflowing gas having intrinsic velocities higher than the escape velocity may then be larger. 
Moreover, the above calculations of the escape velocity only hold for ballistic motions. If instead the outflow keeps on being accelerated (as it is in the case of the driving mechanisms probed in Sect. \ref{ssec:outfl_accel}), a larger fraction of it would be able to escape the galaxy and DM halo potential.
{On the other hand, the ejected gas could encounter mechanical resistance opposed by the gaseous halo, which would have the effect of slowing down the outflow. It is not obvious which of these effects would dominate and we refrain from further speculations.}

Summarising, while on one hand the ionised outflow is observed at distances of kpc to tens of kpc, meaning that it is actually able to escape the galaxy, and its depletion times are quite small ($\lesssim$10$^8$ yr), on the other hand only a negligible fraction of the galaxy gas reservoir {appears to be} actually able to escape the DM halo. The gas that does not leave the DM halo potential may possibly be re-cycled and re-accreted on the galaxy at later times, as invoked by theory (e.g. \citealt{Tumlinson:2017aa} and references therein). 
{Even if the gas is unable to escape the DM halo potential, the outflow could still have a significant feedback effect without this being completely ejective. The material expelled from the galaxy could progressively inject energy into the gaseous halo which could hinder, over time, the gas cooling and its (re-)accretion on the galaxy \citep[e.g.][]{Costa:2020a}.}


\section{Enhanced line velocity widths perpendicular to the radio structure}\label{sec:enhanced_sigma}
The ionised gas velocity dispersion presents a strong enhancement in an elongated region to the NW and SE of the nucleus, along the optical minor axis, perpendicular to the \oiii\ lobes and radio jet, with values $\gtrsim$300 km s$^{-1}$, visible in both the \oiii\ full-profile and second-component maps (Fig. \ref{fig:kinmaps}, {bottom-right and mid-right panels, respectively}). The enhancement region extends over $\sim$5$''$, corresponding to $\sim$8 kpc. Such line width enhancement can also be seen in the VIMOS IFU data in \cite{Harrison:2015a} and is also present in the molecular gas from CO observations (\citealt{Ramos-Almeida:2022a,Audibert:2023a}), though in the molecular phase it only extends on $\sim$1.6 kpc and has lower velocity dispersions ($\sim$100-140 km/s) than in the ionised phase.

Regarding the dominant ionisation mechanism, the gas in this velocity dispersion-enhanced region has line ratios compatible with the presence of shocks according to the {emission-line ratio} diagnostic diagrams (Figs. \ref{fig:bpt} and \ref{fig:bpt_comp}), especially when considering only the first, narrower component (upper panels of Fig.~\ref{fig:bpt_comp}).

We can take advantage of the findings from recent works for interpreting this phenomenon, that is, a velocity dispersion enhancement elongated perpendicular to the radio jets and the AGN ionisation cones observed in radio-quiet local AGN (e.g. \citealt{Riffel:2014a, Shin:2019a, Feruglio:2020a, Cazzoli:2022a, Venturi:2021a}, and references therein).
\cite{Venturi:2021a}, based on the fact that all the sources in which the phenomenon is observed so far host a low-power ($\lesssim$10$^{44}$ erg s$^{-1}$), compact ($\lesssim$1 kpc) AGN jet having low inclinations with respect to their host galaxy discs, concluded that the most likely origin for the observed phenomenon may be the action of the low-power jets interacting with the ISM in the galaxy disc. Supporting this interpretation, simulation of jet-ISM interaction (e.g. \citealt{Wagner:2011a, Mukherjee:2016a,Mukherjee:2018a,Mukherjee:2018b}) predict that low-power ($\lesssim$10$^{44}$ erg s$^{-1}$) jets having low inclinations with respect to the galaxy disc ($\lesssim$45\degree) will have maximal interaction with the disc material and, during their slow propagation through the disc, will perturb and shock the ISM and drive turbulence perpendicular to their direction of motion and to the disc plane, along the direction of minor resistance.

In the case of the Teacup, the same phenomenon may be at work, operated by the kiloparsec-scale jet hosted in the galaxy centre. The jet is found to subtend a small inclination angle relative to the molecular gas disc ($\sim$32\degree), leading to maximal jet-ISM interaction, and the connection between the jet and the perpendicular molecular gas with large velocity dispersions is supported by tailored simulations (\citealt{Audibert:2023a}).
{From the ionised gas velocity dispersion maps in Fig. \ref{fig:kinmaps}, bottom-right and mid-right panels, we see that the maximum values ($\sim$500 km/s) are co-spatial with the head of the jet (the knot HR-B). This strongly points to a causal link between the jet and the large velocity dispersions perpendicular to it.}

{The detection of this phenomenon in the Teacup indicates that it} occurs not only in Seyferts, but also in more radiatively powerful ($L_\mathrm{AGN}$ $\gtrsim$10$^{45}$ erg/s) QSOs, as also reported in \cite{Girdhar:2022aa} and \cite{Ulivi:2023a} in QSOs of similar redshift and AGN luminosity as the Teacup. Moreover, if the large-scale radio lobes are the result of past nuclear activity and the small-scale radio jet consequently represent a re-triggering of jet activity, it follows that the perpendicular velocity dispersion enhancement produced by a compact radio jet in its early phases may occur multiple times during an AGN lifetime, as a result of re-triggering of jet activity. This would help to mix the ISM and, if the turbulent material is lifted perpendicular to the galaxy disc, even the CGM, therefore chemically enriching it.

The fact that the molecular gas is less affected by this phenomenon compared to the ionised gas in terms of magnitude and spatial extension of the velocity dispersion enhancement, as also observed in the case of other sources exhibiting the same phenomenon (\citealt{Diniz:2015aa, Feruglio:2020a, Riffel:2015aa, Shimizu:2019aa, Girdhar:2022aa}), may be due to the higher density and clumpiness of the molecular gas, making it harder to perturb and disperse it.
{On the other hand, by comparing the CO(2-1) and CO(3-2) emission, \cite{Audibert:2023a} also reported enhanced gas temperature in the direction perpendicular to the jet in the Teacup, as also predicted by the simulations.}

\begin{figure}
    \centering
\includegraphics[scale=0.4,trim={2cm 0.5cm 1cm 0.5cm},clip]{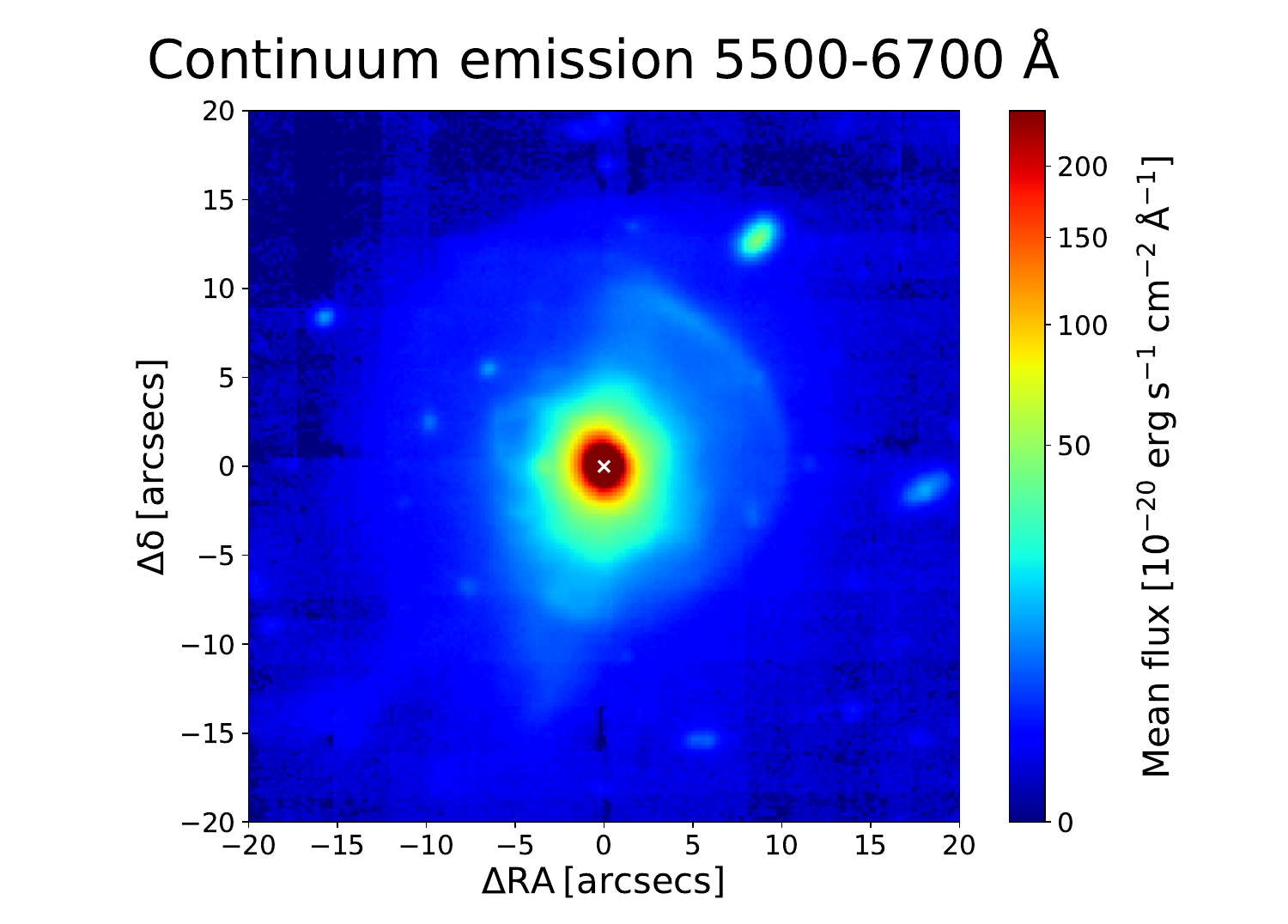}
    \caption{Map of continuum emission, obtained by averaging the (observed) spectral channels of the data cube between 5500--6700 \AA, free from the strongest gas emission lines {(same as in Fig. \ref{fig:rgb}, left panel, here zoomed in the central $40'' \times 40''$)}.}
    \label{fig:stellar_cont}
\end{figure}

\begin{figure*}
\centering
\includegraphics[width=0.32\textwidth]{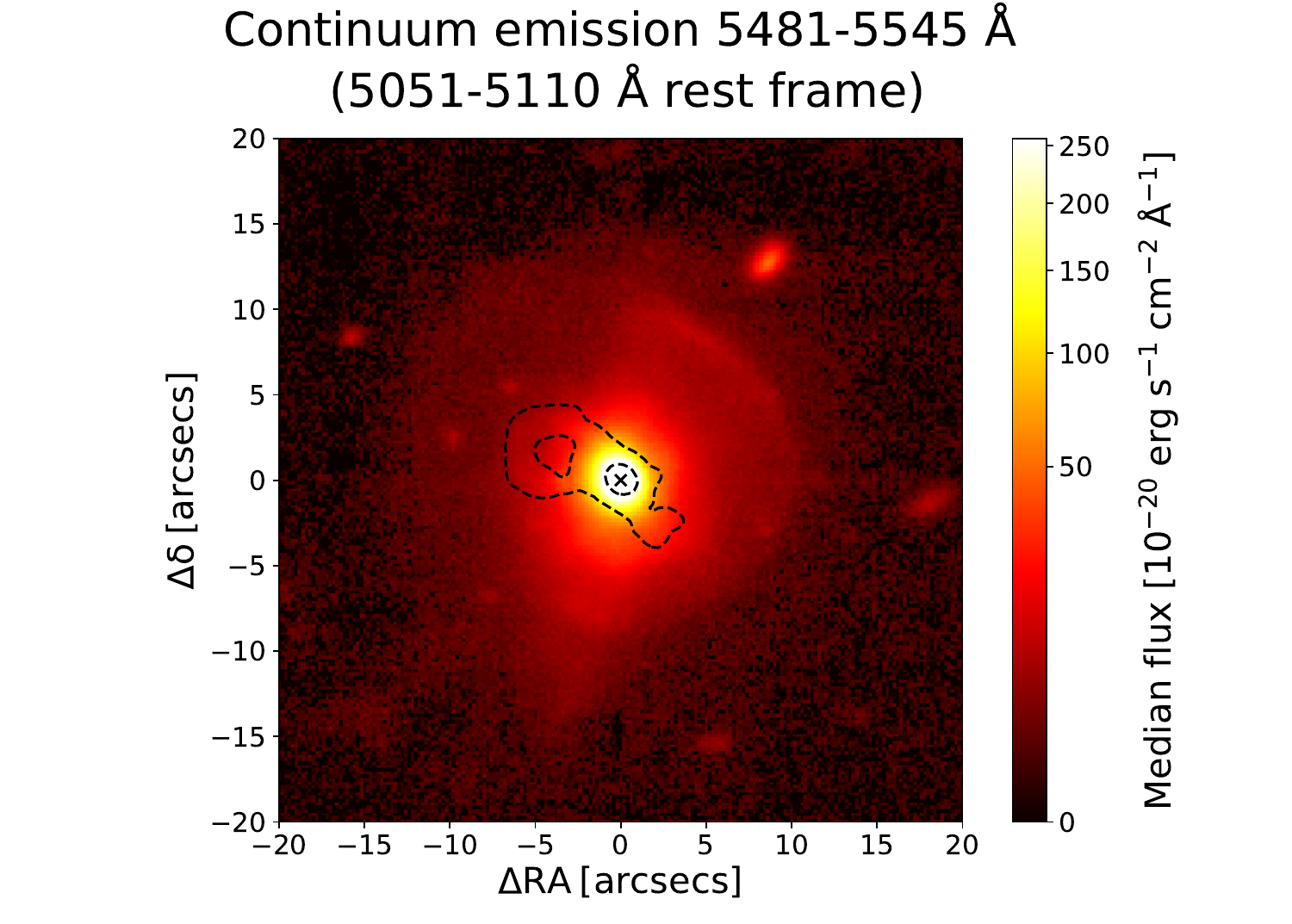}
\includegraphics[width=0.32\textwidth]{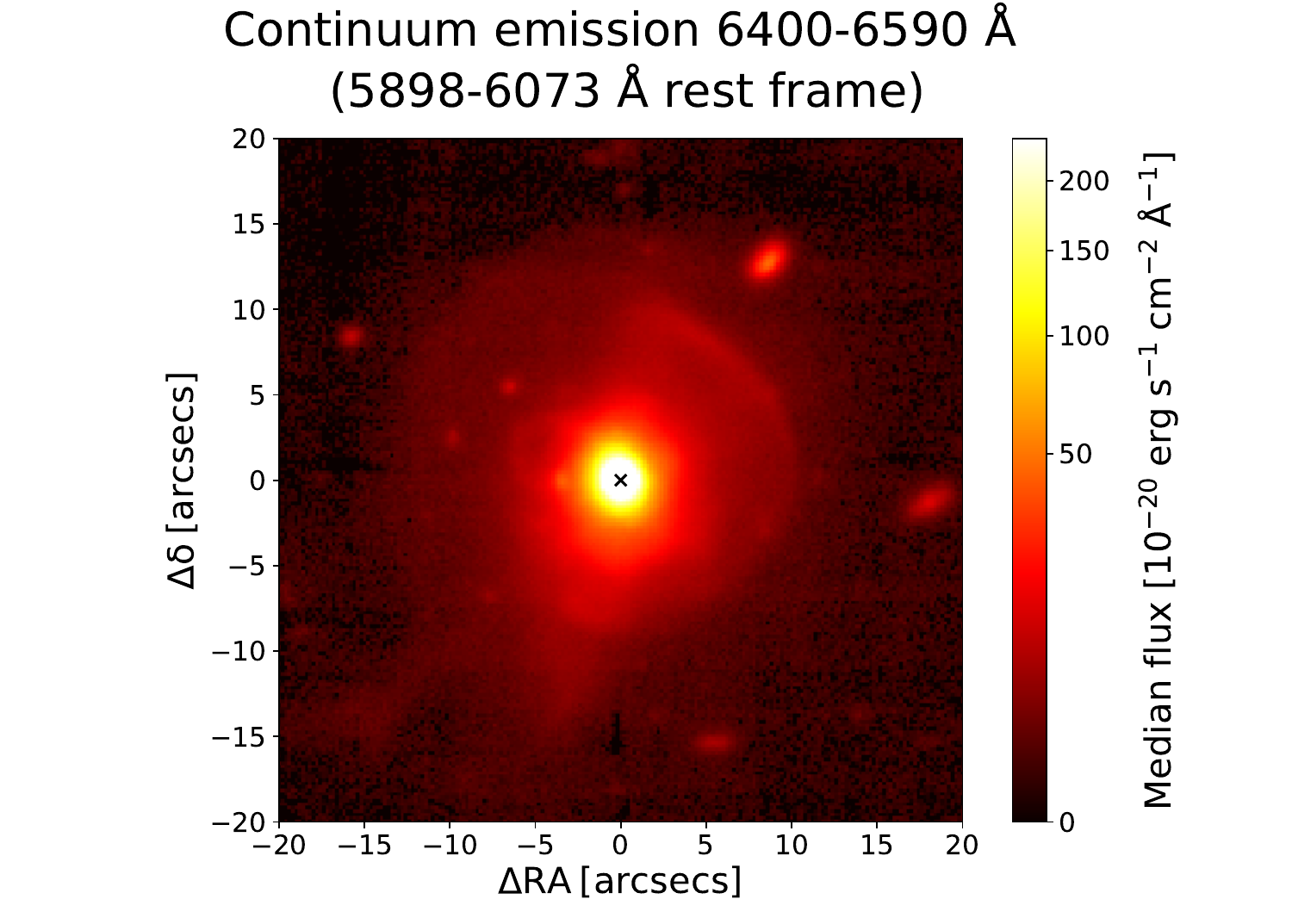}
\includegraphics[width=0.32\textwidth]{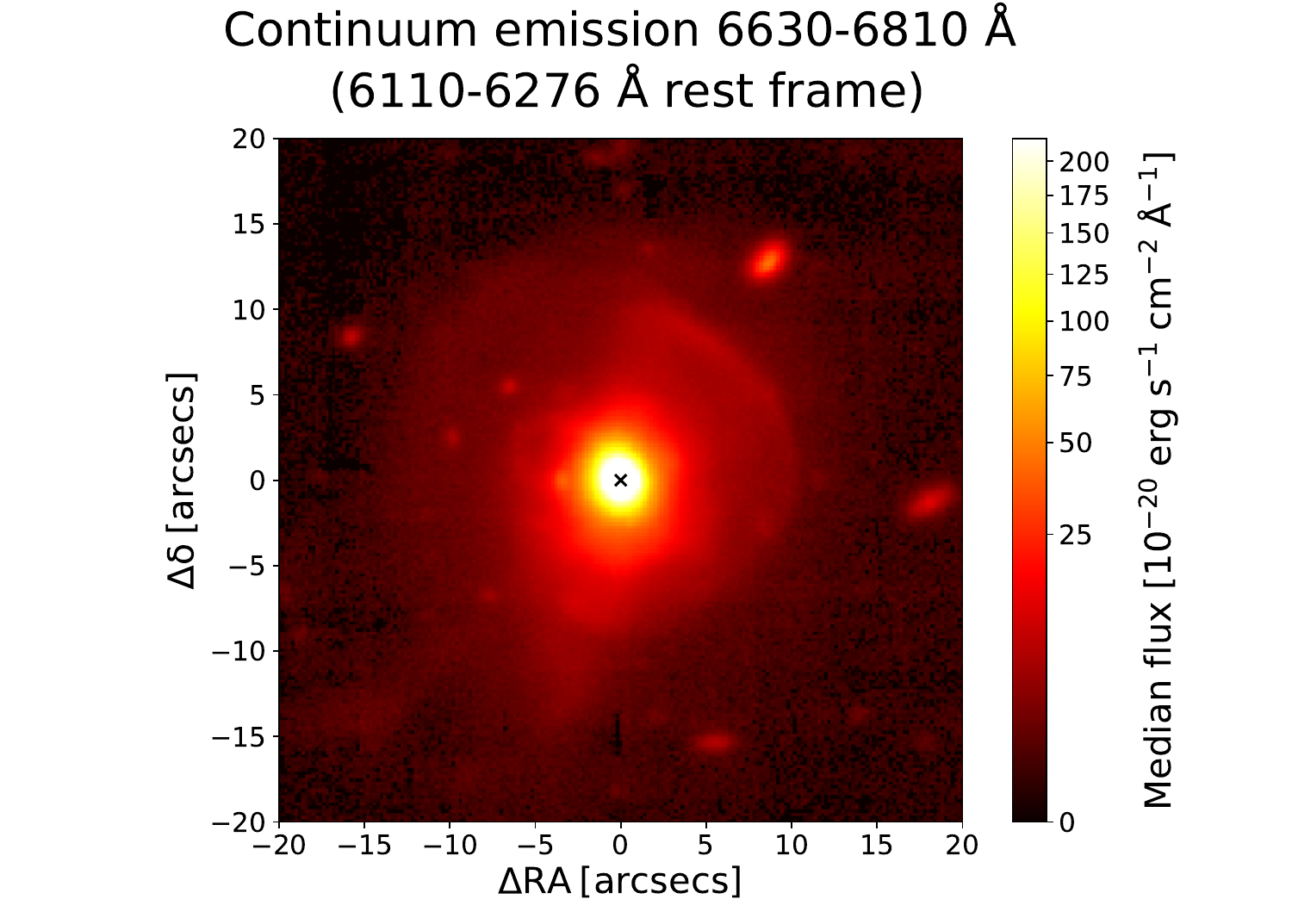}\\
\includegraphics[width=0.32\textwidth]{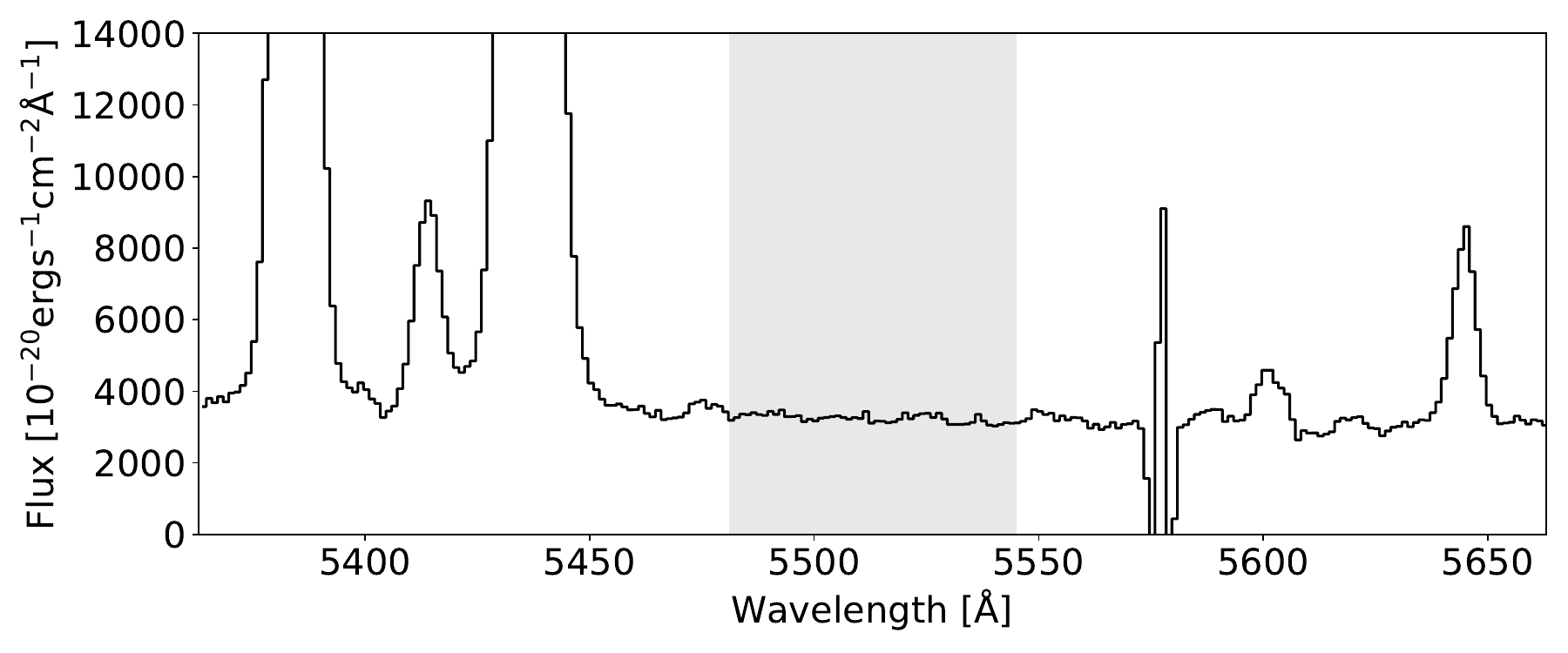}
\includegraphics[width=0.32\textwidth]{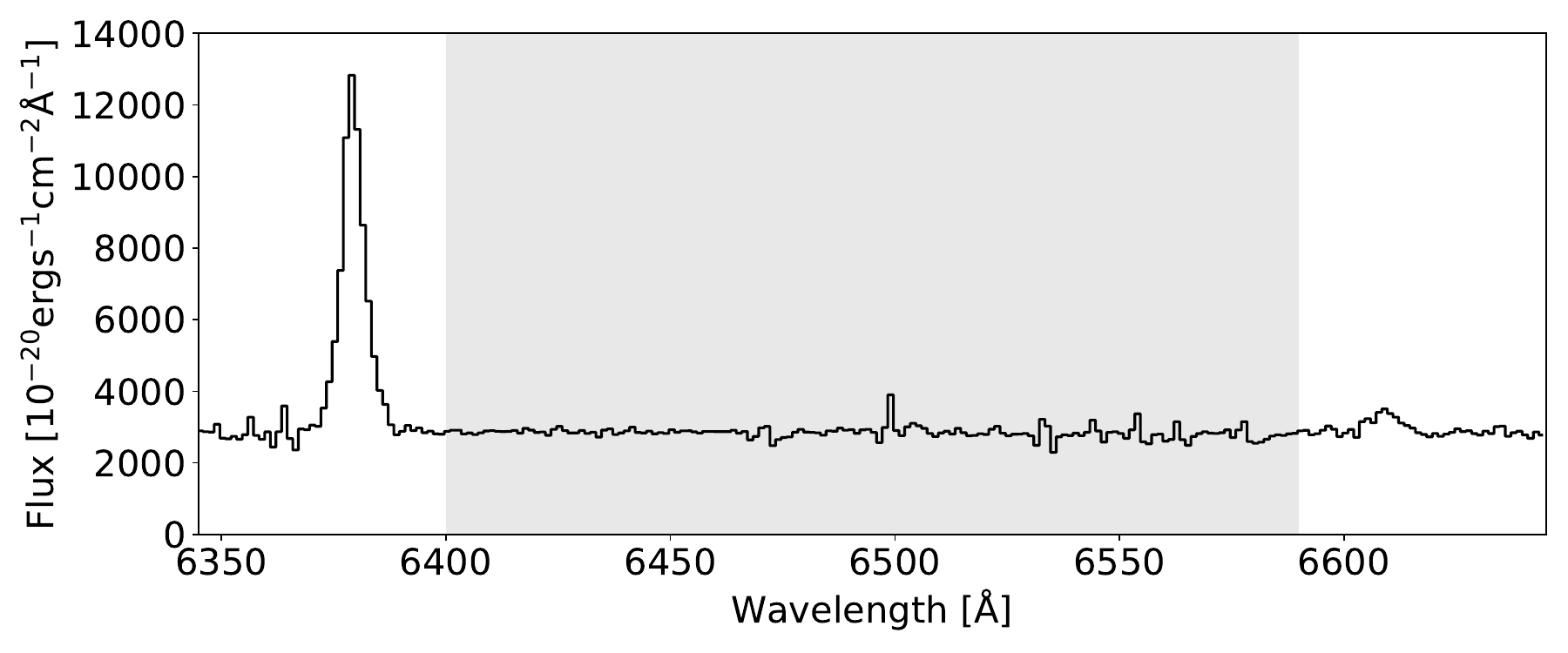}
\includegraphics[width=0.32\textwidth]{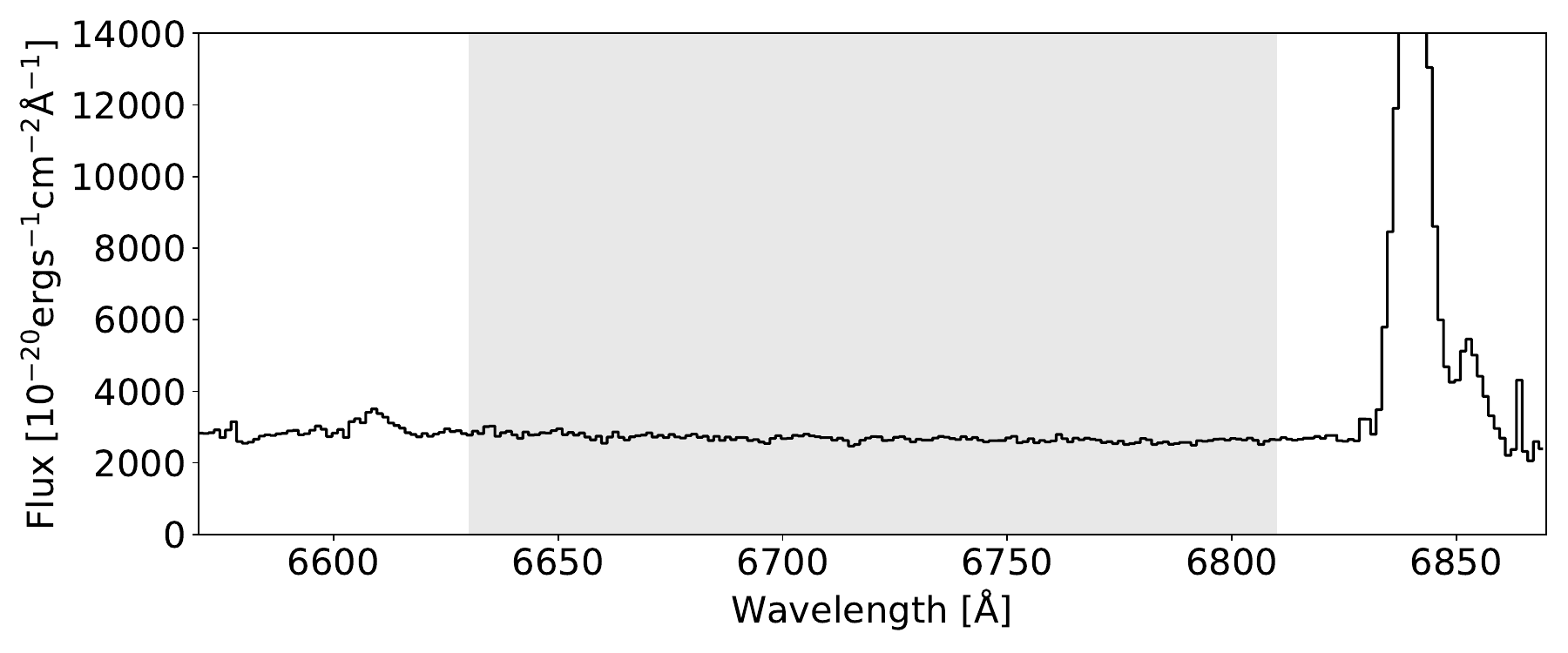}\\
\includegraphics[width=0.32\textwidth]{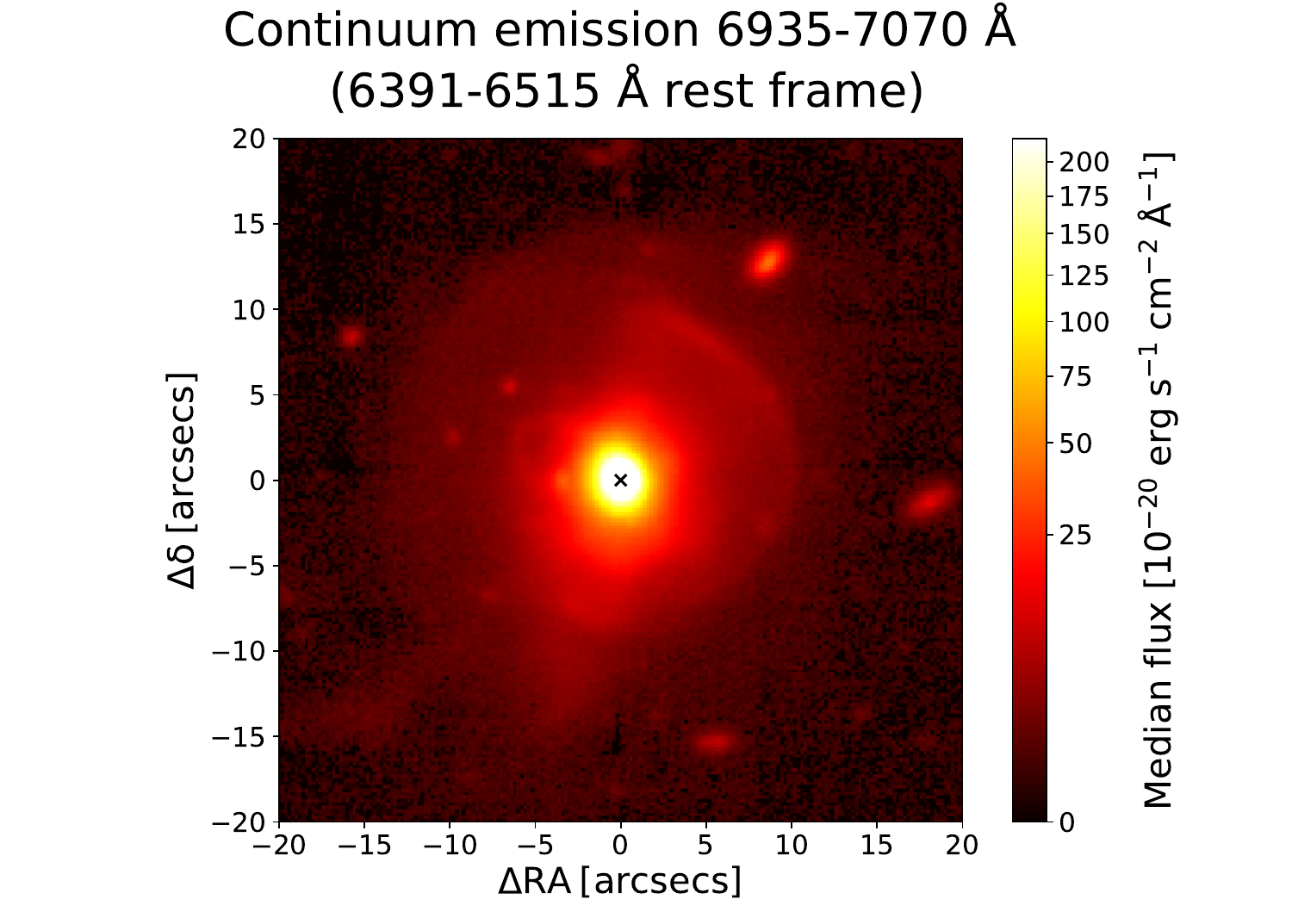}
\includegraphics[width=0.32\textwidth]{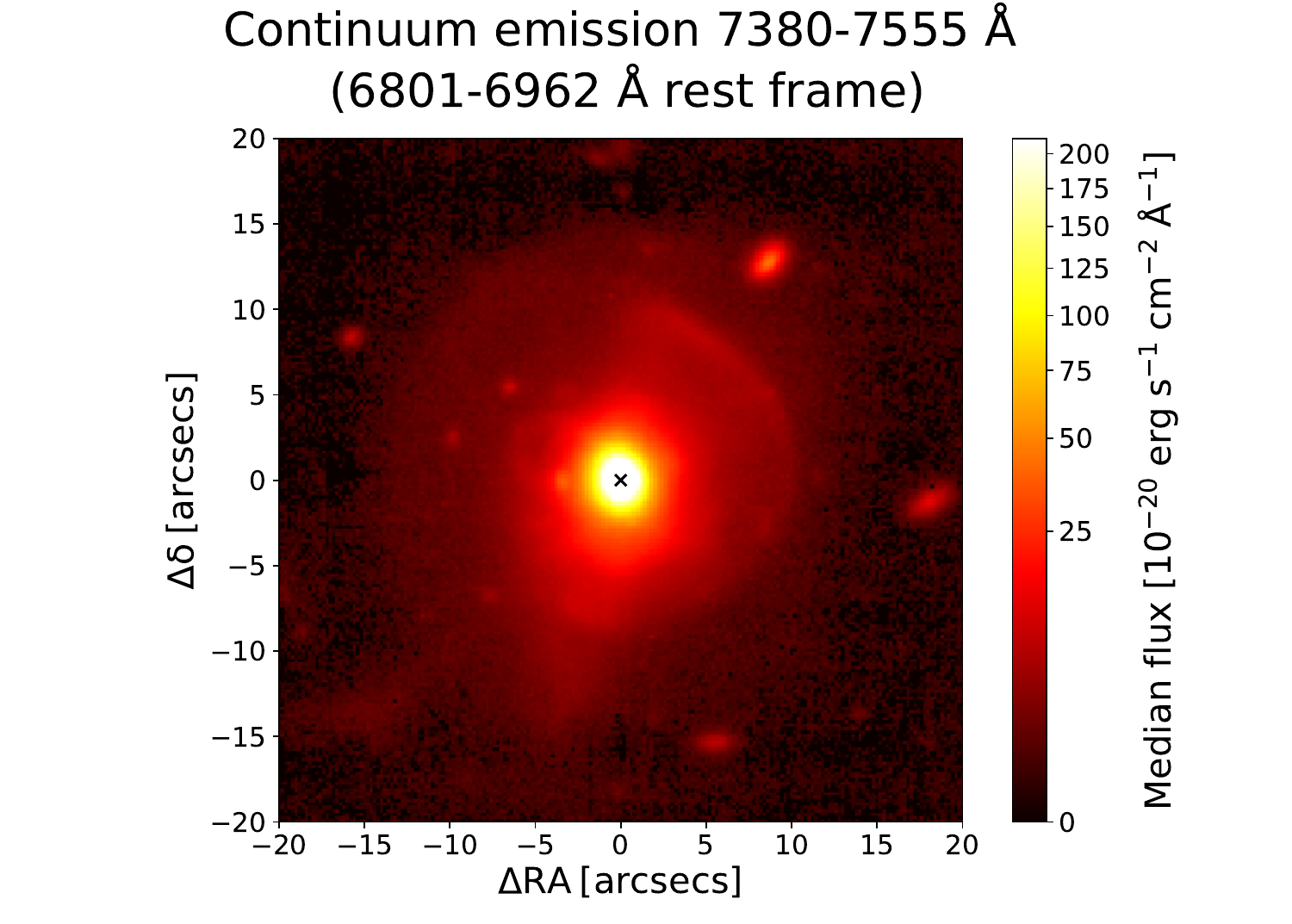}
\includegraphics[width=0.32\textwidth]{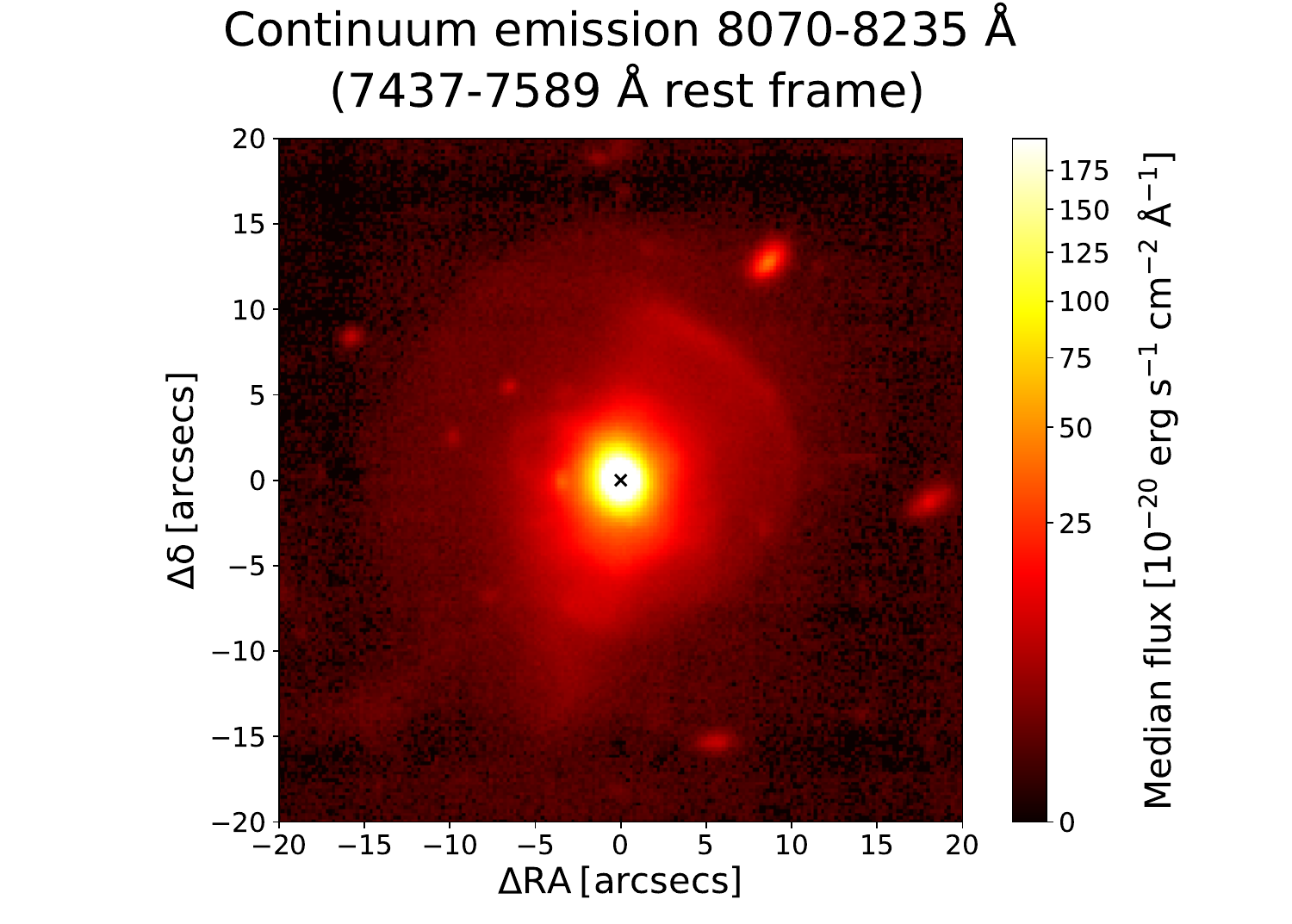}\\
\includegraphics[width=0.32\textwidth]{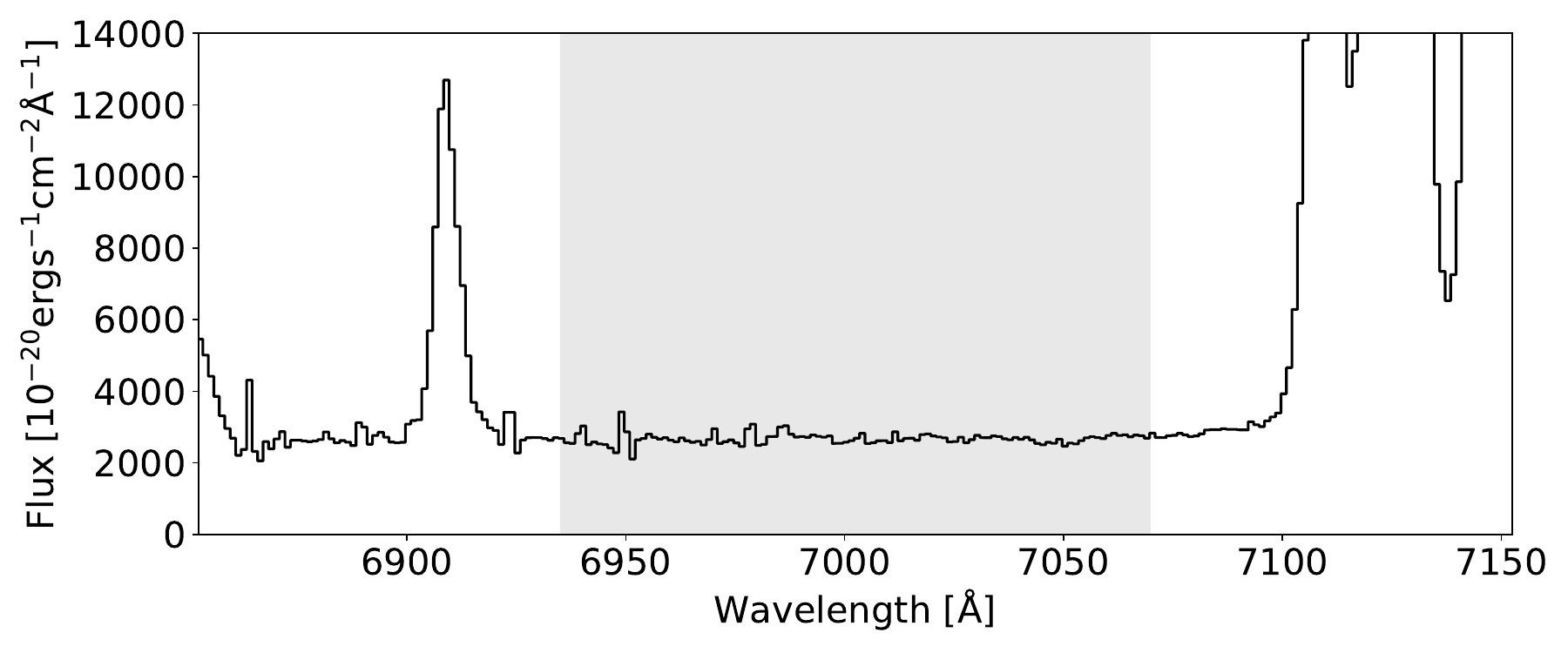}
\includegraphics[width=0.32\textwidth]{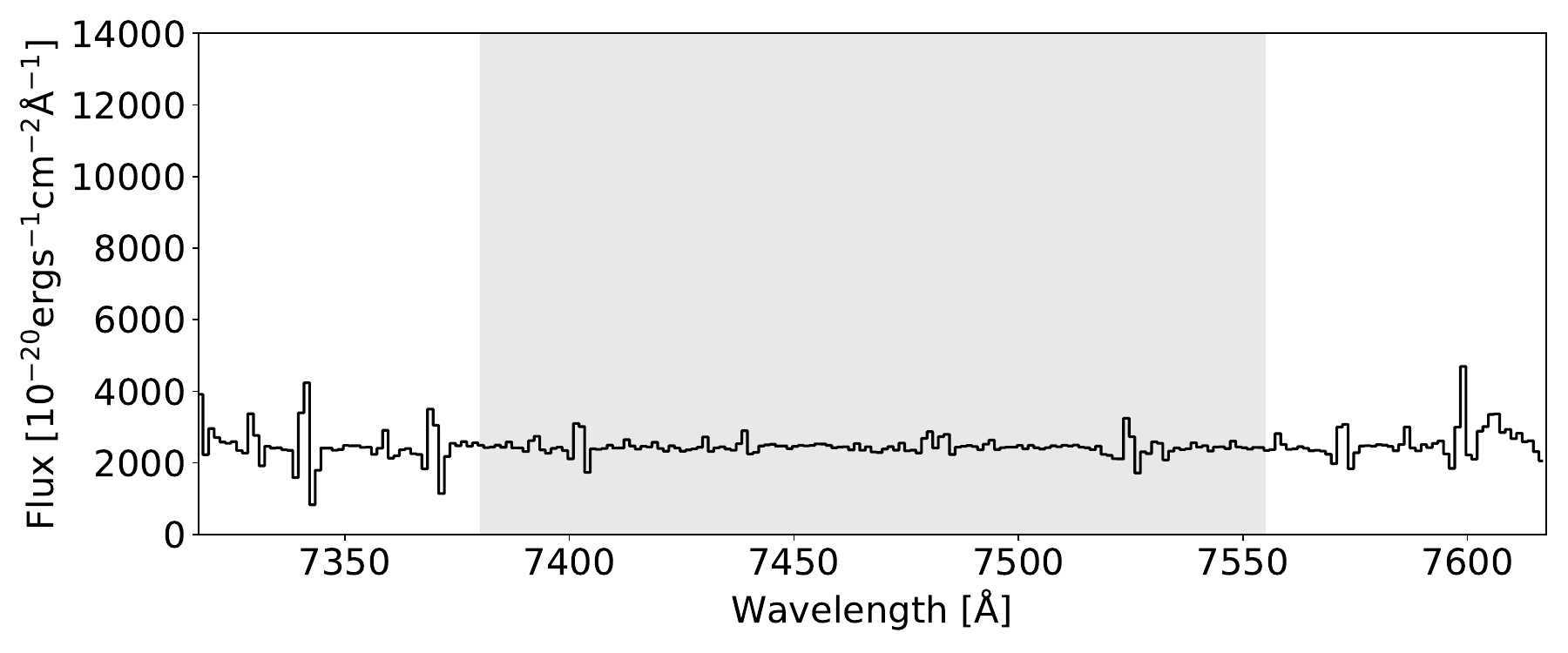}
\includegraphics[width=0.32\textwidth]{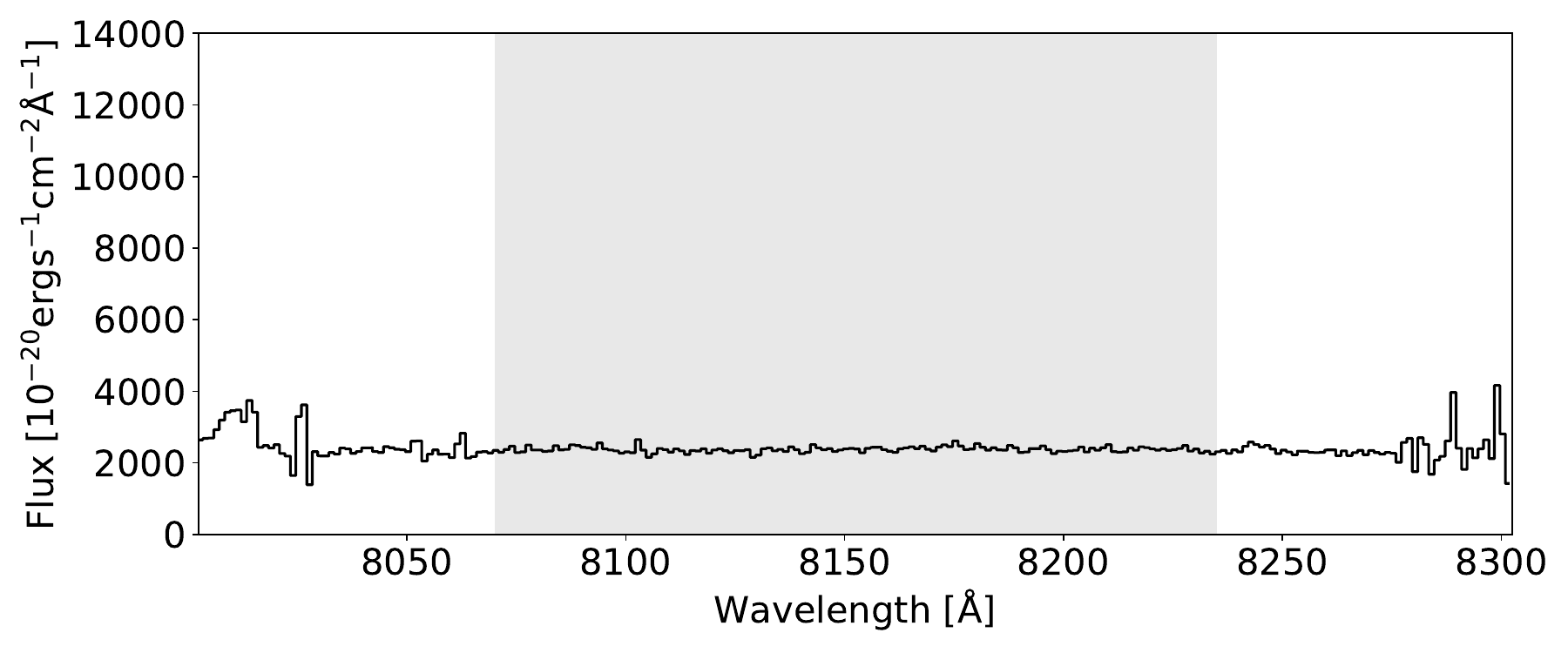}
\caption{Zoomed-in maps of median stellar continuum in selected spectral ranges free of gas emission lines. The spectral range in which the median has been calculated are indicated by the light grey shaded area in the spectra below. 
The spectra are extracted by summing all the spaxels in the handle, as traced by \oiii\ emission, whose contours are reported in top-left image for comparison; the extraction aperture of the spectra is shown in Fig. \ref{fig:stars_handle_merge}, top-left panel, labelled as number 1.
}
\label{fig:stars_handle}
\end{figure*}

\begin{figure*}
\centering
\begin{minipage}[c]{0.5\textwidth}
\centering
\includegraphics[width=0.85\textwidth]{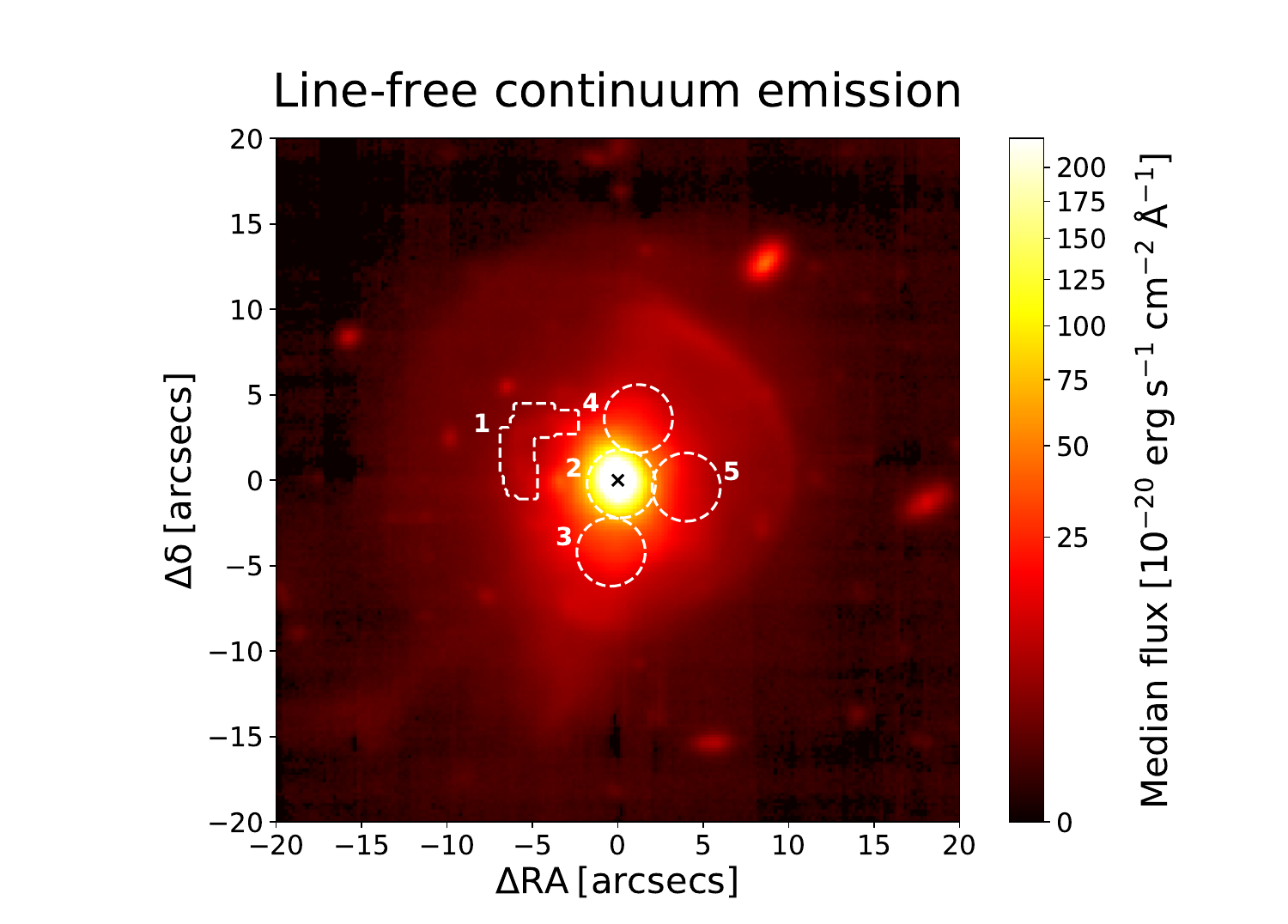}\\
\includegraphics[width=0.85\textwidth]{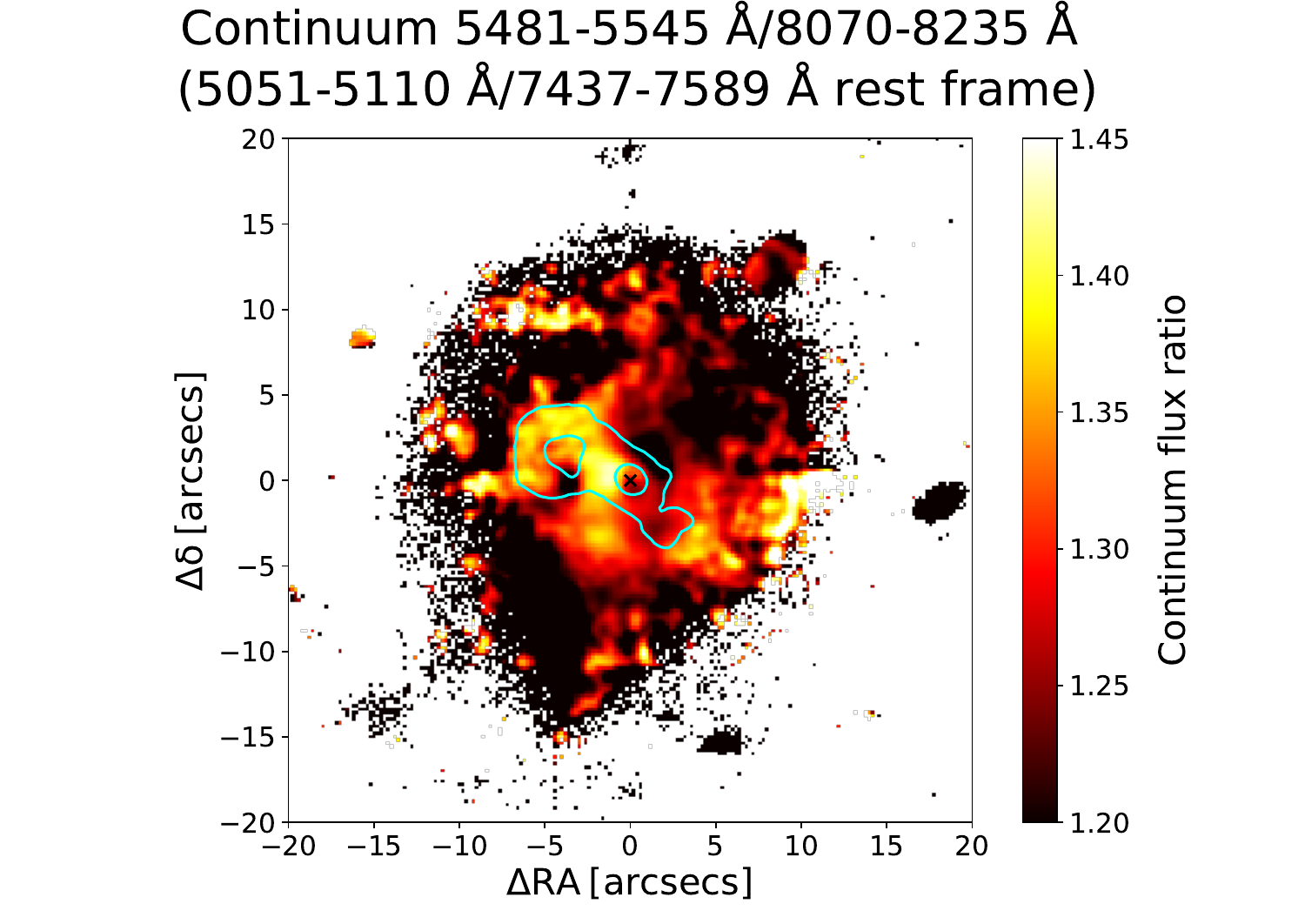}
\end{minipage} 
\begin{minipage}[c]{0.45\textwidth} 
\centering
\includegraphics[width=0.7\textwidth,trim={0 0.5cm 0 0.5cm},clip]{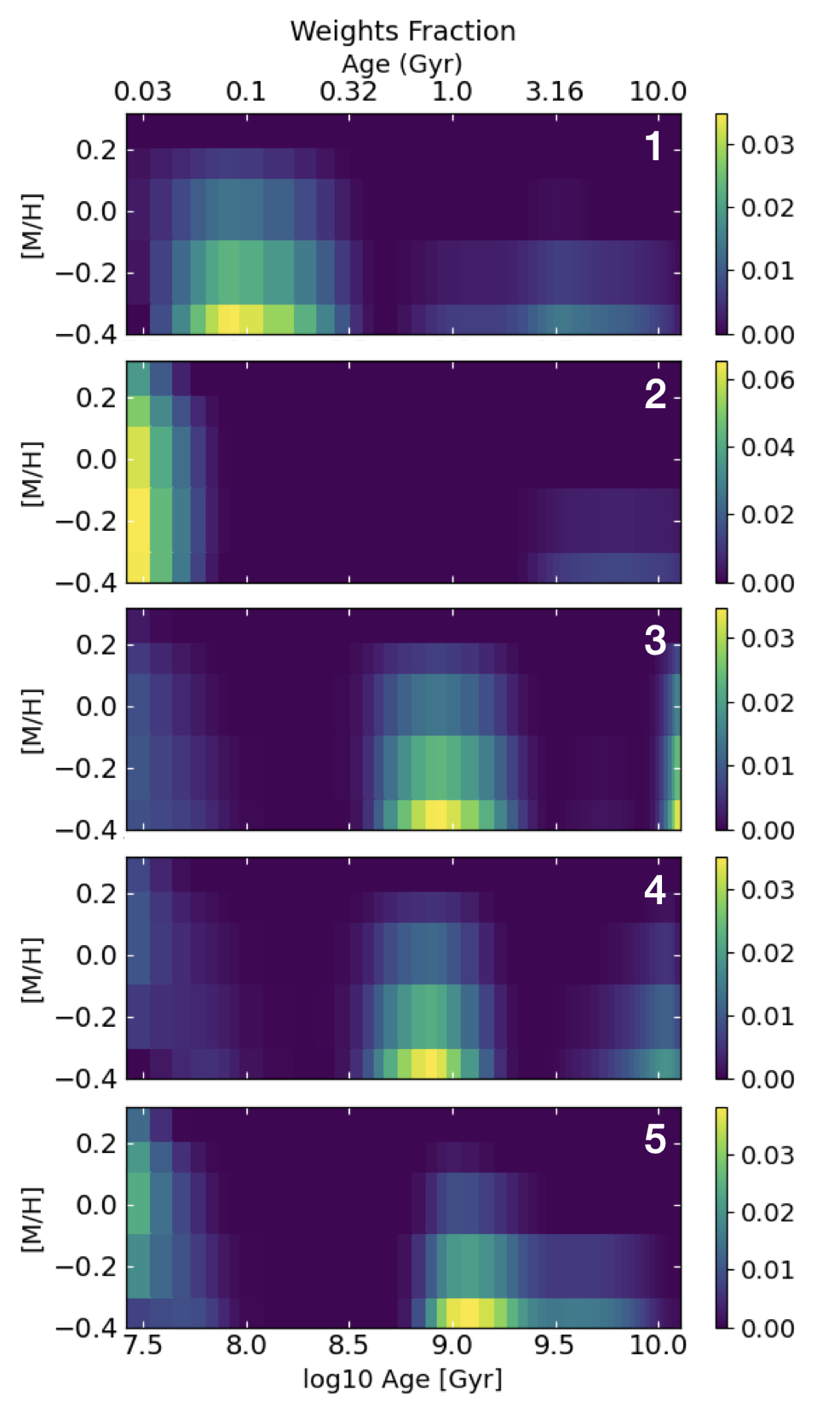} 
\end{minipage}
\includegraphics[width=0.9\textwidth]{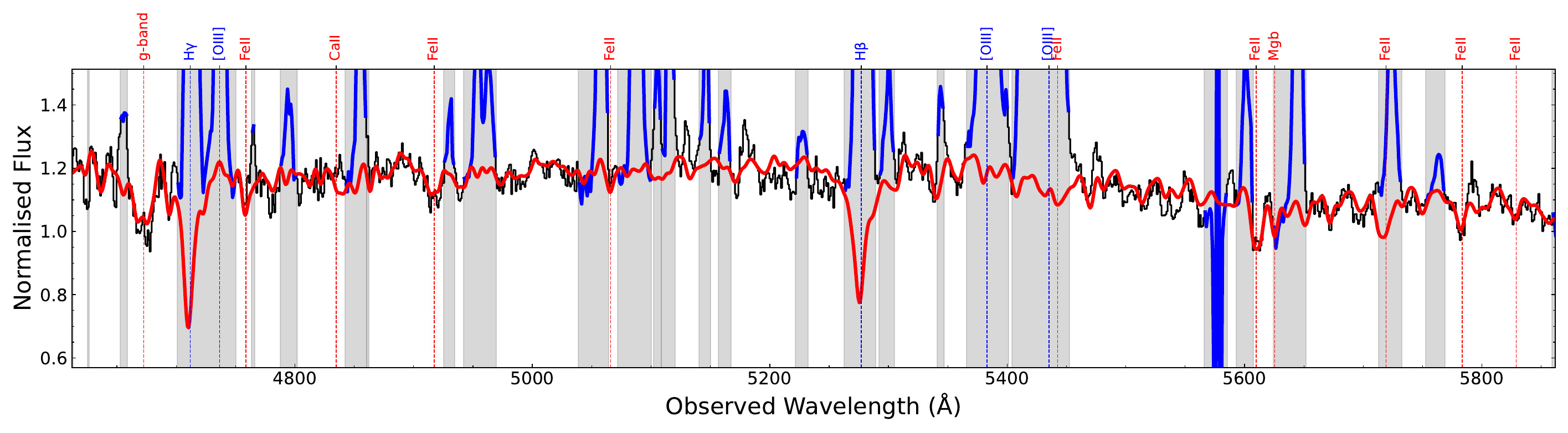}
\caption{Properties of the stellar populations, showing evidence for young stars in the Teacup handle. Top-left: Zoomed-in map of the stellar continuum obtained by performing the median in all the six spectral ranges reported in Fig. \ref{fig:stars_handle} together (light grey shaded areas). 
Mid-left: Map of continuum flux ratio (i.e. colour) between the images corresponding to the bluest and the reddest spectral ranges reported in Fig. \ref{fig:stars_handle} (top-left and bottom-right panels, respectively). Prior to performing the ratio, the continuum images were smoothed with a Gaussian kernel having $\sigma$ = 1.5 spaxels, for a better visual output. The patch having low continuum flux ratio in the southern part of the handle is due to a background red galaxy.
Top-right: {Luminosity-weighted} fractions of stellar populations in the stellar metallicity {(in log scale, with respect to solar metallicity)} versus age plane resulting from the stellar population modelling (details in text) of each numbered aperture shown in top-left panel.
{Bottom: Spectrum (black) extracted from the aperture covering the handle (number 1 in top-left panel) with the modelled stellar continuum overlaid (red); shaded grey vertical bands indicate spectral ranges which were masked in the fitting, and the spectrum within these ranges is marked in blue.}}
\label{fig:stars_handle_merge}
\end{figure*}

\section{Stars in the Teacup handle}\label{sec:stars_handle}
In Fig. \ref{fig:stellar_cont}, we report an image of the stellar continuum emission, in the (observed) spectral range 5500--6700 \AA, free from the strongest gas emission lines.
This shows an excess co-spatial with the prominent eastern ionised gas loop, or Teacup handle. However, the handle is a place of intense gas ionisation where even usually faint emission lines can be quite intense with respect to the continuum. Therefore, the continuum emission reported in Fig. \ref{fig:stellar_cont}, despite the exclusion of the strongest emission lines (such as \oiii, \hb, \oi, and \ha), may be contaminated by other emission lines, that could be responsible for the continuum excess in the handle.

For this reason, we performed a refined extraction of the continuum by carefully selecting different spectral intervals which are completely free even of very faint emission lines in the handle, that we checked visually. To further minimise possible contributions from faint emission lines, the continuum is calculated as the median, instead of the mean, of the signal in the selected {spectral ranges}. The continuum images obtained in this way are shown in Fig. \ref{fig:stars_handle}, together with the spectral ranges selected for continuum extraction (grey shaded areas).
The spectra, 
{extracted from the aperture labelled as number 1 in Fig. \ref{fig:stars_handle_merge}, top-left panel,}
show that no faint line emission is present in the selected spectral ranges at the noise level. Moreover, since the spectra are obtained by collapsing an extended region of {362} spaxels and show no significant presence of weak lines, we can rule out that they contribute on a spaxel level at a much lower S/N. 
The continuum images from all the six different extraction spectral ranges still show the presence of an excess of continuum emission stemming from the Teacup handle. A higher signal-to-noise image, produced by averaging together the continuum from all the six spectral ranges, is shown in Fig. \ref{fig:stars_handle_merge}, top-left panel.

The continuum excess indicates that stars are actually present in the handle, in addition to the underlying large-scale weak stellar continuum of the galaxy. 
{The presence of stars is confirmed by the sharp stellar absorption lines observed in the spectrum extracted from the handle shown in Fig. \ref{fig:stars_handle_merge}, bottom panel, whose $y$-axis is stretched to highlight the absorption features.}

Two are the possible scenarios explaining {the presence of stars in the handle.} 
The first is that the handle is a tidal feature due to the past merger occurred in this system, in which both stars and gas are present, the gas being illuminated by the AGN.
The second, in which the handle is a bubble of gas inflated by the action of radio jet and outflowing gas, as also suggested by the handle being in the same direction of jet and outflow, by its co-spatiality with the eastern radio bubble (see Sect. \ref{sec:muse_maps} and \citealt{Harrison:2015a}), {and by having its vertex on the nucleus (as expected for AGN-inflated bubbles from simulations, see e.g. \citealt{Nelson:2019aa})}, is that the stars have formed directly in the handle due to the action of the jet and outflow, that through the compression of gas clouds triggered the formation of stars (so-called positive feedback).

We checked the colour of the continuum to investigate if a bluer colour is associated with the handle, that would suggest the presence of a younger stellar population and support the second scenario above. To do so, we performed the ratio between the bluest and the reddest line-free continuum images from Fig. \ref{fig:stars_handle}. The map of this blue-vs-red continuum ratio is presented in Fig. \ref{fig:stars_handle_merge}, bottom-left panel. The map shows that the handle is characterised by a higher ratio (i.e a bluer continuum) than the rest of the galaxy. The region within $\sim$2--3$''$ to the E of the nucleus, constituting the base of the handle, is also characterised by a bluer continuum.

To check for the {actual} presence of a younger stellar population in the handle than in the rest of the galaxy, we modelled the stellar continuum in the handle in terms of its stellar population and compared with that obtained in other regions of the galaxy (dashed apertures in Fig. \ref{fig:stars_handle_merge}, top-left panel).
{To do so, we made use of a python-based graphical user interface (GUI) which allows us to extract spectra from single spaxels or multi-spaxel regions (details in \citealt{DAgo:2021aa}). The interface allowed us to interactively probe a variety of regions and to test different fitting parameters to model the spectra with the software  \textsc{ppxf} (\citealt{Cappellari:2004aa,Cappellari:2017a}).}
As done for the stellar continuum fitting of the cube, we employed the E-MILES single-stellar population (SSP) model spectra (details in Sect. \ref{ssec:data_anal}).
{In this case, since we are interested in the properties of stellar populations (rather than in simply modelling and subtracting the stellar continuum as in Sect. \ref{ssec:data_anal}), we adopt a more realistic \cite{Chabrier:2003a} IMF and BaSTI isochrones \citep{Pietrinferni:2004aa,Pietrinferni:2006aa}, which provide a finer sampling of the parameters of the stellar populations (age and metallicity) compared to the Padova+00 isochrones.}
Gas emission lines were masked and the automatic sigma-clipping included in \textsc{ppxf} was also employed to mask additional fainter emission lines. Since our goal in this case is the modelling of the stellar population and not of its kinematics, we did not employ an additive polynomial but only a multiplicative one, to avoid changes in the strength of the absorption line features in the templates, as indicated in \cite{Cappellari:2017a}. We tested different multiplicative polynomial degrees, between 5 and 30. The variation of the degree did not result in a significant change in the dominant stellar population of each region. We thus employed an intermediate value of 15 for the multiplicative polynomial degree.
We used a regularisation parameter of 100 to obtain the smoothest solution among the many degenerate ones that equally reproduce the data (details in \citealt{Cappellari:2017a}).

The resulting luminosity-weighted fractions of stellar populations in the stellar metallicity versus population age plane are reported in Fig. \ref{fig:stars_handle_merge}, top-right panels. 
We see that while other parts of the galaxy (regions number 3, 4, and 5) are dominated by a relatively older stellar population with an age between $\gtrsim$0.5--1 Gyr, the handle (region number 1) and the central part of the galaxy (region number 2) show a significant presence of a younger stellar population, with ages $\lesssim$100--150 Myr, {typical of OB stars.} In particular, in the handle {(as well as in the centre)} this younger component by far dominates the stellar populations.
\cite{Moiseev:2023a} confirm that a younger stellar population is present in the central regions of the Teacup, on scales comparable with those of the handle ($\lesssim$5 arcsec).

This result indicates that star formation activity in the last $\sim$150 Myr gave rise to the population of young stars, which in the case of the handle may have been triggered by the action of jet, outflow, and expanding radio lobe compressing the gas at the edge of the bubble.
Star formation triggered by AGN jets and outflows is predicted by models, due to the compression of the gas clouds, either by direct impact or at the edge of an over-pressurised expanding shocked bubble, increasing their density, leading to their collapse and eventually to star formation (e.g. \citealt{Rees:1989a,Begelman:1989a,De-Young:1989a,Daly:1990a,Mellema:2002a,Silk:2009aa,Gaibler:2012a,Nayakshin:2012aa,Dugan:2014a,Silk:2013aa,Zubovas:2013aa,Nayakshin:2014aa,Bieri:2016a,Fragile:2004a,Fragile:2017a}).

The Teacup handle could therefore represent the expanding bow shock of gas and stars formed in-situ predicted by models of AGN jet-triggered star formation (e.g. \citealt{Begelman:1989a,Gaibler:2012a,Dugan:2014a,Bieri:2016a,Fragile:2004a,Fragile:2017a}). In this context, the enhancement of gas density observed at the edge of the handle in the direction of the jet, radio lobes, and outflow may favour the scenario of gas in the handle being compressed, invoked as a necessary condition to form stars by the above models.
{The timescale of $\sim$100--150 Myr is compatible with the time needed by the bubble to expand to its actual size ($\sim$10 kpc) considering an expansion velocity of $\sim$100 km/s.}

Observational evidence for jet- or outflow-induced star formation has been found in a number of cases so far, for example within the filaments along the jet of Centaurus A (\citealt{Mould:2000a,Rejkuba:2002a,Crockett:2012a,Santoro:2016a}) {and of 4C 41.17 \citep{Bicknell:2000a}, in companions aligned with jets from another galaxy, as in the case of the Minkowski's Object around NGC 541 (\citealt{Croft:2006a,Salome:2015a}) and the companions of PKS2250-41 \citep{Inskip:2008a} and 3C 285 (\citealt{Salome:2015a}),} at the jet termination in NGC 5643 (\citealt{Cresci:2015aa}), in the gas filaments surrounding the giant radio lobes in Coma A (\citealt{Capetti:2022a}), {and at the edge of the outflow in the type-2 QSO Mrk 34 \citep{Bessiere:2022aa}.} See also the reviews in \cite{Miley:2008a}, \cite{Wagner:2016aa}, and \cite{ODea:2021a}.

{To our knowledge, the Teacup represents the first case that evidence for widespread stars formed around a galactic bubble inflated by the action of an AGN is found, confirming the predictions from theory.}




\section{Conclusions}\label{sec:concl}
In this work we have presented a spatially resolved study of the ionised gas in the Teacup {QSO ($z$ $\simeq$ 0.08506; $L_\mathrm{AGN} \sim 2 \times 10^{45}$ erg/s}) employing optical integral field observations from MUSE at VLT.
{This complex galaxy constitutes a great laboratory to study AGN feedback in all of its manifestations, since it hosts $\sim$10 kpc-scale co-spatial optical ionised-gas and radio bubbles, a compact ($\sim$1 kpc), low-power ($\dot{P}_\mathrm{jet}$ $\sim$ $10^{42.5-43.5}$ erg/s) jet, multi-phase (ionised plus molecular) outflow activity, and a giant ionised nebula ($\gtrsim$100 kpc size in total; Fig. \ref{fig:rgb}).}

We produced maps of ionised gas line-emission, kinematics, excitation, and physical properties (density, temperature, and dust extinction) spanning a total physical size of about 100 kpc.
We confirm the presence of an ionised outflow aligned with the optical AGN-ionised lobes (one of which resembles the handle of a teacup, giving the name to the galaxy), with velocities up to $\gtrsim$500 km/s {(Figs. \ref{fig:spectra}, \ref{fig:kinmaps}, and \ref{fig:moutrate})}.
We measured dust extinction from the Balmer decrement, \ha/\hb, reaching values above $A_V$ $\sim$ 1.2 around the nucleus, where it peaks, and with values of $\sim$0.6 or lower in correspondence with the optical emission-line lobes {(Fig. \ref{fig:ext_dens})}. 
We found electron densities reaching $\gtrsim$500 cm$^{-3}$ at the nucleus and above 100 cm$^{-3}$ at the edge of the Teacup handle, while in the rest of the galaxy it is generally {a few tens of cm$^{-3}$ or around or below the threshold value of 10 cm$^{-3}$ under} which the \sii\ doublet diagnostic is not 
{sensitive to}  electron density anymore {(Fig. \ref{fig:ext_dens})}.
We found electron temperatures around 1.3--1.4$\times$10$^4$ K across the galaxy {(Fig. \ref{fig:ext_dens})}.
{The gas ionisation is dominated by the AGN photons (Figs. \ref{fig:bpt} and \ref{fig:bpt_comp}).}

{We detected strongly enhanced  ionised gas velocity dispersions ($\gtrsim$300 km/s) elongated over $\sim$17 kpc perpendicular to the kiloparsec-size radio jet, AGN ionisation lobes, and fast outflows (Fig. \ref{fig:kinmaps}). The same is observed in the cold molecular phase, though with smaller velocity dispersions (by a factor of $\sim$4--5) and spatial extension (by a factor of $\sim$5). The largest ionised gas velocity dispersion ($\sim$500 km/s) is observed co-spatial with the head of the jet, strongly pointing to a causal connection between the two. This phenomenon is similar to what was found in nearby AGN (Seyferts) also hosting compact, low-power jets, interpreted as due to the action of the jets interacting with the galaxy ISM and driving turbulent material perpendicular to their propagation direction.
If the same process is occurring here, this work demonstrates that the phenomenon of line velocity width enhancement in the direction perpendicular to jets can also occur in sources that are more radiatively powerful (QSOs) and more evolved from the radio point of view, such as the Teacup (whose radio bubbles reach $\sim$10 kpc per side).}

{Three ionised gas bubbles at increasing distances from the nucleus can be identified in the Teacup in the NE direction (with less definite counterparts to the SW), each likely corresponding to a past AGN outburst, at $\sim$10 kpc (the handle), $\sim$25 kpc, and $\sim$60
kpc, respectively (Figs. \ref{fig:rgb} and \ref{fig:oiii_flux}). These multiple gas shells resulting from repetitive AGN ejections are very similar to what the most recent simulations of AGN feedback predict.}

The $\sim$10 kpc-scale radio bubbles are misaligned by $\sim$20\degree\ with respect to the small-scale ($\sim$1 kpc) jet and the optical kinematic axes, which are instead aligned {(Figs. \ref{fig:rgb}, \ref{fig:oiii_flux}, and \ref{fig:kinmaps})}. The origin of the misalignment, {whether due to jet precession or jet-bending due to jet-cloud interactions for instance,} is unclear.

We {determined} the properties of the ionised outflow as a function of distance from the AGN {(Fig. \ref{fig:moutrate})}. We observed a decreasing radial trend of the mass outflow rate, kinetic rate, and momentum rate. The mass outflow rate drops from around 40--130 $M_\odot$/yr in the inner 1--2 kpc to $\lesssim$0.1 $M_\odot$/yr at 30 kpc {(i.e. by up to about three orders of magnitude)}, with a small secondary peak at the distance of the eastern ionised-gas loop ($\sim$10 kpc; the handle).
{If using a more extreme approach to estimate the outflow velocity adopted in many outflow studies, the mass outflow rate, kinetic rate, and momentum rate would be a factor $\sim$1.7, 5, and 3 higher, respectively, than those we obtained.}

{We compared the ionised outflow energetics with the power of the nuclear radio jet ($\dot{P}_\mathrm{jet} \sim 10^{42.5-43.5}$ erg/s) and with the AGN radiative output ($L_\mathrm{AGN} \sim 7\times10^{44} - 6\times10^{45}$ erg/s) to assess the dominant outflow driving mechanism. 
Due to the large uncertainties in the jet and AGN luminosities, as well as in the outflow energetics, we could not robustly constrain the outflow driving mechanism from energetic considerations. Nevertheless, we find that direct AGN radiation pressure on dust is more generally compatible with the kinetic coupling efficiencies ($\dot{E}_\mathrm{kin}$/$L_\mathrm{AGN}$) and momentum boosts ($\dot{p}$/$(L_\mathrm{AGN}/c)$) which we estimated for the outflow, compared to an energy- or momentum-conserving outflow scenario due to an expanding shocked bubble driven by an accretion-disc wind. From purely energetic arguments, the jet alone is able to drive the outflow only when simultaneously considering the lower end of the inferred kinetic rates and the upper end of the jet powers.
Another, more likely possibility is that the outflow is driven by the combined action of the jet and AGN radiation. The fact that the head of the kiloparsec-scale jet is co-spatial with the largest ionised gas velocity dispersions {(Fig. \ref{fig:kinmaps})} suggests that the jet has a significant role in driving the outflow.}


The properties found for the ionised outflow in the Teacup are at the high end for ionised outflows in AGN with a similar luminosity.
Moreover, the mass outflow rate in the ionised phase ($\sim$40--130 $M_\odot$/yr) is significantly higher than that in the molecular phase {($\sim$15--40 $M_\odot$/yr; only detected in the inner $\sim$800 pc from the nucleus) on similar scales.} This is in contrast to what is generally observed in other AGN, where the molecular mass outflow rate exceeds the ionised one by a factor of $\sim$10--10$^3$.
The addition of the molecular budget to the ionised one does not substantially change the picture obtained from the ionised phase alone, in terms of the outflow acceleration mechanisms.

The mass-loading factor ($\dot{M}_\mathrm{out}$/SFR) of the ionised (plus molecular) gas is $\sim$3--{16} ($\sim$5--{20}), indicating that the outflow consumes the gas reservoir much faster than star formation processes.
The molecular gas depletion time due to {the multi-phase (ionised plus molecular)} outflows, $\tau_\mathrm{depl}$ = $M_\mathrm{H_2}/\dot{M}_\mathrm{out}$ $\lesssim$ 10$^8$ yr, is short enough for the outflow to be effective in depleting the molecular gas reservoir.
However, {we find that} the fraction of the ionised outflow that is able to escape the dark matter halo gravitational potential is only $\sim$0.4 $\%$. 
Therefore, while the outflow depletion time is short and {the expelled gas} is located at distances of tens of kiloparsecs, meaning that the outflow can escape the galaxy, only a negligible fraction of it is actually able to {leave} the DM halo without being possibly re-accreted at later times.
On the other hand, the expelled material could progressively inject energy into the halo and thus hamper the gas cooling and its (re-)accretion on the galaxy.

Finally, we detected blue-coloured continuum emission co-spatial with the eastern ionised gas loop {(the handle; scales of $\sim$10 kpc)}, whose modelled stellar populations are younger ($\lesssim$100--150 Myr) than in the rest of the galaxy ($\gtrsim$0.5--1 Gyr; {Fig. \ref{fig:stars_handle_merge}}). This constitutes possible evidence for widespread star formation triggered at the edge of the bubble due to the compressing action of the jet and outflow (positive feedback), as predicted by theory. The presence of an enhanced gas electron density at the eastern edge of the ionised gas loop, in the direction of the jet and outflow {(Fig. \ref{fig:ext_dens})}, is consistent with the compression scenario.

\begin{acknowledgements}
We acknowledge funding from ANID programmes FONDECYT Postdoctorado 3200802 (G.V.), FONDECYT Regular - 1190818 (E.T., F.E.B.) and 1200495 (F.E.B., E.T.), CATA-BASAL AFB170002 (G.V., E.T.), ACE210002, and FB210003 (E.T., F.E.B.), and Millennium Science Initiative Program - NCN19\_058 (E.T.), European Union’s HE ERC Starting Grant No. 101040227 - WINGS (G.V.), a United Kingdom Research and Innovation grant, code: MR/V022830/1 (C.M.H.), DLR grant FKZ 50 OR 2203 (D.T.), and the projects ``Feeding and feedback in active galaxies'', with reference PID2019-106027GB-C42, funded by MICINN-AEI/10.13039/501100011033, and ``Quantifying the impact of quasar feedback on galaxy evolution'', with reference EUR2020-112266, funded by MICINN-AEI/10.13039/501100011033 and the European Union NextGenerationEU/PRTR (C.R.A.).
For the purpose of open access, the authors have applied a Creative Commons Attribution (CC-BY) license to any author accepted version arising.
This research has made use of NASA’s Astrophysics Data System Bibliographic Services.
This research has made use of the NASA/IPAC Extragalactic Database (NED), which is funded by the National Aeronautics and Space Administration and operated by the California Institute of Technology.
This research has made use of the Python packages SciPy \citep{scipy2020}, NumPy \citep{numpy2020}, IPython \citep{ipython2007}, Matplotlib \citep{matplotlib2007}, galpy (\citealt{galpy2015}), and Astropy (\url{http://www.astropy.org}), a community-developed core Python package for Astronomy \citep{astropy:2013, astropy:2018}.
This research has made use of the FITS files visualisation tools QFitsView \url{https://www.mpe.mpg.de/~ott/QFitsView/}) and DS9 (\citealt{DS9}).
The National Radio Astronomy Observatory is a facility of the National Science Foundation operated under cooperative agreement by Associated Universities, Inc.
We thank Antonino Marasco for the useful conversation on the implementation of DM halo potentials.
We thank Anna Gallazzi for the insightful discussion on stellar metallicities in galaxies.
\end{acknowledgements}


\bibliographystyle{aa}
\bibliography{bibl_teacup.bib}

\newpage	
\begin{appendix}
\section{Stellar kinematics}\label{sec:stellar_kinem}
\begin{figure*}
\hfill
\includegraphics[scale=0.3,trim={2cm 0.5cm 2.3cm 0.5cm},clip]{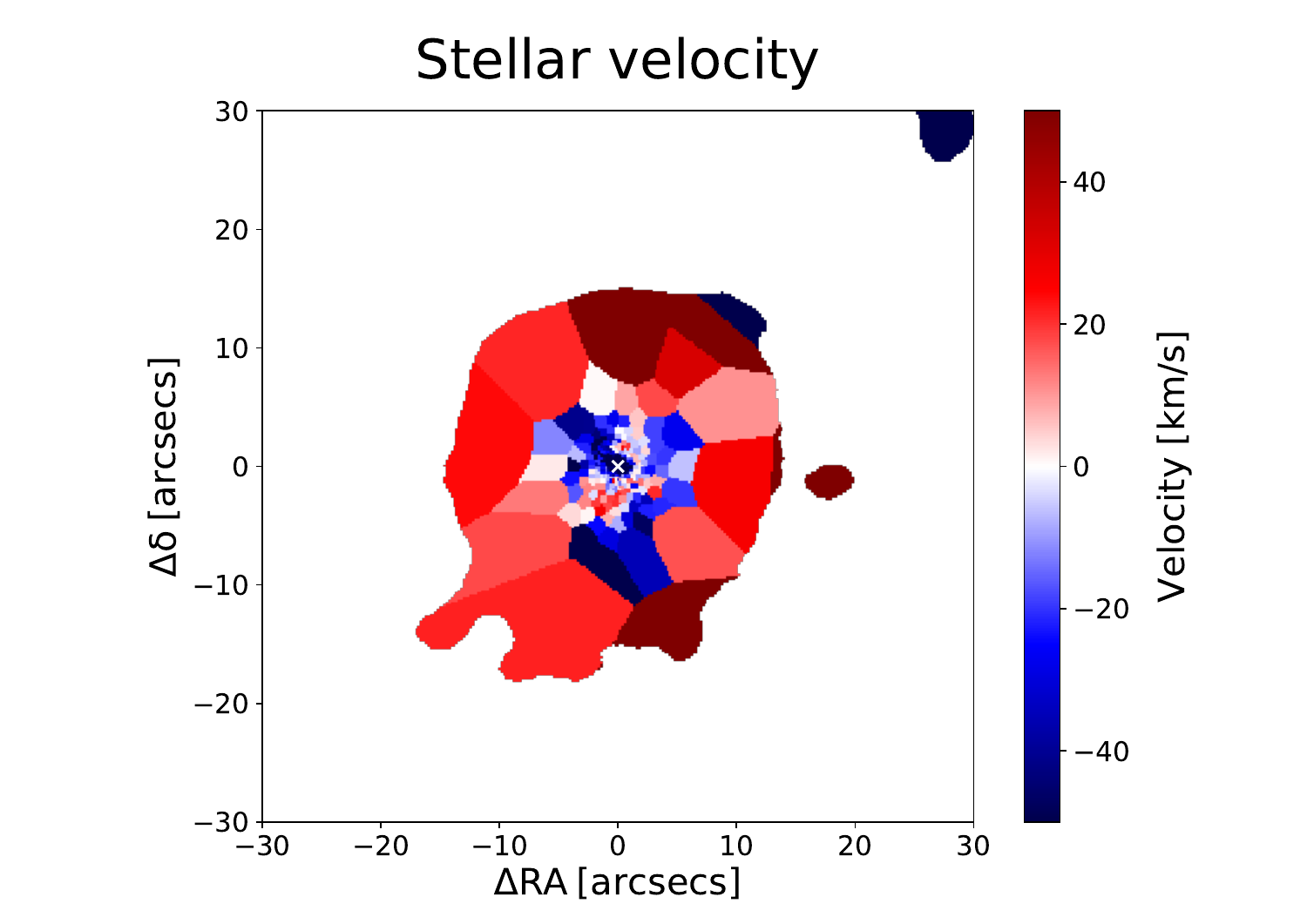}
\hfill
\includegraphics[scale=0.3,trim={2cm 0.5cm 2.3cm 0.5cm},clip]{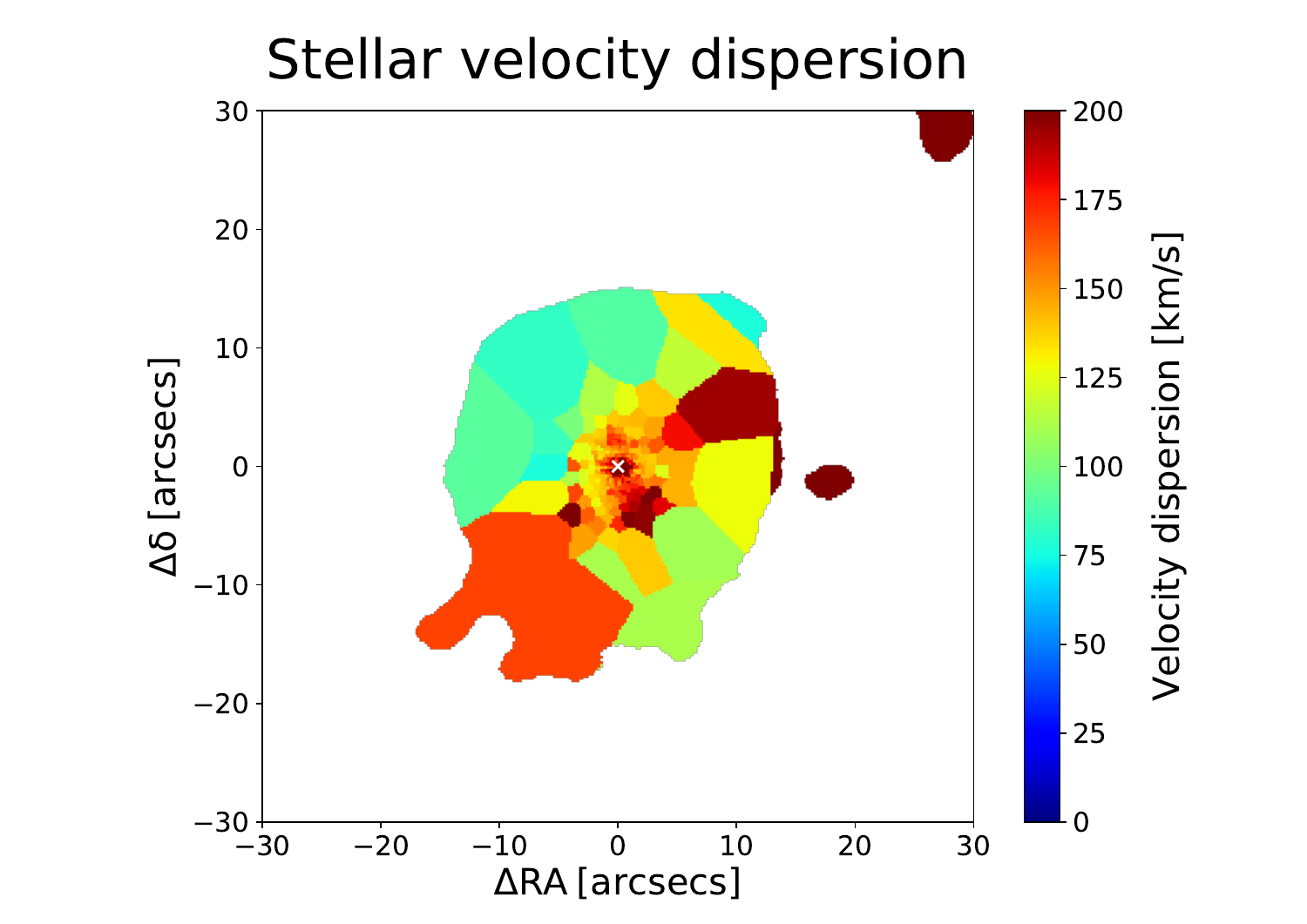}
\hfill\null
\caption{
Maps of stellar velocity (left) and velocity dispersion (right), obtained from the stellar-continuum fitting of the Voronoi-binned data cube (see Sect. \ref{ssec:data_anal}).}
\label{fig:stellar_maps}
\end{figure*}

The stellar velocity and velocity dispersion, obtained from the stellar continuum modelling of the Voronoi-binned cube (described in Sect. \ref{ssec:data_anal}), are reported in Fig. \ref{fig:stellar_maps} (left and right panels, respectively). 
The stellar velocity does not seem to show any clear pattern that could suggest rotation. It shows, however, a general pattern of more redshifted velocities at the outskirts of the stellar emission of the galaxy and lower velocities in the inner, most luminous part of the galaxy. This might be related to the post-merger nature of the source, with the disturbed stellar features of the galaxy (see continuum image in Fig. \ref{fig:stellar_cont}) having different velocities since the galaxy still has to completely relax. The stellar velocity dispersion is also disturbed. It is not radially symmetric and shows an elongation in the same SE-NW direction along which morphological features are seen in the continuum image, in particular extended tails in the SE and S direction and a shell-like feature to the NW.

\section{Additional figures for ionised gas}
\begin{figure*}
    \centering
	\hfill
	\includegraphics[scale=0.3,trim={2cm 0.5cm 2.3cm 0.5cm},clip]{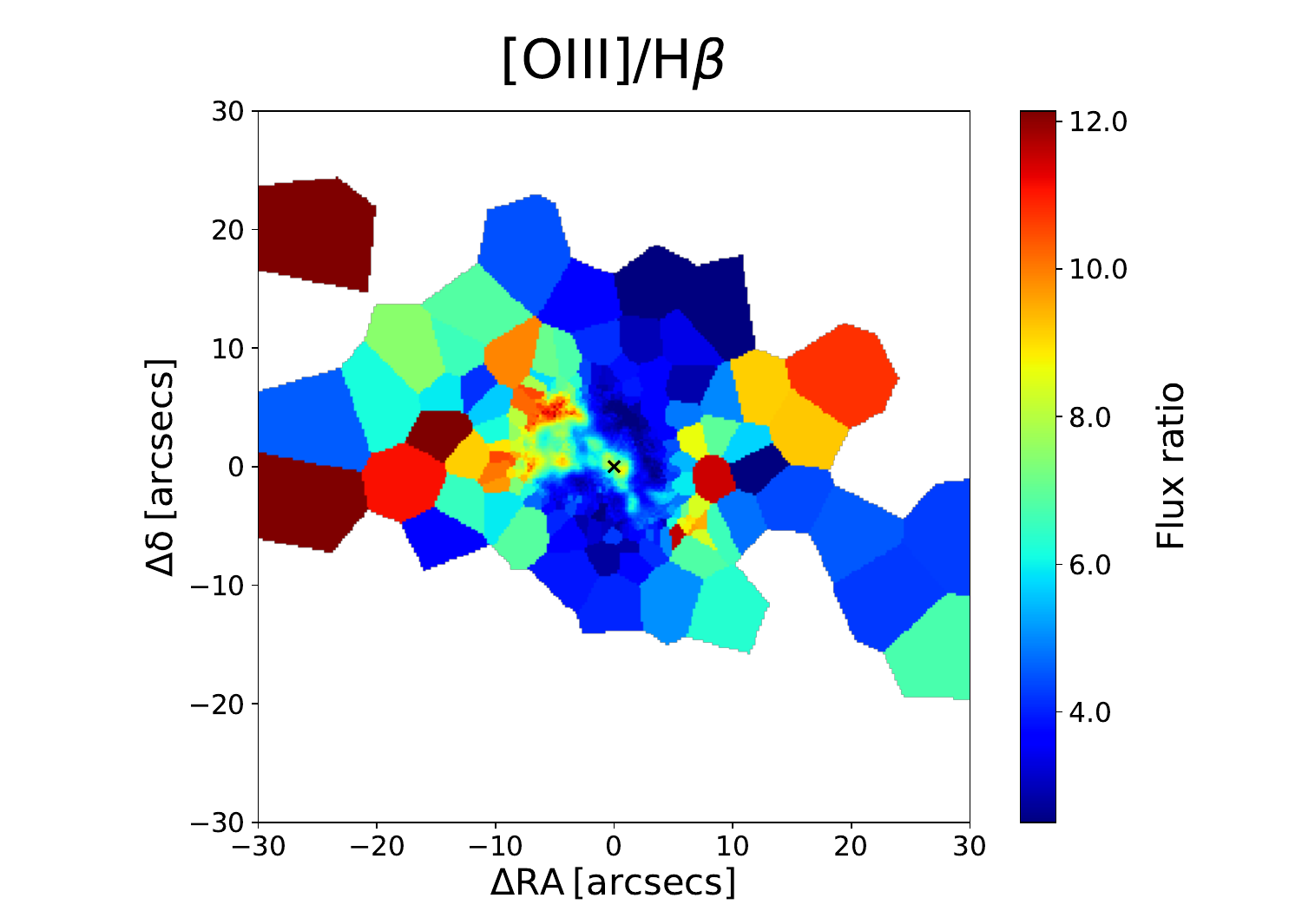}
	\hfill
    \includegraphics[scale=0.3,trim={2cm 0.5cm 2.3cm 0.5cm},clip]{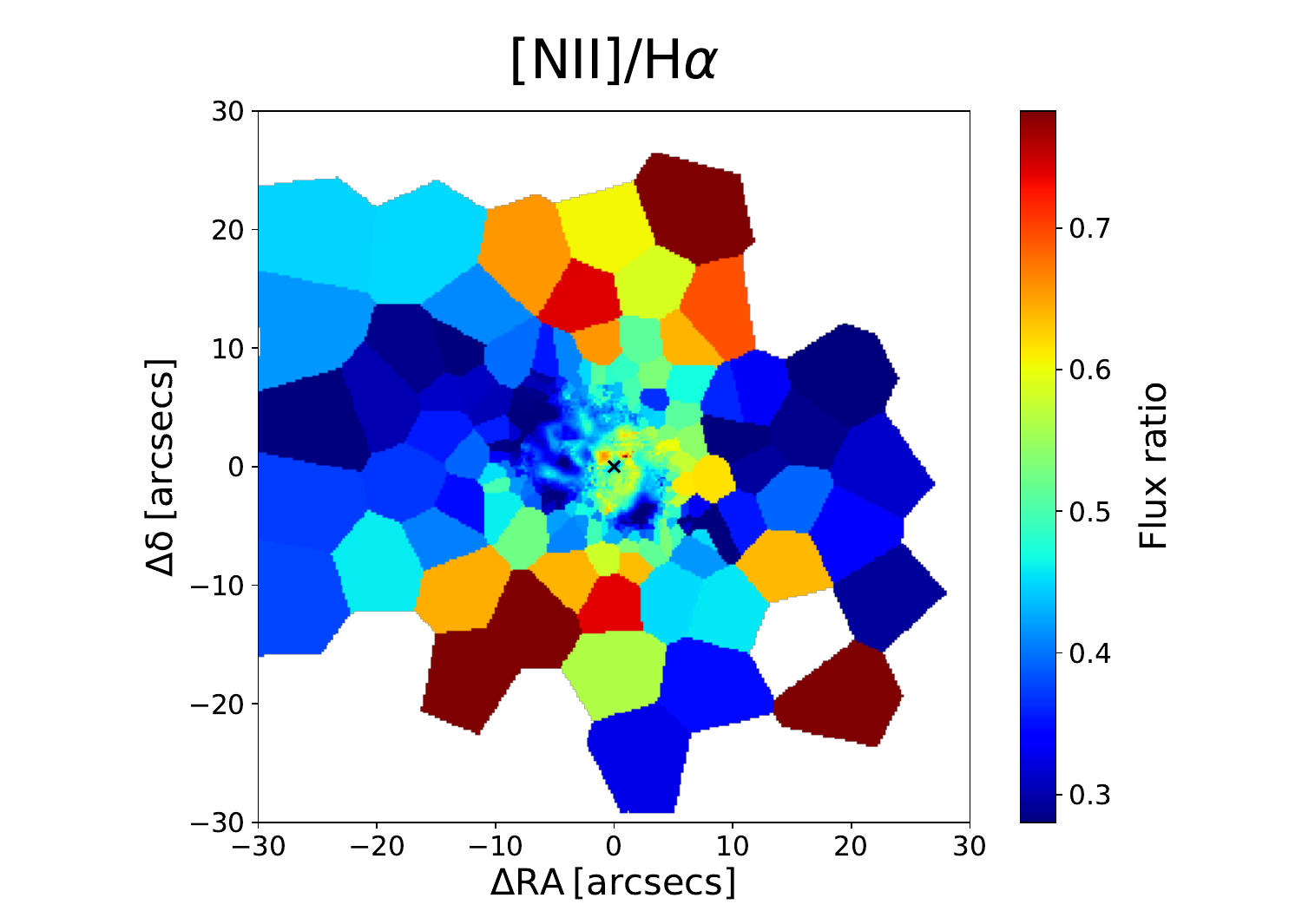}
	\hfill\null\\
    \null\hfill
    \includegraphics[scale=0.3,trim={2cm 0.5cm 2.3cm 0.5cm},clip]{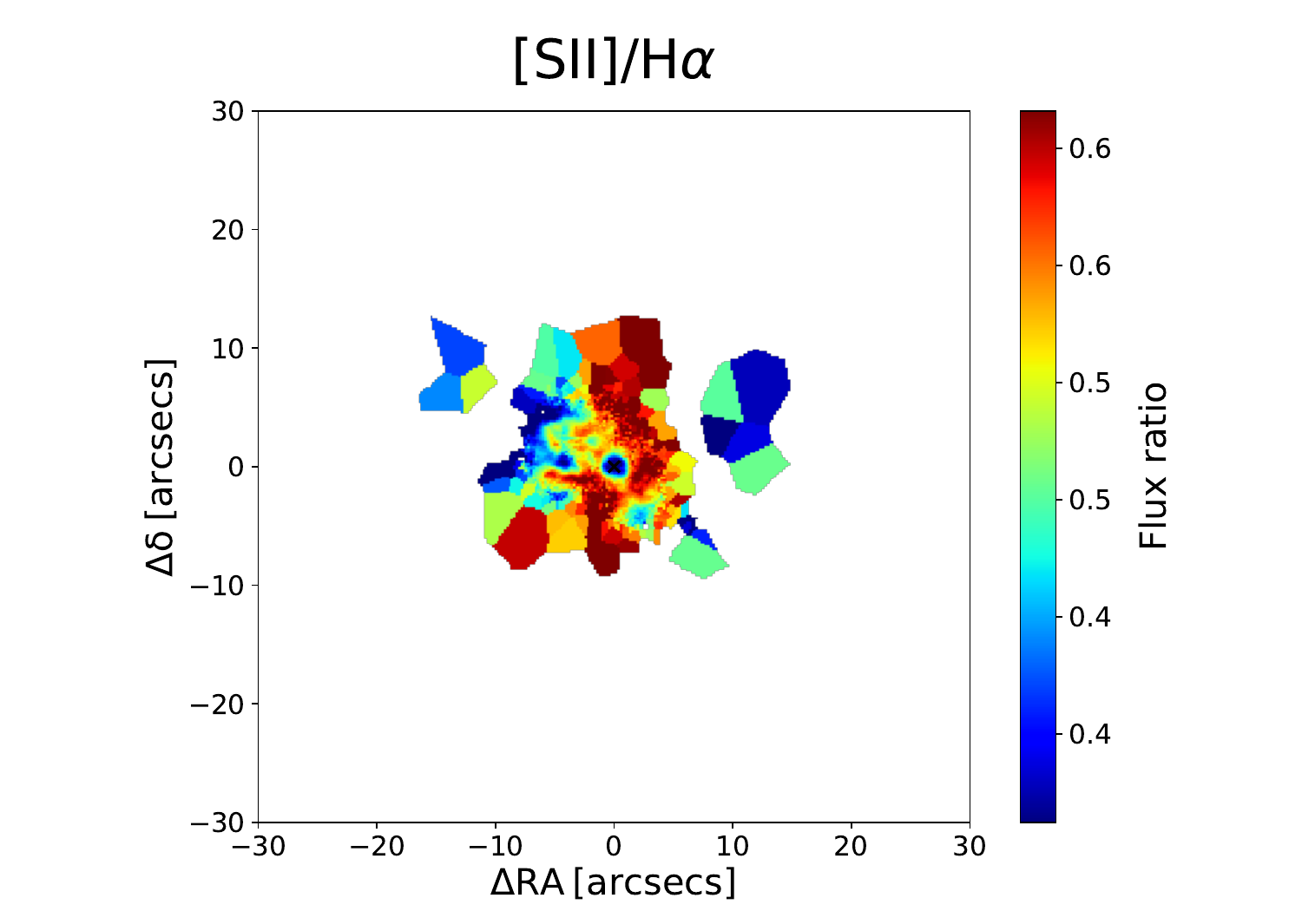}
	\hfill
    \includegraphics[scale=0.3,trim={2cm 0.5cm 2.3cm 0.5cm},clip]{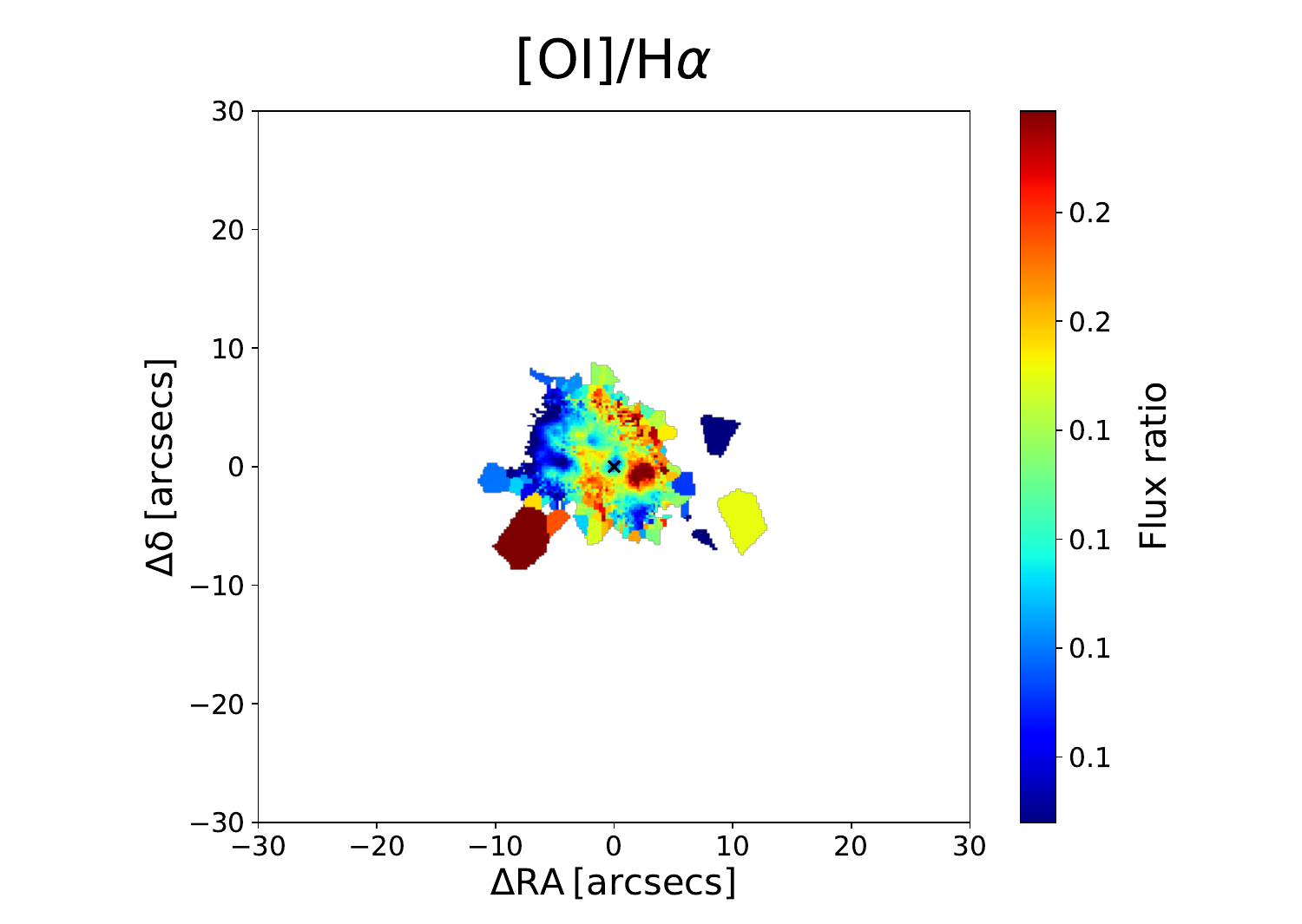}
	\hfill\null
\caption{Emission-line ratio maps obtained from the fit of the stellar continuum-subtracted, Voronoi-binned data cube (see Sect. \ref{ssec:data_anal}). \oiii/\hb\ (top-left), \nii/\ha\ (top-right), \sii/\ha\ (bottom-left), and \oi/\ha\ (bottom-right) maps for the total modelled line profiles.}
\label{fig:line_ratios}	
\end{figure*}

In Fig. \ref{fig:line_ratios}, we show the maps for the \oiii/\hb, \nii/\ha, \sii/\ha, and \oi/\ha\ line ratios, from the stellar continuum-subtracted, Voronoi-binned data cube (see Sect. \ref{ssec:data_anal}).
\oiii/\hb\ is strongest in the direction of the \oiii\ lobes, while the ratios involving low-ionisation lines (\nii, \sii, and \oi) are maximal in the perpendicular direction, where the gas velocity dispersion enhancement is observed (Sect. \ref{sec:enhanced_sigma}).

\begin{figure*}
    \centering
	\null\hfill
	\includegraphics[scale=0.3,trim={2cm 0.5cm 6cm 0.5cm},clip]{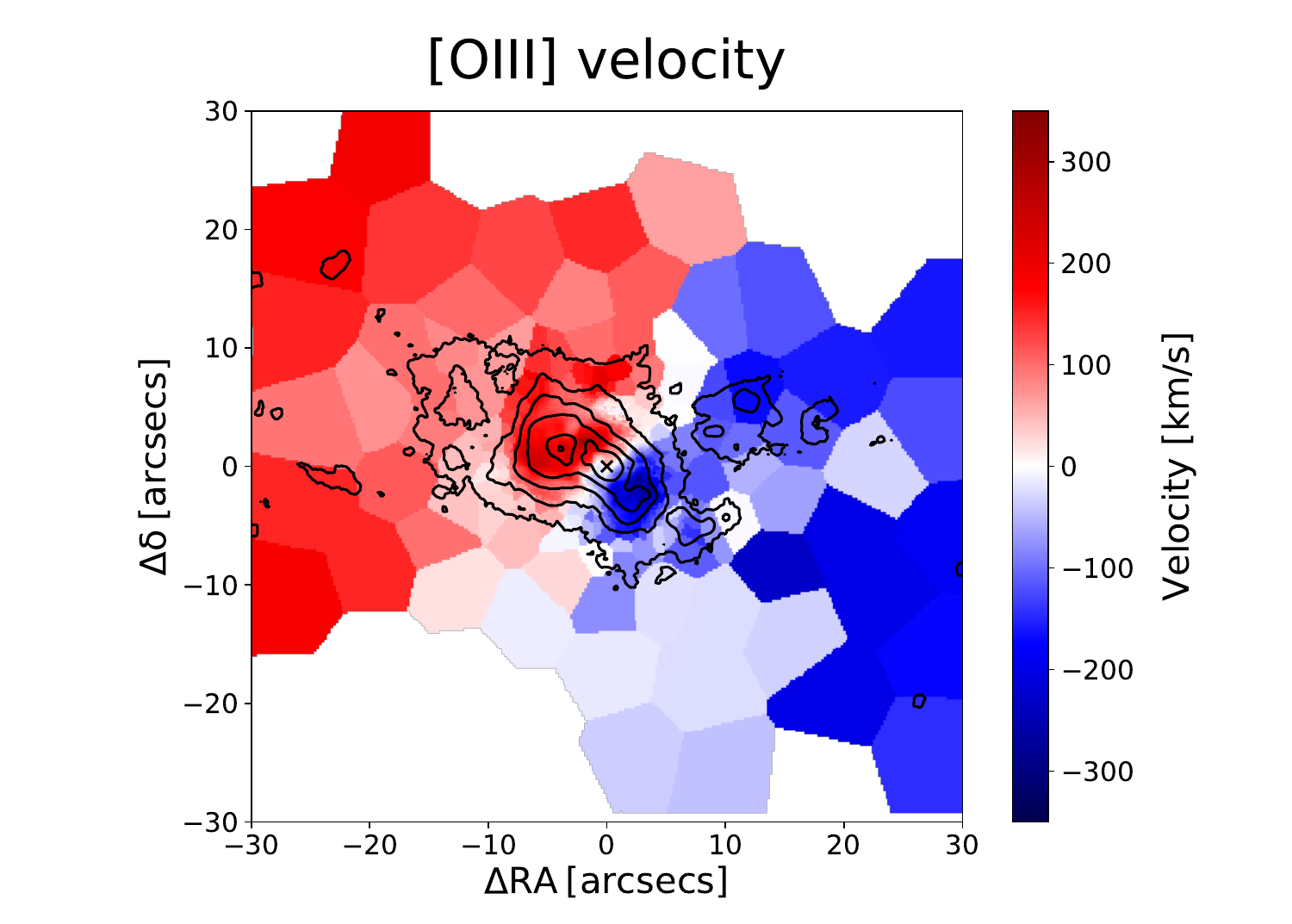}
	\hfill
    \includegraphics[scale=0.3,trim={2cm 0.5cm 6cm 0.5cm},clip]{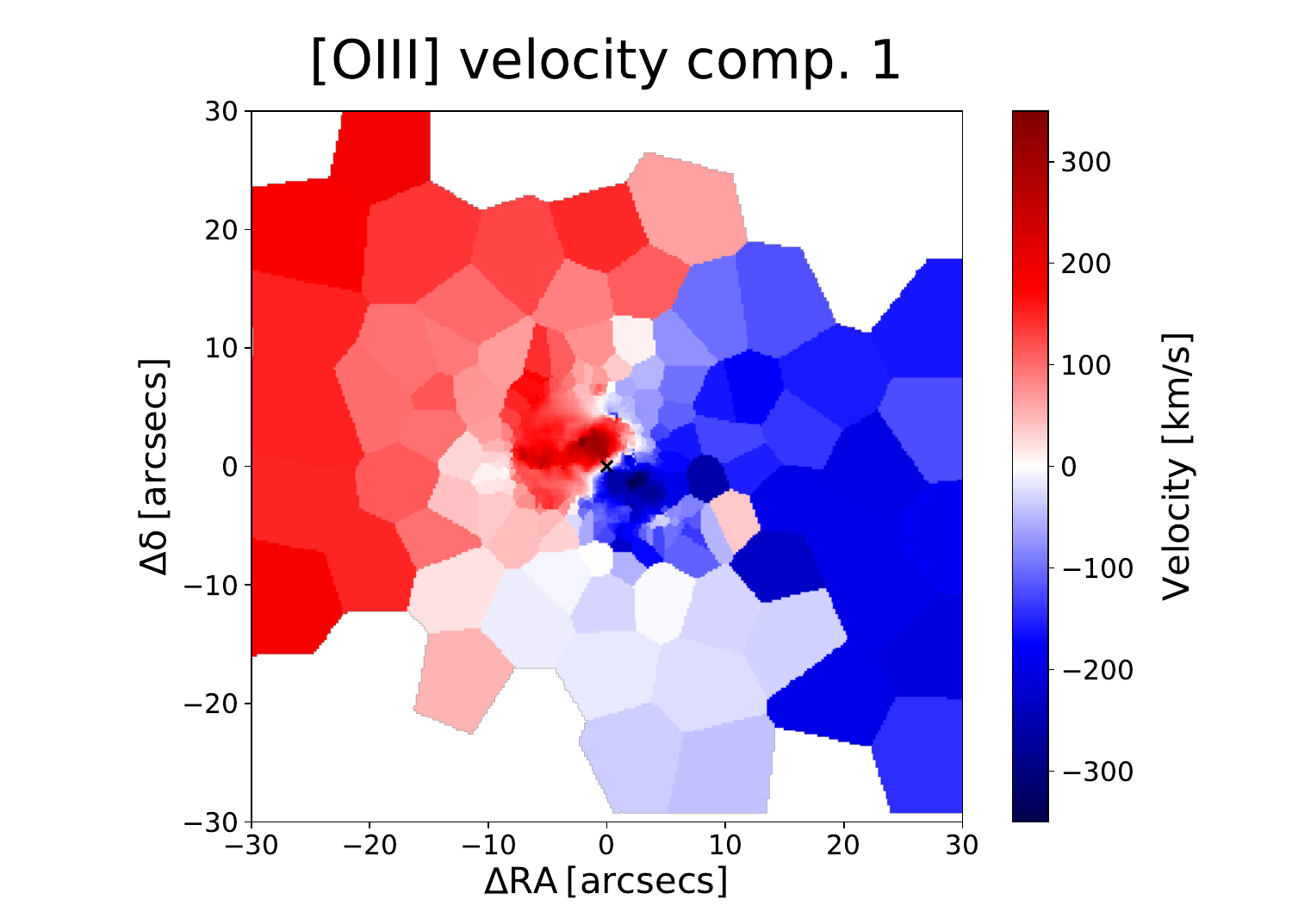}
	\hfill
    \includegraphics[scale=0.3,trim={2cm 0.5cm 1.8cm 0.5cm},clip]{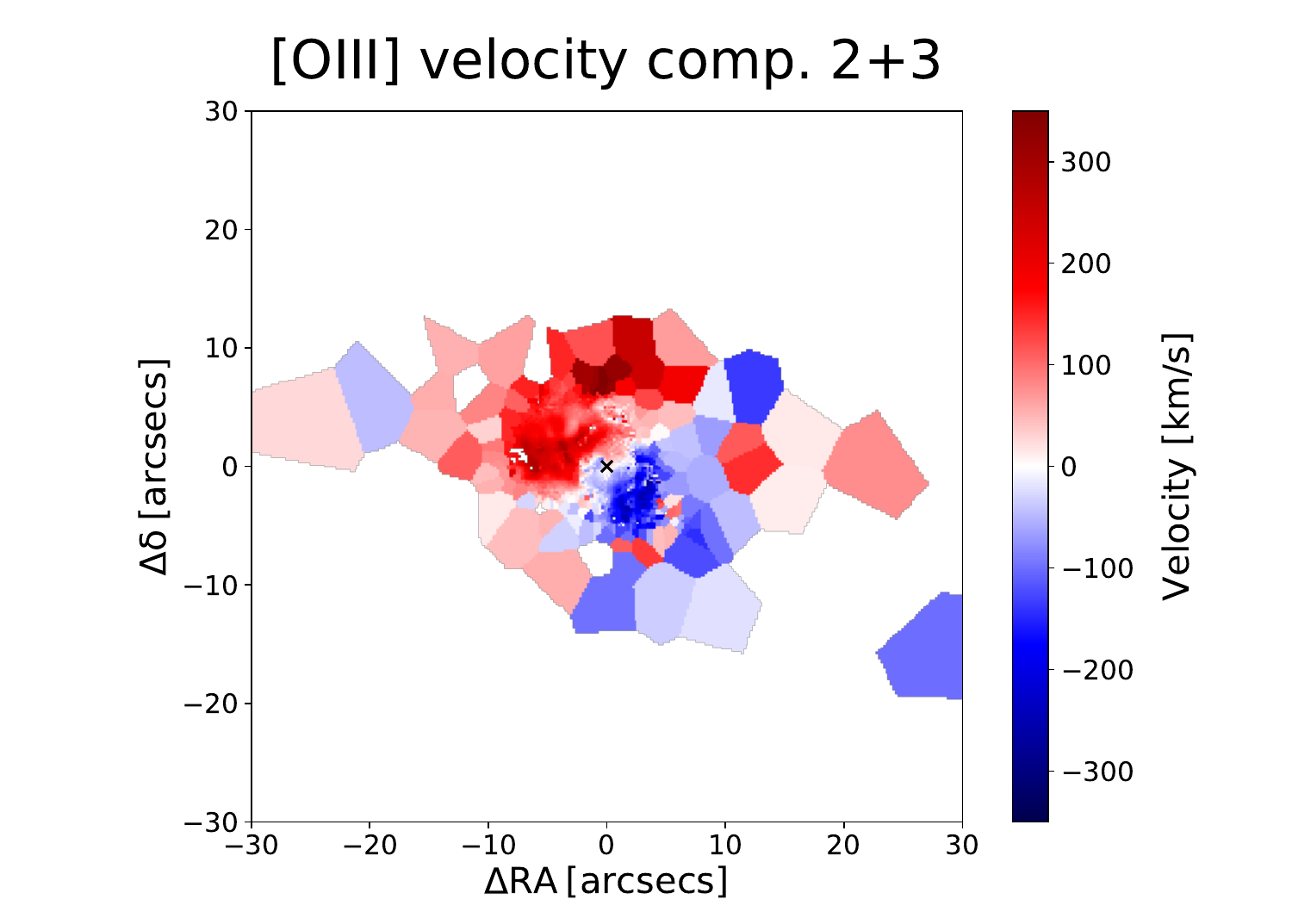}
	\hfill\null
\caption{Ionised gas velocity maps of the Teacup from the fit of the stellar continuum-subtracted, Voronoi-binned data cube (see Sect. \ref{ssec:data_anal}). \oiii\ velocity, obtained as the first-order moment of velocity of the total modelled line profile (left panel), as well as velocity of the first (middle panel) and second plus third Gaussian components (first-order moment of velocity of their summed profile; right panel) employed for the line profile modelling. {In the left panel we report the contours of \oiii\ flux (from Fig. \ref{fig:oiii_flux}, top-left panel) to allow for an easier visual comparison between the Voronoi-binned velocity field and the morphology of the ionised gas emission.}}
\label{fig:kinmaps_binned}	
\end{figure*}

Fig. \ref{fig:kinmaps_binned} displays the maps of the \oiii\ velocity for the total line profile, as well as for the first and the second plus third modelled Gaussian components, from the stellar continuum-subtracted, Voronoi-binned data cube.

\begin{figure*}
    \centering
	\hfill
	\includegraphics[scale=0.3,trim={2cm 0.5cm 2.3cm 0.5cm},clip]{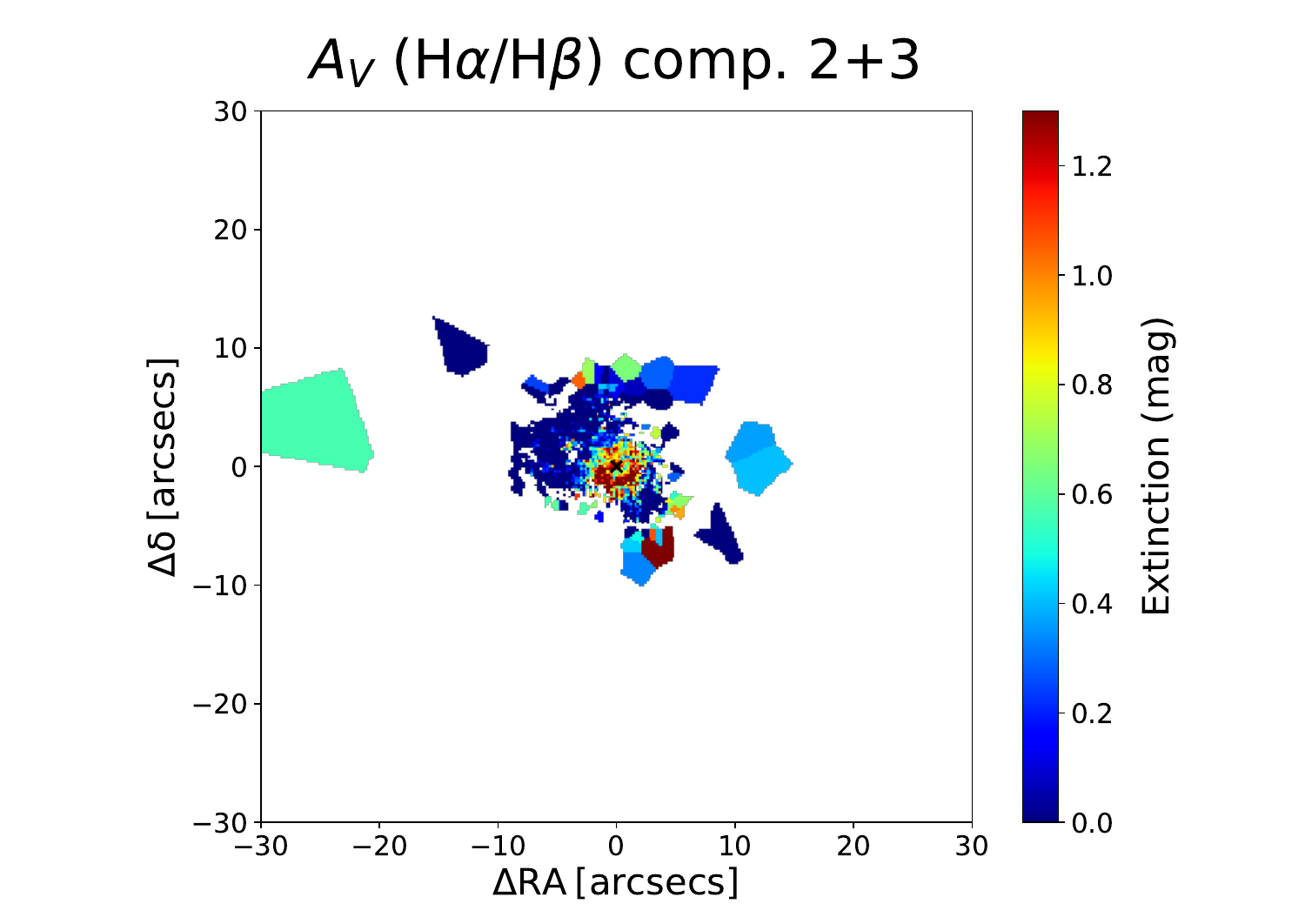}
	\hfill
    \includegraphics[scale=0.3,trim={2cm 0.5cm 2.3cm 0.5cm},clip]{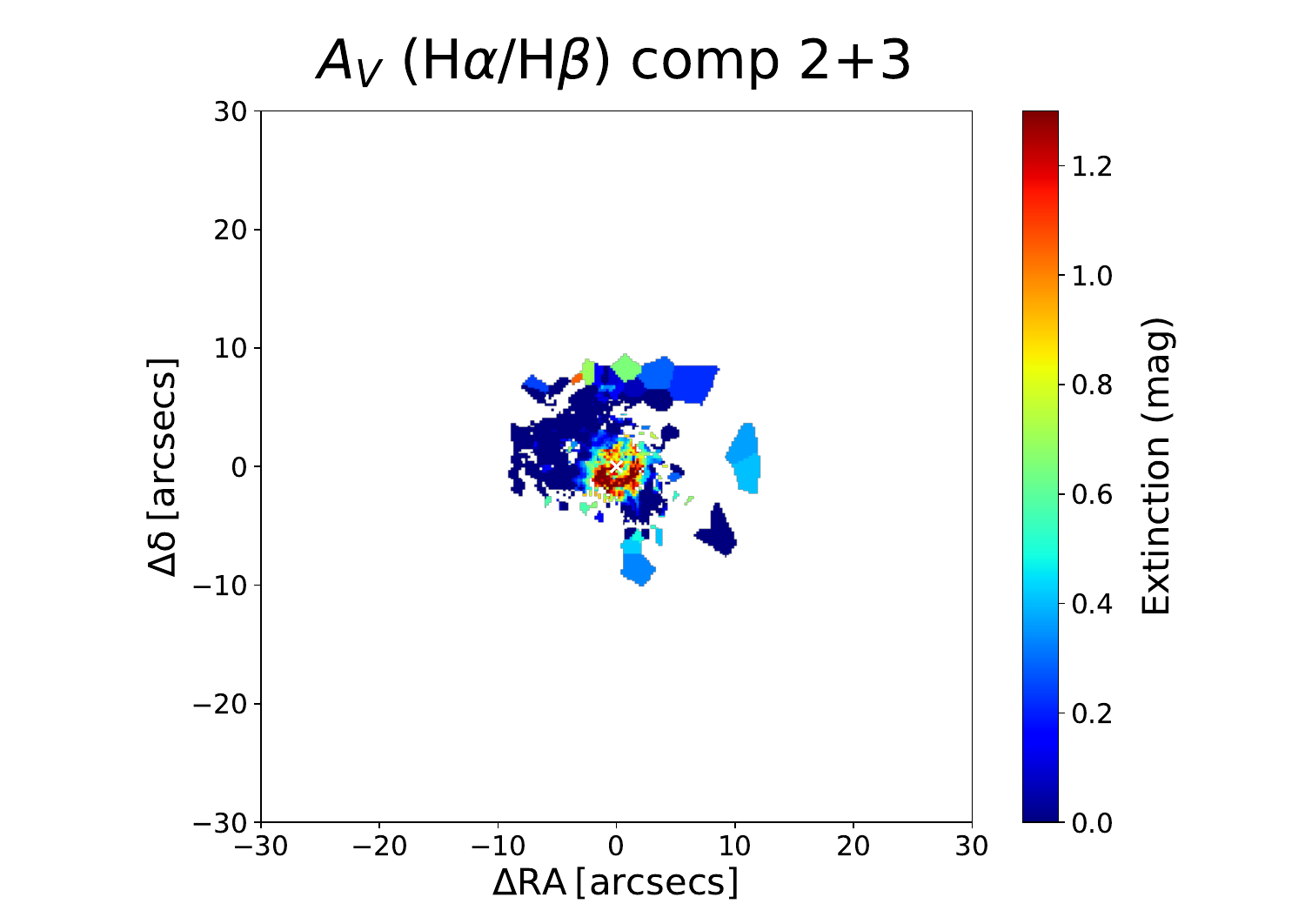}
    \hfill\null\\
    \null\hfill
    \includegraphics[scale=0.3,trim={2cm 0.5cm 2.3cm 0.5cm},clip]{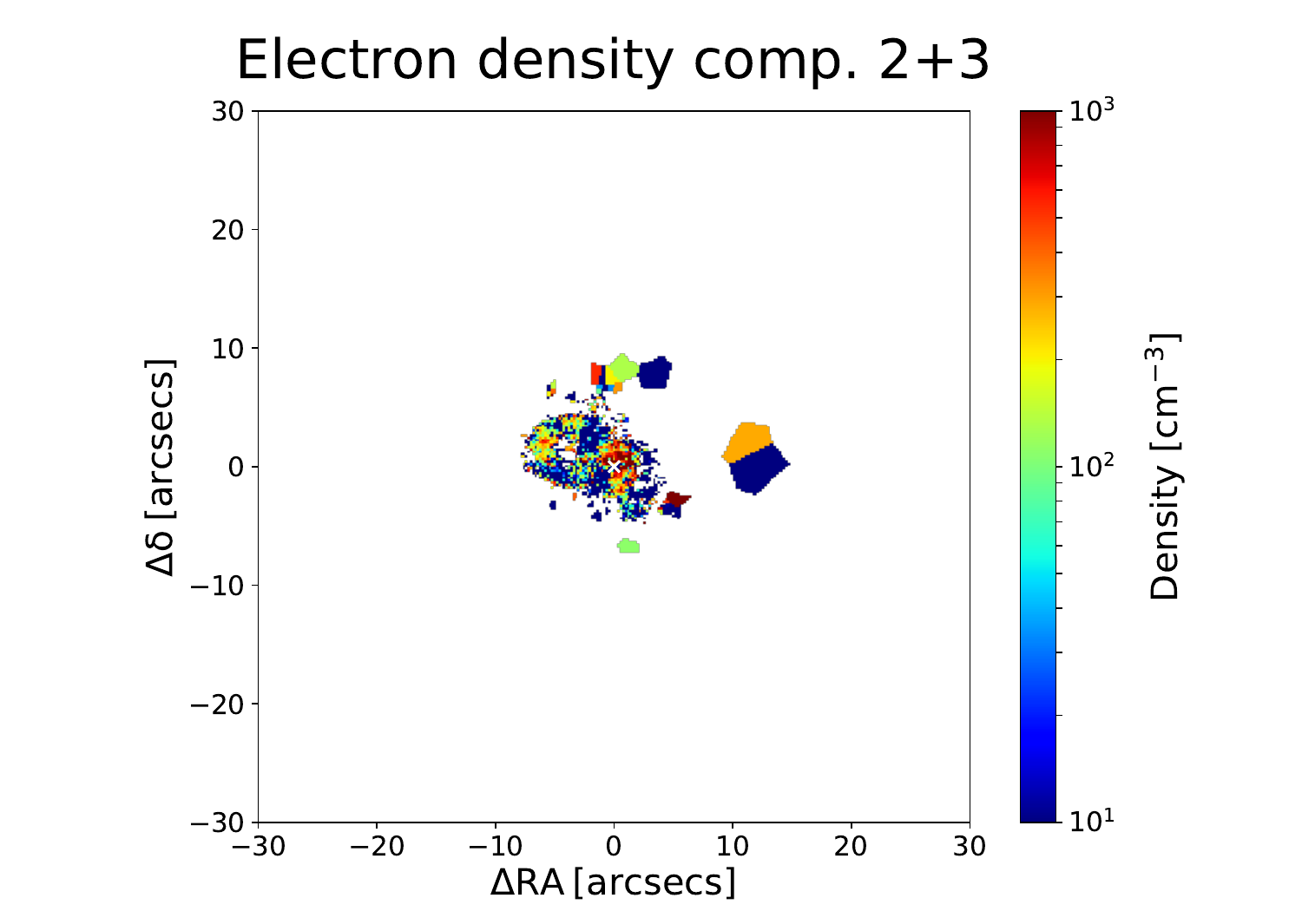}
	\hfill
    \includegraphics[scale=0.3,trim={2cm 0.5cm 2.3cm 0.5cm},clip]{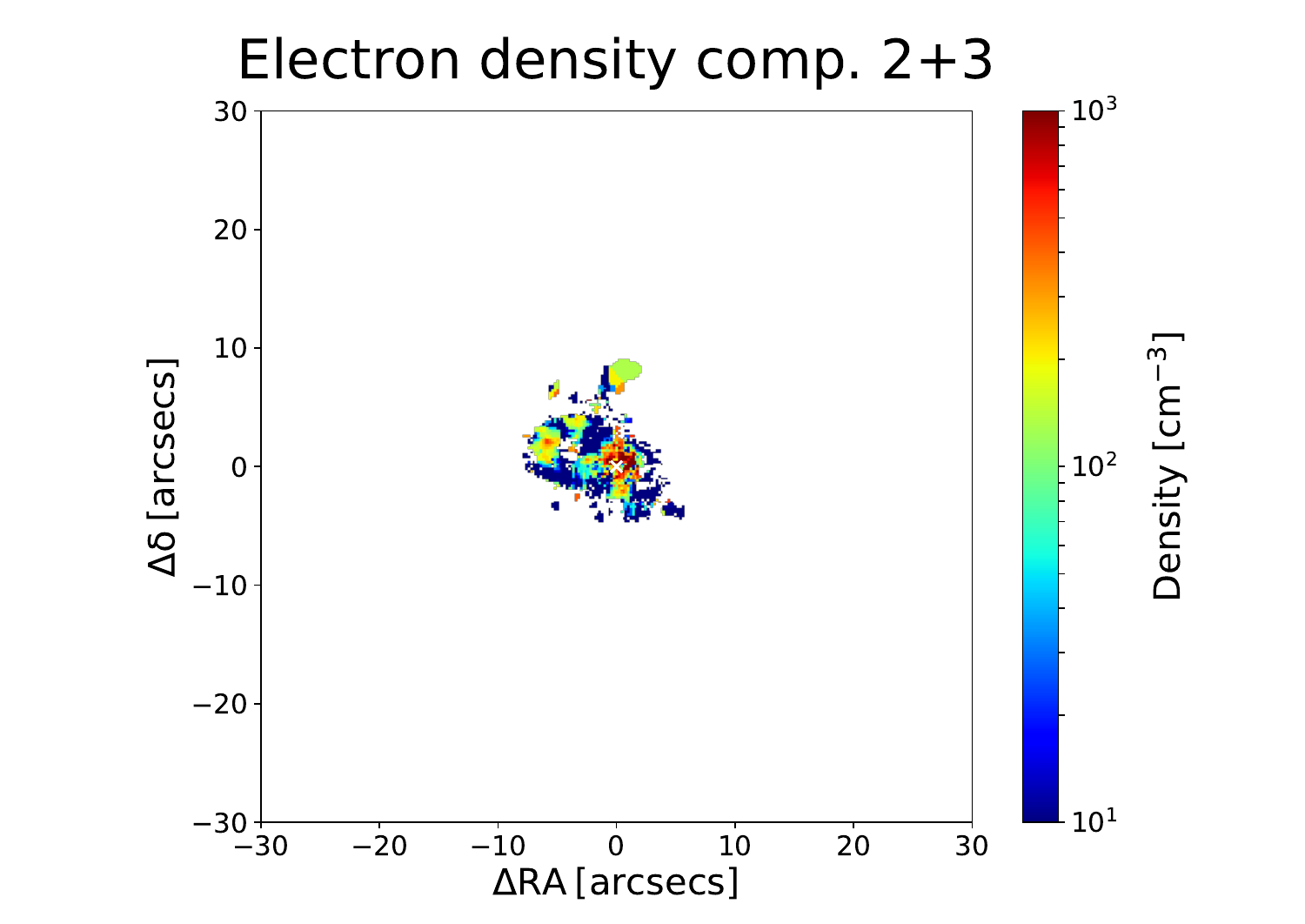}
	\hfill\null
\caption{Maps of extinction from \ha/\hb\ (top-left panel) and electron density from \sii\ ratio (bottom-left panel) of the outflow (second plus third) components from the stellar continuum-subtracted, Voronoi-binned data cube, and polished versions of these maps (top-right and bottom-right panels, respectively) used to derive the mass outflow rate and related quantities (see Sect. \ref{sec:ionised_outfl}).}
\label{fig:Avdens_c2+grid}	
\end{figure*}

Fig. \ref{fig:Avdens_c2+grid} shows the maps of extinction and electron density of the second plus third modelled Gaussian components from the stellar continuum-subtracted, Voronoi-binned data cube, together with their polished versions used to derive the mass outflow rate and related quantities (Sect. \ref{sec:ionised_outfl}).

\section{Outflow electron density radial profile} \label{sec:appendix_dens}
\begin{figure}
\centering
    \includegraphics[scale=0.8]{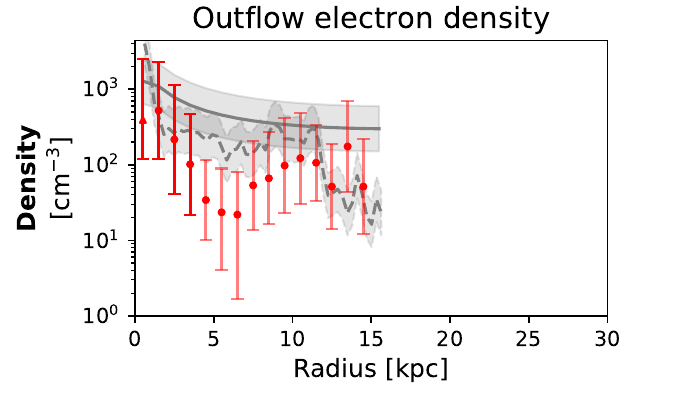}
\caption{Radial profile of the ionised outflow electron density, obtained as the mean of the outflow densities (from the \sii\ flux ratio of second plus third modelled components), weighted by the respective \sii\ ($\lambda$6716+$\lambda$6731) fluxes of the outflow component(s), in each radial bin. We stress that the reported bars are not statistical errors but a combination of the statistical errors and the standard deviation of the values to show the extent of variation of the densities in each radial bin.
{Symbols and bar shading have the same meaning as in Fig. \ref{fig:moutrate}. Solid and dashed grey lines show the electron density profiles inferred from solving Eq. \ref{eq:dens_Q+U}, by employing the $Q$(H) radial profiles from \cite{Gagne:2014a} and \cite{Keel:2017a}, respectively, with a typical uncertainty of $\pm$0.3 dex (shown as light-grey shaded regions).}}
\label{fig:outdens}
\end{figure}

In Fig. \ref{fig:outdens} we show the radial profile of the outflow electron density, calculated as the mean of the values in each radial bin, weighted by the respective fluxes of \sii\ ($\lambda$6716+$\lambda$6731) of the outflow component(s). 
We stress that this radial profile does not fully capture the spread of density values at each radial bin, and should therefore not be strictly taken as a quantitative measurement of the outflow density at each radial bin, but more like a qualitative indication of its variation with distance from the nucleus.

{In order to explore the effects of different density estimates, we compare the electron densities derived from the outflowing portions of the \sii\ emission line ratio to the hydrogen densities derived by solving the ionisation parameter equation (e.g. \citealt{Baron:2019a,Davies:2020a,Revalski:2022aa}). Specifically, we solve for the density at each radius using
\begin{equation}
n_\mathrm{H} = \dfrac{Q(H)}{4 \pi r^2 c\,U}
\label{eq:dens_Q+U}
\end{equation}
\noindent
where $Q$(H) is the number of ionising photons s$^{-1}$ from the AGN, $r$  the distance of the gas from the AGN, $U$  the ionisation parameter, and $c$  the speed of light (see \citealt{Revalski:2022aa} for details). In the case of the Teacup, $Q$(H) is a function of the distance from the nucleus because the AGN luminosity changes over time (see e.g. \citealt{Gagne:2014a,Keel:2017a}). We explore using the different $Q$(H) radial profiles derived by each of the above two studies. In the first case, we adopt $Q$(H) = $C \times 9.781 \times 10^{53}$ s$^{-1}$, where $C = 1$ in the nucleus, and increases by steps of 100/16 over the next 16 radial bins, so the last value at $r = 16$ kpc is $Q$(H) = $9.781 \times 10^{55}$ s$^{-1}$ \citep{Gagne:2014a}. In the second case, we use the variable $Q$(H) profile derived by \cite{Keel:2017a}, which is estimated based on the \ha\ emissivity of the gas.

The ionisation parameter is then approximated from our \oiii/\hb\ ratio at each radius (each obtained as the mean of the ratios at each radius, weighted by the \oiii\ flux, considering the outflow components only), which is relatively insensitive to other physical conditions, following \cite{Revalski:2022aa}.
\oiii/\hb\ decreases from $\sim$8.5 (log($U$) = --3.0) in the nucleus to $\sim$2.5 (log($U$) = --3.4) at larger radii. 
We adopt a smooth monotonic decrease of log($U$) by 0.4/16 dex in each of the 16 radial bins from the nucleus up to 16 kpc.
This direct estimate of log($U$) from \oiii/\hb, though simplistic, is in general agreement with the models of \cite{Gagne:2014a}.

Finally, we solve Eq. \ref{eq:dens_Q+U} for the hydrogen gas density using the $Q$(H) profiles from \cite{Gagne:2014a} and \cite{Keel:2017a}, and show the resulting radial profiles in Fig. \ref{fig:outdens}, after reporting them to electron densities, being $n_\mathrm{H}/n_e$ $\simeq$ 0.85 (e.g. \citealt{Crenshaw:2015aa,Revalski:2018aa}).

We see a general agreement between the various profiles at some radii, with small systematic offsets due to the different assumptions in each technique. Overall, the density profiles obtained from Eq. \ref{eq:dens_Q+U} tend to be a bit larger (by $\sim$0.5 dex on average) than that from \sii.
There is good agreement in the inner regions ($\lesssim$3 kpc), where the outflow is strongest, except in the innermost radial bin, where the outflow electron density from \sii\ was poorly constrained and was therefore not used for quantitative purposes (see Sect. \ref{sec:ionised_outfl}). We stress that, as Fig. \ref{fig:outdens} shows, differences are not systematic in the overall profile. At the same time, uncertainties of this magnitude are expected due to the assumptions made by each individual method.


We emphasise, however, that Fig. \ref{fig:outdens} must not be taken as a strict quantitative comparison between the density estimates at each radius from different techniques, but rather as an overall qualitative one, given the uncertainties and different assumptions involved in each of them. For example, even the same method (solving Eq. \ref{eq:dens_Q+U}) gives quite different densities, depending on the employed $Q$(H) radial profile (from either \citealt{Gagne:2014a} or \citealt{Keel:2017a}; solid and dashed lines, respectively).
A more detailed comparison of different gas density measurement techniques is beyond the scope of this work.}

\section{Very broad Pa$\alpha$ line emission}\label{ssec:broad_Pa}
Tentative detection of a very broad (FWHM $\sim$ 3000 km\,s$^{-1}$) Pa$\alpha$ component across the {base of the} NE bubble and the SW fan was reported by \cite{Ramos-Almeida:2017a}, {who did not detect it in the other near-IR emission lines, though.} We do not detect such {a broad} component in the optical emission lines. 
Being the flux of its peak about 10$\%$ the peak of the main body of the line, this broad component should be visible in the MUSE optical spectra, as it would be much stronger than the noise level, assuming that the above relative proportion between the broad component and the main body of the line holds also in the optical. 
{Nevertheless, the centroid velocity of these very broad components, with respect to that of the peak of the narrower component, have the same sign that we also observe for the broad components in the ionised gas, that is, blueshifted to the NE and redshifted to the SW of the nucleus (see panels 1 and 3 in Fig. \ref{fig:spectra}), which only reach FWHMs of $\sim$1300 km/s, though.
Therefore, the very broad Pa$\alpha$ components reported in \cite{Ramos-Almeida:2017a} could be real, rather than being only an instrumental artefact in the SINFONI data, but their extreme broadness could be the result of the high noisiness of the data, affecting especially the line wings, which would have artificially broadened the modelled component.}

\end{appendix}

\end{document}